\documentclass[12pt, nofootinbib]{article}

\usepackage{amssymb}
\usepackage{amsmath,bm}
\usepackage{amssymb}
\usepackage{graphicx}
\usepackage{amsfonts}         
\usepackage{fancybox}

\usepackage{enumitem}

\usepackage{slashed}

\usepackage{yfonts} 
\usepackage{epstopdf}
\usepackage[usenames,dvipsnames]{xcolor}
\usepackage[pdftex, bookmarks={false}, pdfauthor={John McGreevy}, pdftitle={Yay, physics!}]{hyperref}
\hypersetup{colorlinks=true, linkcolor=BrickRed, citecolor=Violet, filecolor=OliveGreen, urlcolor=RoyalBlue, filebordercolor={.8 .8 1}, urlbordercolor={.8 .8 0}}
\usepackage{soul}
\setstcolor{Red}

\usepackage{wrapfig}

\definecolor{darkgreen}{rgb}{0,0.4,0}
\definecolor{darkred}{rgb}{0.4,0,0}
\definecolor{darkblue}{rgb}{0,0,0.4}
\definecolor{lightblue}{rgb}{.6,.6,0.9}
\newcommand{\ttref}[1]{\textcolor{darkgreen}{\tiny{#1}}}
\newcommand{\greencom}[1]{\textcolor{darkgreen}{\footnotesize{#1}}}
\newcommand{\redcom}[1]{\textcolor{darkred}{\footnotesize{#1}}}
\newcommand{\bluecom}[1]{\textcolor{darkblue}{\footnotesize{#1}}}

\def\bl{\textcolor{darkblue}}

\def\redt{\textcolor{darkred}}

\newcommand{\cobl}{\color{darkblue}}

\newcommand{\cor}{\color{red}}
\newcommand{\cog}{\color{darkgreen}}
\newcommand{\cob}{\color{black}}

\definecolor{uglybrown}{rgb}{0.8,  0.7,  0.5}

\definecolor{palatinatepurple}{rgb}{0.41, 0.16, 0.38}
\definecolor{celebrationcolor}{rgb}{0.75,  0.0,  0.9}

 \usepackage{mdframed}

\usepackage{framed}
\definecolor{shadecolor}{rgb}{0.90,0.90,0.90}

\usepackage{wasysym}

\def\incite#1{\cite{#1}}

\def\parsnip{-.22in}

\def\parfig#1#2{
\parbox{#1\textwidth}
{\includegraphics[width=#1\textwidth]{#2}}
}

\def\subsubsubsection#1{{\bf #1}}

\def\breakS{\vfill\eject}

\newcommand{\body}{}
\newcommand{\blankpage}{}
\def\chapter[#1]#2{}

\newcommand{\titleHACK}[1]{\vbox{\center\LARGE{#1}}\vspace{5mm}}
\renewcommand{\author}[1]{\vbox{\center#1}\vspace{5mm}}
\newcommand{\address}[1]{\vbox{\center\em#1}}

\renewcommand{\date}[1]{\vbox{\center#1}}

\numberwithin{equation}{section}

\renewcommand{\theequation}{\arabic{section}.\arabic{equation}}

\def\nd{{ \vphantom{\dagger}}}

\input epsf
\newcommand{\vev}[1]{\langle #1 \rangle}

\newlength{\extraspace}
\newlength{\extraspaces}
\def\be{\begin{equation}}
\def\ee{\end{equation}}

\newcommand{\bea}{\begin{eqnarray}}
\newcommand{\eea}{\end{eqnarray}}

\def\eps{\epsilon}
\def\half{{1\over 2}}

\def\tr{{\rm tr}}

\def\Im{{\rm Im\hskip0.1em}}

\def\bra#1{\left\langle#1\right|}
\def\ket#1{\left|#1\right\rangle}

\def\vev#1{\left\langle{#1}\right\rangle}
\def\Dslash{\rlap{\hskip0.2em/}D}

\def\CA{{\cal A}}

\def\CF{{\cal F}}
\def\CG{{\cal G}}
\def\CH{{\cal H}}

\def\CL{{\cal L}}

\def\CO{{\cal O}}

\def\CT{{\cal T}}

\def\II{\relax{I\kern-.10em I}}

\def\IZ{\mathbb{Z}}
\def\IB{\relax{\rm I\kern-.18em B}}

\def\ID{\relax{\rm I\kern-.18em D}}
\def\IE{\relax{\rm I\kern-.18em E}}
\def\IF{\relax{\rm I\kern-.18em F}}
\def\IG{\relax\hbox{$\inbar\kern-.3em{\rm G}$}}
\def\IGa{\relax\hbox{${\rm I}\kern-.18em\Gamma$}}
\def\IH{\relax{\rm I\kern-.18em H}}
\def\II{\relax{\rm I\kern-.18em I}}
\def\IK{\relax{\rm I\kern-.18em K}}

\def\inbar{\,\vrule height1.5ex width.4pt depth0pt}

\def\IR{\mathbb{R}}

\def\sdtimes{\mathbin{\hbox{\hskip2pt\vrule
height 4.1pt depth -.3pt width .30pt\hskip-2pt$\times$}}}

\def\gs{g_s}
\def\lp10{\ell_p^{10}}
\def\lp11{\ell_p^{11}}
\def\R11{R_{11}}

\def\frac#1#2{{#1 \over #2}}

\def\norm#1{\vrule\hskip-1.2pt \vrule \hskip1.2pt #1  \hskip1.2pt \vrule\hskip-1.2pt \vrule}
\def\norm#1{|\hskip-1.5pt | \hskip1.5pt #1  \hskip1.5pt |\hskip-1.5pt |}

\def\Ione{\hbox{$1\hskip -1.2pt\vrule depth 0pt height 1.53ex width 0.7pt
                  \vrule depth 0pt height 0.3pt width 0.12em$}}

\def\norm#1{\vrule\hskip-1.2pt \vrule \hskip1.2pt #1  \hskip1.2pt \vrule\hskip-1.2pt \vrule}
\def\norm#1{|\hskip-1.5pt | \hskip1.5pt #1  \hskip1.5pt |\hskip-1.5pt |}

\newdimen\tableauside\tableauside=1.0ex
\newdimen\tableaurule\tableaurule=0.4pt
\newdimen\tableaustep
\def\phantomhrule#1{\hbox{\vbox to0pt{\hrule height\tableaurule width#1\vss}}}
\def\phantomvrule#1{\vbox{\hbox to0pt{\vrule width\tableaurule height#1\hss}}}
\def\sqr{\vbox{%
  \phantomhrule\tableaustep
  \hbox{\phantomvrule\tableaustep\kern\tableaustep\phantomvrule\tableaustep}%
  \hbox{\vbox{\phantomhrule\tableauside}\kern-\tableaurule}}}
\def\squares#1{\hbox{\count0=#1\noindent\loop\sqr
  \advance\count0 by-1 \ifnum\count0>0\repeat}}
\def\tableau#1{\vcenter{\offinterlineskip
  \tableaustep=\tableauside\advance\tableaustep by-\tableaurule
  \kern\normallineskip\hbox
    {\kern\normallineskip\vbox
      {\gettableau#1 0 }%
     \kern\normallineskip\kern\tableaurule}%
  \kern\normallineskip\kern\tableaurule}}
\def\gettableau#1 {\ifnum#1=0\let\next=\null\else
  \squares{#1}\let\next=\gettableau\fi\next}

 \def\eqnn#1{\xdef #1{(\secsym\the\meqno)}\writedef{#1\leftbracket#1}%
 \global\advance\meqno by1\wrlabeL#1}
 \def\eqna#1{\xdef #1##1{\hbox{$(\secsym\the\meqno##1)$}}
 \writedef{#1\numbersign1\leftbracket#1{\numbersign1}}%
 \global\advance\meqno by1\wrlabeL{#1$\{\}$}}
 \def\eqn#1#2{\xdef #1{(\secsym\the\meqno)}\writedef{#1\leftbracket#1}%
 \global\advance\meqno by1$$#2\eqno#1\eqlabeL#1$$}

\global\newcount\itemno \global\itemno=0

\def\itemaut#1{\global\advance\itemno by1\noindent\item{\the\itemno.}#1}

\def\bfe{{\bf e}}

\def\({\left(}
\def\){\right)}

\def\sidebarb{\color{darkblue}
\vskip.15in
\hrule 
\vskip.05in
}

\def\sidebare{
\vskip.1in
\hrule 
\vskip.05in
\cob
}

\def\grad{\vec \nabla}
\def\vB{\vec B}
\def\vE{\vec E}

\def\cc{{\bf c}}

\def\ii{{\bf i}}

\def\dd{\text{d}}

\def\aa{{\bf a}}

\def\AA{{\bf A}}
\def\BB{{\bf B}}

\def\HH{{\bf H}}
\def\KK{{\bf K}}

\def\UU{{\bf U}}

\def\WW{{\bf W}}
\def\rrho{{\boldsymbol\rho}}
\def\ssigma{{\boldsymbol\sigma}}
\def\ttau{{\boldsymbol\tau}}

 \def\XX{{\bf X}}
 \def\ZZ{{\bf Z}}

\def\DDD{\boldsymbol{\Delta}}

\def\lsim{\mathrel{\mathstrut\smash{\ooalign{\raise2.5pt\hbox{$<$}\cr\lower2.5pt\hbox{$\sim$}}}}}
\def\gsim{\mathrel{\mathstrut\smash{\ooalign{\raise2.5pt\hbox{$>$}\cr\lower2.5pt\hbox{$\sim$}}}}}

\def\overleftrightarrow#1{\vbox{\ialign{##\crcr
     $\leftrightarrow$\crcr\noalign{\kern-0pt\nointerlineskip}
     $\hfil\displaystyle{#1}\hfil$\crcr}}}
     
     \def\overleftarrow#1{\vbox{\ialign{##\crcr
     $\leftarrow$\crcr\noalign{\kern-0pt\nointerlineskip}
     $\hfil\displaystyle{#1}\hfil$\crcr}}}

\def\eg{{\it e.g.}}
\def\ie{{\it i.e.}}

\def\gG{\textsf{G}}

\def\gSU{\textsf{SU}}
\def\gU{\textsf{U}}

\def\gs{\text{gs}}

\usepackage[yyyymmdd,hhmmss]{datetime}
\newif{\ifeq}           
\eqtrue                 
\newcounter{lecturecounter}

\makeindex

\begin{document}


\chapter[Quantum matter]{TASI lectures on quantum matter \\  (with a view toward holographic duality)}
\titleHACK{TASI lectures on quantum matter \\ (with a view toward holographic duality)}

\author{John McGreevy}


\address{Department of Physics\\
University of California at San Diego\\
{\tt mcgreevy@physics.ucsd.edu}}

\begin{abstract}
These are notes from my lectures at TASI 2015.
The goal is to provide context for
the study of strongly-correlated quantum many-body systems
using quantum field theory, and possibly string theory.
\end{abstract}


\renewcommand{\contentsname}{}

\setcounter{tocdepth}{2}    
\tableofcontents

\body




\vfill\eject

My assignment in these lectures is to speak about applications
of AdS/CFT to condensed matter physics.
As the lectures proceed, it may possibly come to your attention
that they do not contain so very much discussion
of strings and gravity and AdS and the bulk and D-branes and things like that.
I assure you, however, that I am not shirking my responsibilities.  
I propose that a useful (if criminally immodest) comparison to what is going on here is 
the following.
Consider the answer given at a congressional hearing by Robert R.~Wilson
(founder of Fermilab) when he was asked what 
good is particle physics for the defense of the country.  
His answer: 
{\it It has nothing to do directly with defending our country,  except to make it worth defending.}

So to mix all my metaphors into a big pot here: 
Most of what  I have to say here 
has nothing to do directly with attacking 
condensed matter physics with holography,
except to make it worth attacking.

\vskip.1in
\hrule
\vskip.1in

{\bf Context.}  In these lectures we're going to think about 
(holographic perspectives on)
physical systems with {\it extensive} degrees of freedom.

\hskip-.23in
\parbox{.6\textwidth}{
To illustrate this definition, 
consider the picture at right.
Assume space is covered in patches labelled $x$.  
Define the system in two steps.
Spread some quantum sod over these patches:
the Hilbert space is $\CH = \otimes_x \CH_x $.
Each patch has 
some small-dimensional 
Hilbert space $\CH_x$, such as a single qbit (a two-state system).
}~~\parfig{.4}{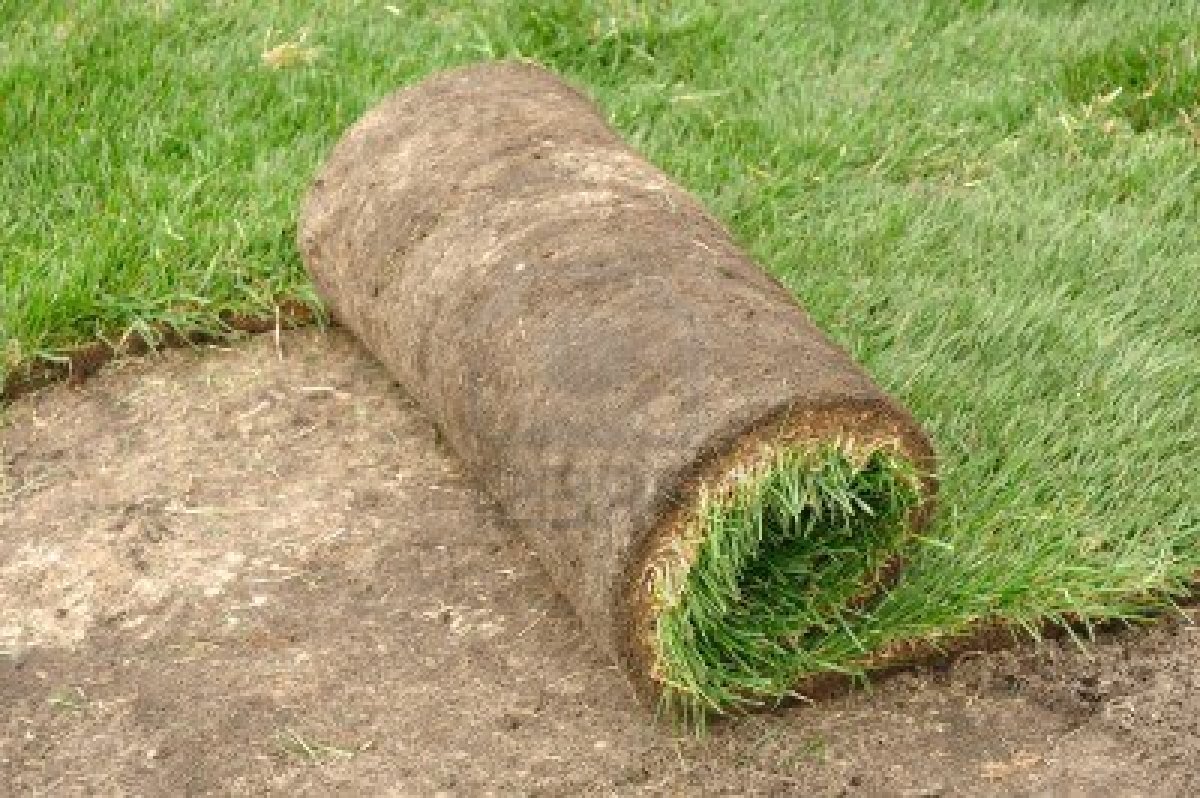}

Next, couple together the patches of sod:
the Hamiltonian is $ H = \sum_x H_x$.  
The support of $H_x$, the `hamiltonian motif,'  is localized near $x$:
$H_x$ acts as the identity operator far from $x$.
This motivic structure rules out many horrible pathologies.

We'll also assume an IR (infrared) cutoff -- put the whole system in a big box of linear size $L$
(if you force me, I'll pick periodic boundary conditions, but that doesn't fit so well with my landscaping metaphor).


A   more accurate name 
might be {\bf Regulated QFT} (quantum field theory).
Condensed matter provides many realizations of such systems, 
but (alas) not all condensed matter physics is QFT\footnote{We will
discuss a necessary condition for the applicability of QFT below,
as well as some (exotic!) models which fail this criterion.}.
This includes QFT and also lattice models, classical fluids, and many other interesting systems.
Sometimes such systems can be understood directly 
in terms of some weakly interacting particle picture
\bluecom{(aka normal modes)}.
(An example is $D=3+1$ quantum electrodynamics in its Coulomb phase.)
There are many interesting systems 
for which such nearly-gaussian variables are not available.  Then what do we do?  Maybe ...

{\bf Holography.}  
Holographic duality is a wonderful discovery 
\cite{1999IJTP...38.1113M, 1998PhLB..428..105G, 1998AdTMP...2..253W}
which (in a certain regime)
solves certain strongly-interacting quantum field theories
in terms of simple classical field theories in one higher dimension.
This sounds like a powerful tool which we should try to exploit 
to solve some strong-coupling problems.
Which ones can we solve this way?

\sidebarb
{\bf Limitations of classical holography.}  
I am not going to review the bottom-up approach to holographic duality here; 
for that, see \cite{McGreevy:2009xe}.  For our present purposes,
the most important fact is: the classical gravity limit is large $N$.
Let's remind ourselves why this is inevitably so.

The holographic principle
says that 
in a system with gravity, the maximum entropy 
that we can fit in some region of space is 
proportional to the area of the boundary of that region.

\hskip-.23in
\parbox{.7\textwidth}{
The maximum entropy is in turn proportional to the number of degrees of freedom.
The maximum entropy of our regulated QFT above 
is the log of the number of states in $\CH$; this is 
proportional to the number of sites (patches) --  namely 
$$S_\text{max} = \log  \dim \(  \CH_x^{\({L\over \varepsilon}\)^d } \) =  \( {L\over \varepsilon}\)^d \log \dim \CH_x~.
$$ ($d$ is the number of space dimensions, and $\varepsilon$ is the lattice spacing, 
and in the figure I've called the {\it log} of the local Hilbert space dimension {\color{darkred} $N^2$}.)
}~~~~
\parfig{.2}{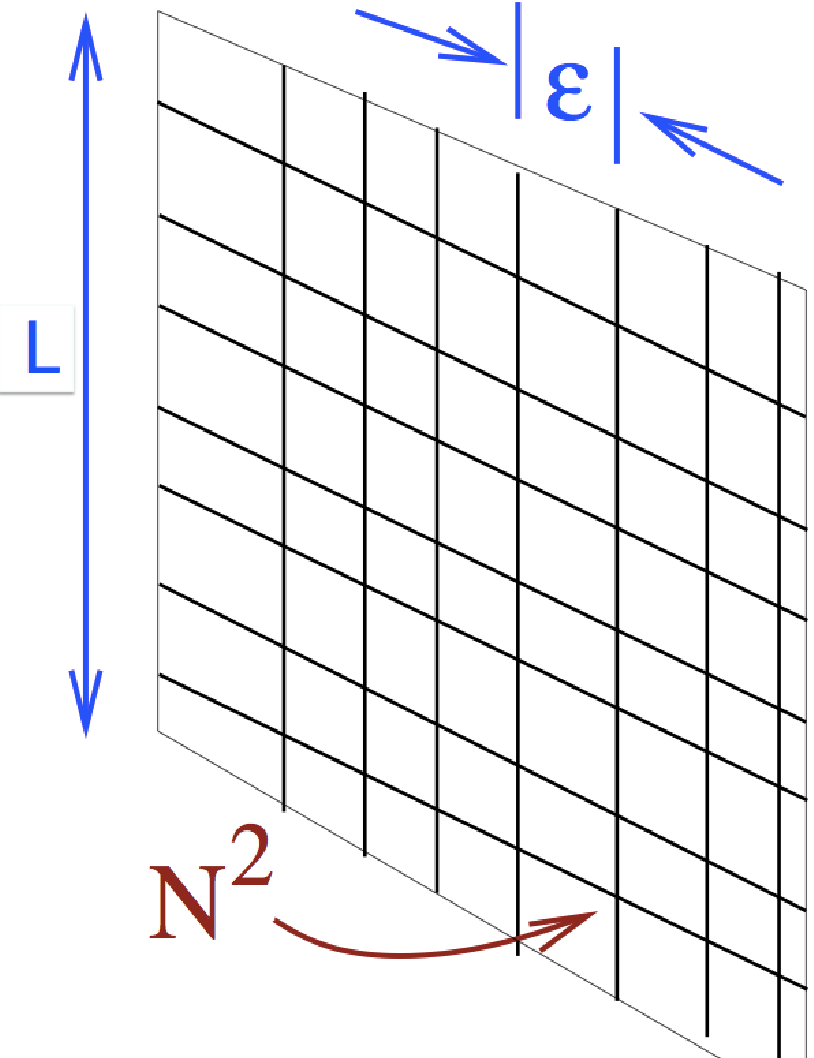}

\begin{shaded}
{\bf Holographic counting.}  In this gray box, we sequester a review of the holographic counting of degrees of freedom,
in a conformal field theory with a classical gravity dual.
The holographic principle says that the maximum entropy 
of a state of a system with Einstein gravity is the area of the boundary in Planck units $ S = {A \over 4 G_N}$.
At fixed time, the $AdS$ metric is : $ds_{AdS}^2 = L_{AdS}^2 { dz^2 + d\vec x^2 \over z^2} $,
so the area of the boundary is 
$$ A = \int_{bdy,~ z~ fixed} \sqrt g d^{d-1} x = \int_{\IR^{d-1}} 
d^{d-1} x 
\({L_{AdS}\over z}\)^{d-1} |_{z\to 0} $$

\parbox{.6\textwidth}{
Imposing an IR cutoff $L$ by $x \equiv x+L$ 
and a UV cutoff $z > \eps$, this is 
$$A =\int_0^L d^{d-1}x{L_{AdS}^{d-1} \over z^{d-1}} |_{z=\epsilon}  =  \left({L L_{AdS}\over \epsilon}\right)^{d-1}$$}
~~
\parfig{.3}{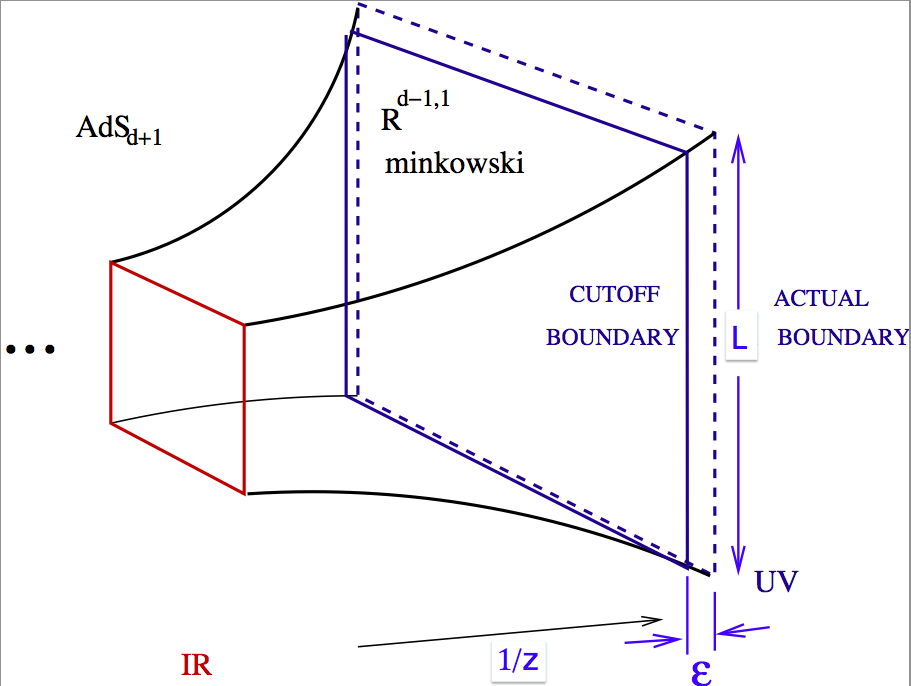}

Therefore, $\displaystyle{ {A \over 4 G_N} \sim  {L_{AdS}^{d-1} \over 4 G_N}  \left({L \over \epsilon}\right)^{d-1} }$.
Combining this with the counting in the extensive quantum lattice model, we must identify
\be\label{eq:large-N-disaster}
 \boxed{{L_{AdS}^{d-1} \over G_N}  = \redt{N^2} }   ~~~~\dim \CH_x \equiv e^{\redt{N^2}}.\ee
Recalling that $G_N$ plays the role of $\hbar$ in the Einstein action, $S \sim {1\over G_N } \int \sqrt{g} R$, we learn that 
gravity is classical if the dual QFT has lots of degrees of freedom (dofs) per point:  $\redt{N^2} \gg 1$.
Thinking of classical physics as the saddle point of a putative path integral, 
the essential holographic dictionary is 
$$
 Z_{QFT}[{\rm sources}] 
  \approx e^{ - \textcolor{darkred}{N^2} 
 S_{{\rm bulk}}[{\rm boundary~conditions~at~}r\to \infty]}|_{{\rm extremum~of~}S_{{\rm bulk}}} 
 $$
\begin{center}
classical gravity \greencom{(sharp saddle)} $\leftrightsquigarrow$ many dofs per point, $\textcolor{darkred}{N^2} \gg 1$
\end{center}

\end{shaded}

This means that holographic systems 
(with classical gravity duals) are twice in the  thermodynamic limit!
Once for large volume and once at each point in space separately.  This is really weird!
This means that it is possible to have sharp phase transitions in finite volume 
(\eg~\cite{Wadia:1980cp, Gross:1980he, Headrick:2010zt} and many many others), 
and to break continuous symmetries in two spacetime dimensions \cite{Anninos:2010sq}.

So: which condensed matter questions might be most usefully approached using holographic duality?
The large-$\color{darkred}N$ problem means that 
the systems which we can solve with this wonderful tool are not microscopically realistic
models of any material.
Because of this, my answer to the above question is: 
the ones where ordinary techniques fail the most desperately.

\sidebare


\section{States of matter, classified by level of desperation}
\label{subsec:intro}


Perhaps the most basic question we can ask about such a system is: 
how many degrees of freedom are there at the lowest energies (lower than any interesting scale in the problem, in particular in the Hamiltonian)?    
There are essentially three possibilities:
\vskip-.5in
\begin{enumerate}
\item {\bf None.}
\item {\bf Some.}
\item {\bf A lot.}
\end{enumerate}
A more informative tour through that list goes like this.  First let me make the 
assumption that the system has (at least discrete) translation invariance, so we can label the excitations by momentum.
(Relaxing this assumption is very interesting and I plan to do so in the last lecture.)
\begin{enumerate}
\item {\bf None:}  Such a system has an energy gap (`is gapped'): the energy difference $\Delta E = E_1 - E_0$ between 
the first excited state and the groundstate is nonzero, even in the thermodynamic limit.
Note that $\Delta E$ is almost always nonzero in finite volume.
(Recall, for example, the spectrum of the electromagnetic field in a box of linear size $L$: $E_n \sim {n \over L} $.)
The crucial thing here (in contrast to the case of photons) is that this energy stays finite even as $L \to \infty$.

The excitations of such a system are generally massive particles\footnote{Ref.~\incite{2013PhRvL.111h0401H}
proves a version of this statement.  I think it is worth thinking about loopholes here.}.

\item {\bf Some:} An example of what I mean by `some' is that the system can have 
excitations which are massless particles,
like the photon.  

The lowest energy degrees of freedom occur at isolated points in momentum space: 
$ \omega(k) = c \sqrt{\vec k\cdot \vec k} $ vanishes at $ \vec k=0$.

In this category I also put the gapless fluctuations at a critical point; 
in that case, it's not necessarily true that $ \omega \sim k^\text{integer}$ 
and those excitations are not necessarily {\it particles}.
But they are still at $k=0$\footnote{or some other isolated points in momentum space.}.  

\item {\bf A lot:}  What I mean by this is Fermi surfaces,
but importantly, not just free fermions or adiabatic continuations of free fermions (Landau Fermi liquid theory).  
Such systems exist, for example
in the half-filled Landau level and in the strange metal regime of cuprate superconductors.

\end{enumerate}

Let's go through that list one more time more slowly.

\hskip-.23in
\parbox{.65\textwidth}{Let's reconsider the case of gapped systems.
Different gapped states are different if we can't deform the hamiltonian to get 
from one to the other without closing the gap.}
\parfig{.29}{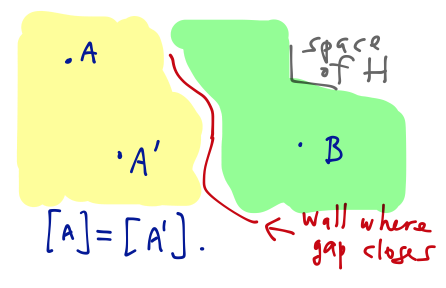}

\footnote{Note that the closing of the gap
does not by itself mean a quantum critical point: at a first order transition, just the lowest two levels
cross each other.}You might be bothered by the the fact that it is hard to imagine checking that there is no way around the wall.
It is therefore important to find sharp characterizations of such states,
like integer labels,
which cannot change smoothly.
This is the very definition of topology.
An important goal in condensed matter physics is to 
figure out labels that can be put on states which can distinguish them in this way
as distinct phases of matter.

Even the lowest-energy (even below the gap) physics of gapped systems can be deeply fascinating.
Such a thing is a (unitary) topological field theory: it is a theory of groundstates,
and it can provide a way to distinguish states of matter.
For example, it may be that the {\it number} of groundstates depends on the topology of the space
on which we put the system.  This phenomenon is called topological order.

Another (distinct!) possibility is that even if the system in infinite space has an energy gap, 
if we cut the space open, new stuff can happen; 
for example there may be gapless edge modes.  

\hskip\parsnip
\parbox{.8\textwidth}{
Both of these phenomena happen in quantum Hall systems.
A reason to think that an interface between the vacuum and a gapped 
state of matter which is distinct from the trivial one
might carry gapless modes
is that the couplings in the hamiltonian are forced to pass through
the wall where the gap closes.  
(In fact there are important exceptions to this conclusion, which you can learn about 
\eg~here \cite{Dennis:2001nw, Kitaev-2011}.)
}\parfig{.18}{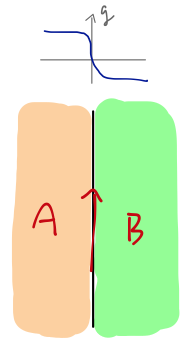}

An energy gap (and no topological order or special edge modes) should 
probably be the generic expectation for what happens if 
you pile together a bunch of degrees of freedom and couple them in some haphazard
(but translation invariant!) way.
At the very least this follows on general grounds of pessimism: if 
you generically got something interesting by doing this,
physics would be a lot easier (or more likely: we wouldn't find it interesting anymore).
{\bf Gaplessness is something special that needs to be explained.}\footnote{Why is this?  
Even a finite $\CO(L^0)$ degeneracy of the groundstate needs to be explained.  This is because 
if our hamiltonian $H_0$ happens to annihilate two states $\ket{\psi_{1,2}}$ 
(set the groundstate energy to zero), 
and we perturb $H$ with any $\Delta H$ such that $ \bra{ \psi_1 } \Delta H \ket{ \psi_2} $, 
the degeneracy will be split.  For example, if $\psi_{1,2}$ are related by a symmetry,
and we only allow symmetric perturbations, then the matrix element will be forced to vanish.
This is a toy model of Goldstone's theorem: the groundstate degeneracy is swept out
by the orbit of the broken symmetry.
}
Here is a list of some possible reasons for gaplessness (if you find another, you should write it down):
\begin{enumerate}
\item \label{item:goldstones} broken continuous symmetry (Goldstone bosons)
\item tuning to a critical point -- notice that this requires some agent to do the tuning, 
and will only occur on some subspace of the space of couplings
of nonzero codimension.
\item \label{item-gauge} continuous unbroken gauge invariance (photons)\footnote{According to 
\cite{Gaiotto:2014kfa}, this is a special case of item \ref{item:goldstones} 
for `one-form symmetries'.}
\item Fermi surface (basically only in this case do we get gapless degrees of freedom
at some locus of dimension greater than one in momentum space)
\item edge of topological phase: non-onsite realization of symmetry, anomaly inflow.
\item CFT with no relevant operators.  I am not sure if there are examples of this 
which are not examples of item \ref{item-gauge}.
\item a symmetry which forbids mass terms.  This is called `technical naturalness'.   An example is 
unbroken chiral symmetry, which forbids fermion masses.  I will put supersymmetry in this category.
\end{enumerate}  
Each entry in this list is something to be understood.\footnote{It is worth noting that 
masslessness of the graviton is a mystery
not obviously solved by an element of this list.  I'll mumble about this a bit at the end of this document.}
If you encounter a gapless model and it does not fit into this list
then I will bet you \$5 that it is fine tuned, meaning that its creator simply didn't add enough terms to the Hamiltonian.

\sidebarb
Notice that the discussion above was very low-energy-centric:
I only talked about groundstates, and their low-lying excited states.
In some sense this is cowardice.
This misses out on a perhaps more fundamental dichotomy between ergodic states versus glassy/many-body localized states.
\sidebare

\noindent {\bf Refinement by symmetry.}  Another important axis along which we may organize states of matter is by symmetry.
Specifically, we can label them according to the 
symmetry group $G$ that acts on their Hilbert space and commutes with the Hamiltonian.
You will notice that here I am speaking about what are called {\it global symmetries},
that is, symmetries (not redundancies of our labelling, like gauge transformations).

There are many refinements of this organization.
We can ask how the symmetry $G$ is {\it realized}, in at least 
three senses:  
\begin{enumerate}
\item most simply, what representations of the group appear in the system?
\item is the symmetry preserved by the groundstate?  If not, this is 
called `spontaneous symmetry breaking'. 
\item is it `on-site'?  Alternatively, is it `anomalous'? What are its anomaly coefficients?
I'll postpone the explanation of these terms.  The keyword associated with them
is SPT (symmetry-protected topological) phases.
\end{enumerate}

\noindent {\bf Entanglement.}  Another important classification axis is the amount of entanglement.
Why do we hear about this so much?  Is it a fad?  No:

Unentangled means product states.
This means mean field theory is correct.
They can be distinguished only by symmetries
acting independently on each site.
This problem is approximately solved (from the point of view of the experiment-free discussion
of condensed matter physics we are having).

So: highly-entangled and mean-field are antonyms.
The description in terms of weakly-interacting 
waves above an ordered groundstate
breaks down when the entanglement matters.
The frontier of our understanding is new states of matter where quantum mechanics is essential,
not just a correction that can be included perturbatively.
(This is one version of what it means for two states to be in the same phase: 
perturbation theory in the difference of hamiltonians works.)
This is now a big industry (\eg~\cite{Wen:2012hm, 2013NJPh...15b5002G}) and I will try to give some flavor of it.
The states of interest here are distinguished instead by patterns of quantum entanglement.
Furthermore, since such new states of matter are distinguished by 
different new kinds of orders,
the phase transitions which separate them 
necessarily go beyond those described by fluctuations
of local symmetry-breaking order parameters.
This leads to new RG fixed points and maybe new CFTs.

\def\grs{{\cog{s}}}

A recent 
refinement
of the entanglement axis 
(into a single useful number) is the development of 
$\grs$-sourcery \cite{s-sourcery}.
This is the subject of \S\ref{sec:s-sourcery}.

\breakS
\section{Gauge fields in condensed matter}
\label{sec:gauge-fields}

In this section I will present a particular point of view on 
the emergence of gauge fields in condensed matter.  
For other useful perspectives, I recommend
\eg~\cite{RevModPhys.78.17, fradkin2013field, Lee:2010fy}.

{\bf Topological order} means deconfined, emergent gauge theory.
(For a `gentle' review of this subject, see \cite{Wen:2012hm}.)
Some sharp symptoms of this phenomenon (in the gapped case) are the following: 
\begin{enumerate} 
\item
\label{item1} Fractionalization of the quantum numbers
of the microscopic particles:
That is, emergent quasiparticle excitations carry quantum numbers 
{(statistics, spin, charge)} which
are rational fractions of those of the constituents.

\item \label{item2} Groundstate degeneracy which depends on the topology of space:
The fractional statistics of the quasiparticles  (point \ref{item1}) can imply a groundstate degeneracy on \eg\ the torus: 
Pair-create quasiparticle-antiquasiparticle pair, move them around a spatial cycle,  then re-annihilate. 
This process $\CF_x $ maps one groundstate to another.
But $\CF_x$ does not commute with $\CF_y$, by the anyonic statistics.
The space of groundstates must represent the algebra of these operators.

\item Long-ranged entanglement: correlations between
regions of space which produce universal deviations from 
(in fact, deficits relative to) the area law for the entanglement entropy of a subregion of the system.
This means that a state with topological order is far from
 a product state.
 (I'll come back to say more about the area law expectation in 
\S\ref{subsec:area-law}.)


\end{enumerate}

These symptoms are closely interrelated.
Here is a more precise argument 
that ${2} \implies {3}$,
due to Ref.~\incite{AshvinTarunTurner} :
Recall $S(A) \equiv - \tr \rrho_A \log \rrho_A$,
the EE (entanglement entropy) of the subregion $A$ 
in the state in question.  In such a state \cite{KP0604,PhysRevLett.96.110405, 2005PhRvA..71b2315H},
$\displaystyle{ S(A) = \Lambda 
\ell(\partial A)
- \gamma }$ \redcom{($\Lambda$ is the UV cutoff on wavenumber)}.
The deficit relative to area law,
$\gamma$, is called the ``topological entanglement entropy" (TEE)
and is proportional to the $ \log \( \# \text{torus groundstates} \) \geq 0$.
We'll give a physical realization in \S\ref{sec:TC} below.
A nonzero TEE should be contrasted with the behavior for a state without
long-range entanglement:
$$S(A) = \oint_{\partial A} s d \ell   
=
\oint \(\Lambda + b K + c K^2 + ... \)
  = \Lambda \ell(\partial A)
 + \tilde b + {\tilde c \over \ell(\partial A)}+...$$
 In the first step, we use the fact that the entanglement is localized at the boundary between 
 the region and its complement. In the second step we parametrize
 the local entropy density functional in a derivative expansion;
 $K$ is the extrinsic curvature of the boundary.
Since the total system is in a pure state, $ S(A) = S(\bar A) \implies b =0 $,
the extrinsic curvature cannot contribute.
This means that the subsystem-size-independent term 
is universal, and cannot be changed by changing the UV regulator.

The precise manner of appearance of the topological entanglement entropy (TEE) \cite{Kitaev:2005dm, PhysRevLett.96.110405} 
is special to $D=2+1$.  Generalization
to other dimensions is discussed here \cite{AshvinTarunTurner}.
A precise (gauge-invariant) definition of topological order 
is elusive in $D>2+1$\footnote{The best definition of topological order 
in general dimension is, I think, an interesting open question.  In 2+1 dimensions,
nontrivial transformation under adiabatic modular transformations 
seems to capture everything \cite{Wen:2012hm}; 
topology-dependent groundstate degeneracy is a corollary.}.

The poster examples of this set of phenomena are {\it fractional} quantum Hall states (experiments!),
and (more theoretically) discrete gauge theory.  
Both have a description in terms of gauge fields, the former crucially involving Chern-Simons terms.
The fractional charge and statistics come from holonomies of these gauge fields,
and the groundstate degeneracy comes from their Wilson lines.


\subsection{Effective field theories of quantum Hall insulators.}

\subsubsection{Electromagnetic response of gapped states in $D=2+1$ (quantum Hall)}
\label{sec:3dU1}

It will be helpful to say a bit more at this point about 
the two roles of topology mentioned above.  Let's think about 
a gapped state of matter made of some stuff in $D=2+1$,
out of which we can construct a conserved $\gU(1)$ current $j_\mu$.  
This means we can couple this current to an external, background, non-dynamical
gauge field $\CA_\mu$, by adding to the action functional like so:
$$S[\text{the~stuff}, \CA] =  S[\text{the~stuff}] + \int j^\mu \CA_\mu . $$
Here we'll treat $\CA$ as a background field that we control\footnote{Notice that 
what we've done here is {\it not} gauging the $\gU(1)$ symmetry.  
We are not changing the Hilbert space of the system.
The gauge field is just like a collection of coupling constants.}.
Integrate out the stuff to see the EM response:
$$ e^{\ii S_\text{eff}[\CA] }\equiv \int [D\text{stuff}] e^{ \ii S[\text{stuff}, \CA]} .$$
Terms quadratic in $\CA$ encode linear response:
$$ \vev{jj} = { \delta^2 \over \delta \CA \delta \CA} \log Z |_{\CA=0} $$

Because the stuff is gapped, $S_\text{eff}$ is local.
In a derivative expansion we can guess $S_\text{eff}[\CA]$:

$\CA$ is a gauge field, which is something that we can add to a derivative
to make it a covariant derivative.  Therefore $ \CA$ has dimension 1.
\be\label{eq:SeffA2d} S_\text{eff}[\CA]  = \int \(  \underbrace{ 0 \cdot \CA^2 }_\text{no symmetry breaking} + {\nu \over 4 \pi} \CA \wedge \CF +  
{1\over g^2}  \CF_{\cdot \cdot}
\CF^{\cdot\cdot} \). \ee
($\CF = \dd \CA$.)
With time-reversal symmetry (and only one gauge field), $\nu=0$.
Maxwell is irrelevant.
(Actually, without Lorentz invariance we can have non-vacuum dielectric constant and magnetic permittivity $ \eps, \mu$,
but this won't affect our story.)

The Kubo formula says that the Hall conductivity is: 
$$ \sigma^{xy} = \lim_{\omega \to 0} {1 \over\ii \omega} 
\underbrace{
\vev{j^x j^y} }_{= { \delta \over \delta \CA_x(k) }  { \delta \over \delta \CA_x(k) } S_\text{eff}[\CA] } |_{k=0} 
 = \nu { e^2 \over h}  = {\nu \over 2\pi}. $$

Next we'll show that under our assumptions $\nu$ is quantized.
So different values of $\nu$ are distinct states, since an integer can't change continuously.  
(Note that there could be other distinctions -- states with the same $\nu$ could be distinct.)

The argument for this actually has a stronger consequence \cite{Senthil:2012tm}
for bosons: For a boson SPT, $\sigma_{xy}$ must be an {\it even} multiple of  $e^2 \over h$.

\sidebarb
\hskip-.23in\parbox{.9\textwidth}{The following  argument implies 
a localized quantum of magnetic flux, 
depicted at right,
has exchange statistics $ \pi \nu $.}
\parfig{.09}{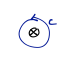}

Thread $2\pi$  worth of localized magnetic flux through
some region of the sample (as in the $\otimes$ at right).  
This means
$$ 2 \pi = \Delta \Phi  = \int \dd t \partial_t \( \oint_{R | \partial R = C } \dd \vec a \cdot \vB 
\) 
\buildrel{\text{Faraday}} \over {=} - \int \dd t  \oint_ C \vE \cdot \dd \vec \ell 
~~\buildrel{ j_r = \sigma_{xy} E_\varphi} \over {=} ~~
- { 1\over \sigma_{xy} } 2 \pi  \underbrace{\int \dd t j_r }_{= \Delta Q } $$
which says that the inserted flux sucks in 
an amount of charge
$$ \Delta Q = \sigma_{xy} . $$
This object is a localized excitation of the system -- it can move around, it's a particle.
But from the Bohm-Aharonov effect, it has statistics angle $ \pi \sigma_{xy}$
(beware: the factor of two is confusing).  
\sidebare

If we assume no fractionalization,
all particles including this one must have the same statistics as the microscopic constituents.
For a non-fractionalized state made from fermions, this means $ \nu \in \IZ$.
For bosons, no fractionalization implies $\nu \in 2 \IZ$ 
\cite{Senthil:2012tm}.

\noindent {\bf Roles of topology.}  
Quantum Hall insulators provide examples which are topological 
in two distinct ways.  The Hall conductivity (apply small electric field in $x$ direction,
measure current in $y$ direction, take ratio)
$$ \sigma^{xy} = {p\over q} {e^2 \over h} $$
is a rational number -- $ p, q \in \IZ$ -- despite  
\greencom{(in fact because of)} disorder.
\bl{IQHE:} $q=1$, happens for free electrons.  
$ p \in \IZ$ because of topology of single-particle orbits.
This does {\it not} exhibit topological order.  
This is an example of a `topological insulator'.

\bl{FQHE}: $q >1 $, requires interactions, topological order.  $q \in \IZ$ because of topology of 
{\it many-body} wave function.  
The electron {\it fractionalizes}: excitations have charge $1/q$, fractional statistics.

 \subsubsection{Abelian Chern-Simons theory}  
 
  \label{sec:abelian-CS}
  
 I want to explain an example of how properties \ref{item1} and \ref{item2} can be realized
 in QFT, 
 using the EFT (effective field theory) 
 that describes the 
 canonical examples of topologically-ordered states\footnote{For more detail see the textbook by Wen \cite{wen04}.
A great review of this subject 
with similar emphasis 
can be found in \cite{Zee:1996fe, zeeQFT}.}: (abelian) {\it fractional} quantum Hall states in $D=2+1$.

The low-energy effective field theory is Chern-Simons-Witten gauge theory, whose basic action is:
\be\label{eq:CSEFT} S_0[a_I] = \sum_{IJ} {K_{IJ}\over 4\pi} \int a_I \wedge \dd a_J  
\ee
$a^I$ are a collection of abelian gauge fields\footnote{Where did they come from?  We'll discuss some possibilities below.  One way to motivate their introduction is 
as follows \cite{Zee:1996fe, zeeQFT}.  By assumption, our system has a conserved $\gU(1)$ current,
satisfying $ \partial_\mu j^\mu = 0 $. 
In $D=2+1$, we can {\it solve} this equation by introducing a field $a$ and writing
$$ j^\mu = \eps^{\mu\nu\rho} \partial_\nu a_\rho .$$
The continuity equation is automatic if $j$ can be written this way (for nonsingular $a$) 
by symmetry of the mixed partials.   
(The equation could also be solved by a sum of such terms, as we write below).
Then we must guess what dynamics should govern $a$.  Here we just add all terms allowed
 by the symmetries.
}.
Notice that we wrote this action in a coordinate-invariant way without needing to mention a metric.
This is a topological field theory.

Two more ingredients are required for this to describe the low-energy EFT of a quantum Hall state: 

(1) We must say how the stuff is coupled to the EM field.
Notice that these gauge fields imply conserved currents $ j^I_\mu = \eps_{\mu\nu\rho} \partial_\nu a_\rho^I $.
This is automatically conserved by antisymmetry of $\eps_{\mu\nu\rho}$, as long as $a$ is single-valued.
In its realization as the EFT for a quantum Hall state,
a linear combination of these currents is coupled to the external EM field $\CA_\mu$: 
$$ S_{EM}[a_I, \CA] = \int \CA^{\mu} t_I j^I_\mu ~.$$

(2) Finally, we must include information about the (gapped) quasiparticle excitations 
of the system. This is encoded by adding (conserved) currents minimally coupled to the CS gauge fields: 
$$ S_{qp} =   \int a_I j^I_{qp } . $$

Let's focus on the case with a single such field 
(this describes \eg~the Laughlin state of electrons at $\nu = 1/3$ for $k=3$).

Notice that the Maxwell term is irrelevant.  Let's add it back in and look at the spectrum 
of fluctuations with the action:
$$ L = {k \over 4 \pi} \eps a \partial a + M^{-1} ( \partial a )^2 .$$
The gauge boson propagator has a pole when $ 0 = { p^2 \over M}  +  {k\over 4 \pi} \eps p $, 
but recall that there is only one polarization state in $D=2+1$; 
the propagating mode is the one at $ | \vec p | = M {k \over 4 \pi} $.  Massive.

Now let's show item \ref{item1}, fractional statistics.
In this case, the quasiparticles are anyons of charge $e/k$.
The idea of how this is accomplished is called flux attachment.
The CS equation of motion is $ 0 = {\delta S \over \delta a} \sim - f_{\mu\nu}  {k \over 2\pi} + j^{qp}_\mu$,
where $j^{qp}$ is a quasiparticle current, coupling minimally to the CS gauge field.
The time component of this equation $ \mu = t$ says $ b = { 2\pi  \over k}\rho $ --
a charge gets $2\pi/k$ worth of magnetic flux attached to it.
Then if we bring another quasiparticle in a loop $C$ around it, the phase of its wavefunction changes by
$$ \Delta\varphi_{12} = q_1 \oint_C a = q_1 \int_{R, \partial R = C} b = q_1 { 2\pi \over k } q_2 . $$
Hence, the quasiparticles have fractional braiding statistics.

Now \ref{item2}:  $\#$ of groundstates = $|\text{det}(K)|^{\text{genus}}$.
Simplest case: $K = k$. $\CF_x = e^{ \ii \oint_{C_x} a } $.
According to the CS action, $a_x$ is the canonical momentum of $a_y$.
Canonical quantization then implies that
these flux-insertion operators satisfy a Heisenberg algebra:
$ \CF_x \CF_y = \CF_y \CF_x e^{ 2\pi i /k} $.
If space is a Riemann surface with $g$ handles (like this:
\parbox{.3\textwidth}{\includegraphics[width=.3\textwidth]{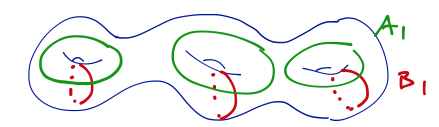}}
), then there 
are $g$ pairs of such operators, 
so $g$ independent Heisenberg algebras,
all of which commute with the Hamiltonian, 
and hence
$ k^g$ groundstates.

{\bf Exercise:} Do the (gaussian!) path integral over $a$ 
to produce an effective action for $\CA$ 
of the form \eqref{eq:SeffA2d}
with a rational Hall coefficient $\nu$.

\subsection{Where did those gauge fields come from? part 1: solvable example}

\label{sec:gt}
\label{sec:TC}

How could such gauge field degrees of freedom emerge from some 
simple sod model?  
Here is a paradigmatic (solvable) example \cite{2003AnPhy.303....2K, WenTC}.

It is in fact a generic quantum spin system in two dimensions (with no symmetries), in 
a certain regime of couplings -- with large ring-exchange terms.
In this region of parameters, such a model is often called 
`the toric code', and it is closely related to $\mathbb{Z}_2$ gauge theory.
This model has a long history,
and the general topography of its phase diagram was first understood by \cite{FS7982},
who showed that in this model the phase resulting from condensing the electric excitations
is adiabatically connected to the phase resulting from condensing the magnetic excitations.
Its study was re-invigorated by Kitaev's work \cite{2003AnPhy.303....2K}
proposing its use as a quantum computer or a zero-temperature topological quantum memory.

\def\ttau{ \ssigma}

{\bf An example of a spin system which emerges gauge theory.} 

The example we'll discuss is realization of $\IZ_2$ lattice gauge theory, 
but beginning from a model with no redundancy in its Hilbert space.
It is called the toric code, for no good reason.

To define the Hilbert space, put a qbit on every link.  

\hskip-.23in\parbox{.79\textwidth}{
A term in the hamiltonian is associated with each 
site $ j \to  A_j \equiv \prod_{l\in i} \ssigma^z_l $
and with each
plaquette $p \to B_p \equiv \prod_{l\in \partial p} \ssigma^x_l $.
$$\HH = - \sum_j A_j  - \sum_p B_p.$$
}
~\parfig{.19}{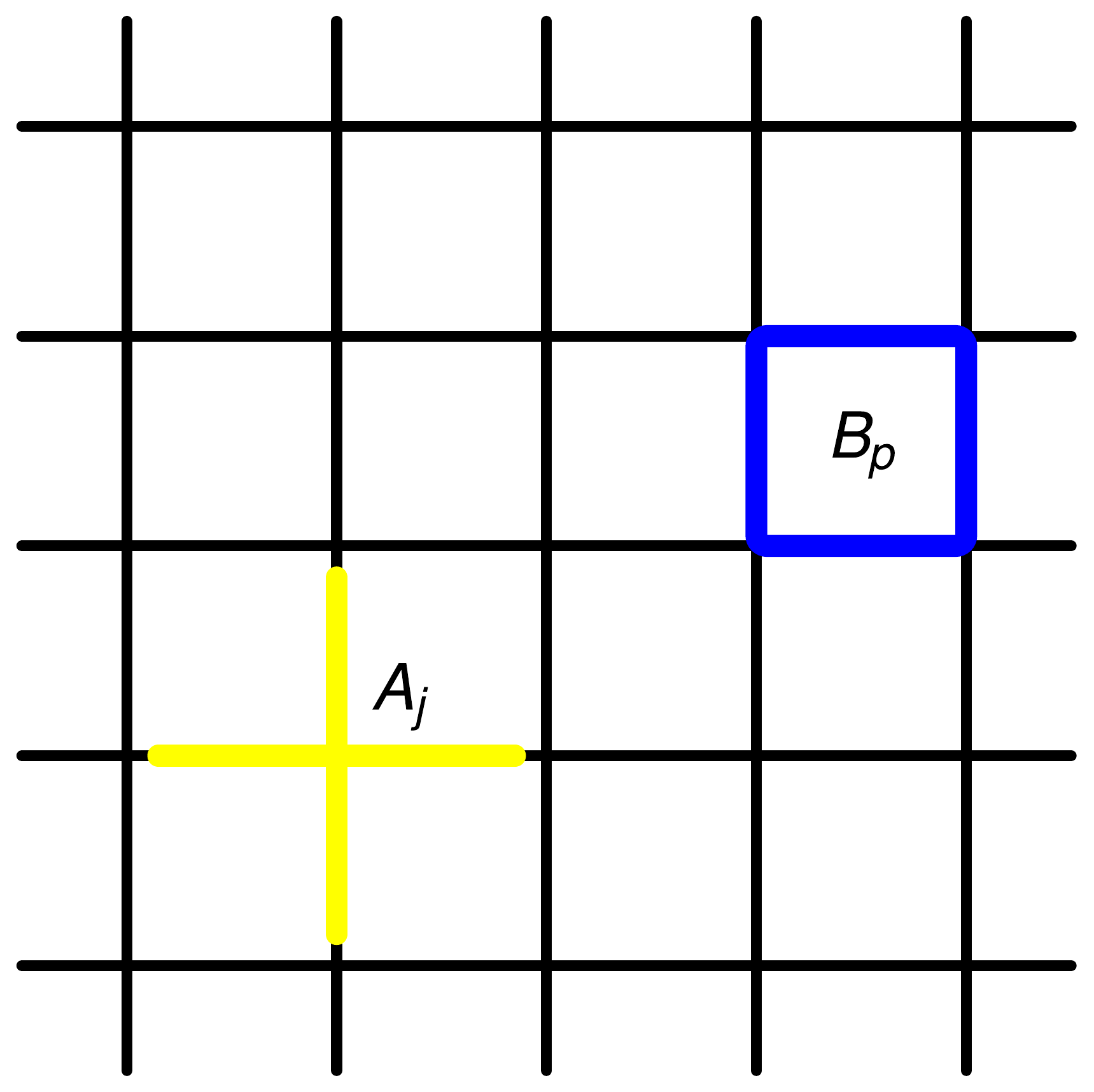}
These terms all commute with each other because they all share
an even number of $\ssigma^z_l$s and  $\ssigma^x_l$s (which anticommmute).
That means we can diagonalize the Hamiltonian by 
minimizing one term at a time.

\hskip-.23in
\parbox{.59\textwidth}{Which states satisfy the `star condition' $A_j = 1$?  
In the $\ttau^x$ basis there is an extremely useful visualization: 
we say a link $l$ of $\hat\Gamma$ is covered with a segment of string 
(an electric flux line)
if $ \bfe_l = 1$ (so $\ttau^x_l = -1$) and is not covered if $ \bfe_l=0$ (so $\ttau^x_l = +1$): 
$\parbox{1cm}{\includegraphics[width=1cm]{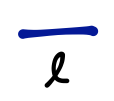}} \equiv 
\ssigma^z_{\ell} = - 1 $.  
In the figure at right, we enumerate the possibilities for a 4-valent vertex.  
$A_j=-1$ if a flux line ends at $j$.
}
~~\parbox{.39\textwidth}
{\includegraphics[width=.4\textwidth]{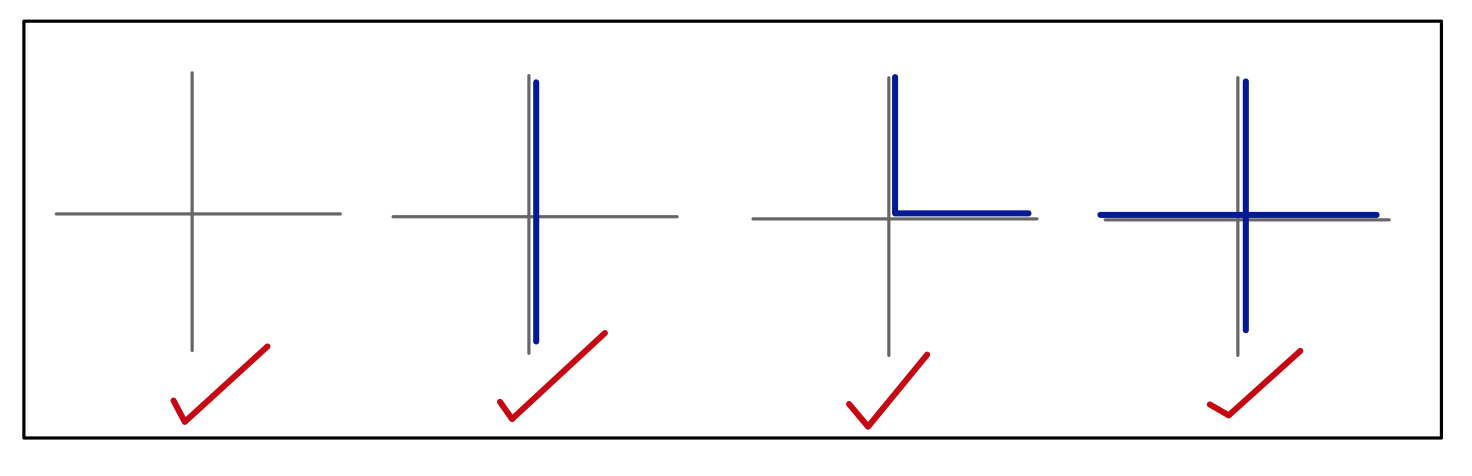}\\
\hskip.1\textwidth \includegraphics[width=.25\textwidth]{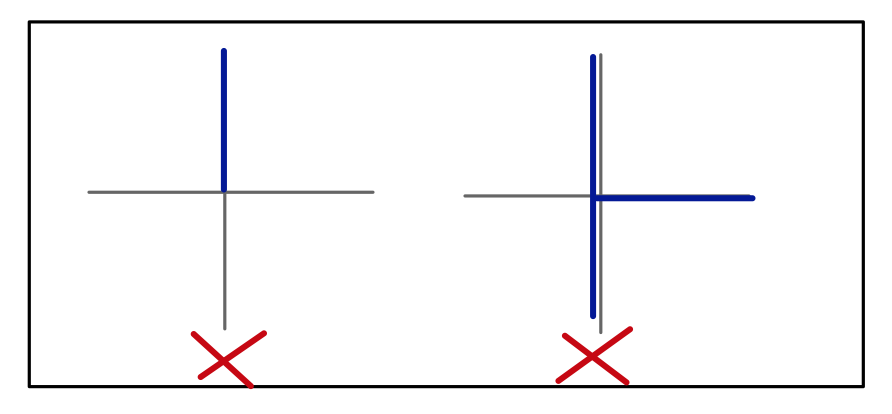}
}

So the subspace of $\CH$ satisfying the star condition is 
spanned by closed-string states, of the form
$\sum_{\cobl  \{ C \} } \Psi(C) \ket{\cobl C}$.

Now we look at the action of $B_p$ on this subspace of states:

\hskip-.23in
\parbox{.34\textwidth}{
\includegraphics[width=.35\textwidth]{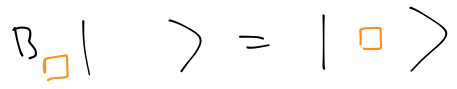} \\
\includegraphics[width=.35\textwidth]{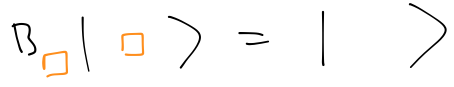} \\
\includegraphics[width=.35\textwidth]{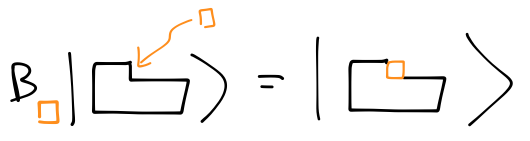}
}
~~~\parbox{.64\textwidth}{
$$ B_p \ket{C} = \ket{ C + \partial p } $$
The condition that $B_p \ket{\gs} = \ket{\gs} $ is a homological equivalence.
In words, the eigenvalue equation $ \BB_{\Box} = 1 $ 
says $ \Psi(C) = \Psi(C')$ 
if $ C' $ and $ C $ can be continuously deformed into each other by attaching
or removing plaquettes.
In the appendix \S\ref{sec:p-form-appendix}, we make
this connection with homology more explicit.
}

If the lattice is simply connected -- if all curves are the boundary of some region
contained in the lattice --
then this means the groundstate
$$ \ket{\gs} = \sum_C \ket{C} $$
is a uniform superposition of all loops.

{\bf Topological order.} 
If the space has non-contractible loops, 
then the eigenvalue equation does not determine the relative
coefficients of loops of different topology!
On a space with $2g$ independent non-contractible loops,
there are $2^{2g}$ independent groundstates.

No local operator mixes these groundstates.  
This makes the topological degeneracy stable to local perturbations of the Hamiltonian.
They are connected by the action of  
$V, W$ -- Wilson loops:
$$ W_C = \prod_{\ell \in C} \ssigma^x, ~~~ V_{ \check C} = \prod_{\ell \perp \check C} \ssigma^z .$$
They commute with $ \HH_\text{TC}$ and don't commute with each other
(specifically $W_C$ anticommutes with $ V_{\check C}$ if 
$C $ and $ \check C$ intersect an odd number of times).
These are the promised operators (called $\CF_{x,y}$ above)
whose algebra is represented on the groundstates.

\def\circsum{\relax\hbox{$\sum\kern-1.1em{\circ}\kern0.5em$}}
\def\BB{{\bf B}}
\def\bb{{\bf b}}
\breakS
\sidebarb
{\bf Gauge theory notation.}  
Why do I call them Wilson loops?
To make it look more like gauge theory familiar from high energy physics, 
regard the non-gauge-invariant variable $\ttau^x$ as 
$$ {}`` ~~~~~~\ttau^x_{\bar i \bar j} = e^{ \ii \int_{\bar i}^{\bar j} \vec a \cdot d \vec s } ~~~~~~ {}''$$
the holonomy of some fictitious continuum gauge field integrated along the link.
More precisely, let
$$ \ttau^x_{\bar i \bar j} \equiv e^{ \ii \pi \aa_{\bar i \bar j}  }, ~~~\aa_{\bar i, \bar j} = 0, 1 . $$
Then the plaquette operator is
$$ \BB_{\Box} = \prod_{l \in \Box} \ttau^x_l 
 ``~~ = ~e^{ \ii \oint_{\partial \Box} \vec a \cdot d \vec l }  ~~"  
= e^{ \ii \pi 
\circsum_{\partial \Box} 
\aa } 
\buildrel{\text{Stokes}}\over{=} e^{ \ii \pi \bb_{\Box} } ~, $$
where $\bb_{\Box}$ is the (discrete) magnetic flux through the plaquette $\Box$.
In the penultimate expression, the symbol $\circsum$ is intended
to emphasize that we are summing the $\aa$s around the closed loop.

In the Hamiltonian description of gauge theory, the field momentum for $ \vec \aa$ 
is the electric field $\vec \bfe$.  
So, we call
$$ \ttau^z_l \equiv e^{ \ii \pi \bfe_l } $$
The star operator is
$$ \AA_+ = \prod_{l \in + } \ttau^z_{l} = e^{ \ii \pi \sum_{l \in + } \bfe_{ij} } 
\equiv e^{ \ii \pi \Delta \cdot \bfe } $$
which is a lattice divergence operator.
The constraint is
$$ 1 = \prod_{l \in+} \ttau^z_l ~~~\leftrightarrow~~~
\Delta \cdot \bfe = 0~ \text{mod}~2. $$
This is {\it binary electrodynamics}, electrodynamics mod two.  
Electric charges are violations of the Gauss' Law constraint: 
if 
$$ \( \Delta \cdot \bfe \)(i) = 1 ~\text{mod} ~2 $$
at some site $i$,  we say there is a $\IZ_2$ charge at site $i$.
Notice that this is {\it not} something we can do in the spin system: 
such a site is the {\it end} of a domain wall.

\sidebare

{\bf String condensation.}  Notice that the deconfined phase of the gauge theory 
involves
 the condensation of the electric flux strings, in the sense that 
the operators $ \BB_\Box$ which create these strings
have a nonzero groundstate expectation value:
$$ \bra{\gs} \BB_\Box \ket{\gs} \buildrel{g=\infty}\over{=} 1 .$$
As with an ordinary condensate of bosons, away from the zero-correlation-length limit 
($g=\infty$),
the condensate will not be exactly $1$, 
since finite $g$ suppresses configurations with electric flux.
But within the deconfined phase it will be nonzero.

{\bf Defects.} There are two kinds of defects: violations of $A_s = 1 $
and violations of $ B_p=1$.
Notice that the former kind of defects would be strictly forbidden in `pure gauge theory'
since $A_s = 1$ is the Gauss' law constraint.  
So pure $\IZ_2$ gauge theory is the limit where the coefficient of $A_s$ goes to infinity.

The defects are created by the endpoints of open Wilson lines. 
Again there are two kinds:
$$ W(C) = \prod_{\ell \in C} \ssigma^x, ~~~ V({ \check C}) = \prod_{\ell \perp \check C} \ssigma^z .$$
Here $C$ is a curve in the lattice, and $\check C$ is a curve in the dual lattice.
Endpoints of $W(C)$ violate $ A_s$ and endpoints of $V({\check C})$ violate $B_p$.

\sidebarb
\hskip-.23in
\parbox{.69\textwidth}
{Consider the cylinder.  There is one nontrivial class of loops; call a representative $\gamma$.
Let $ \eta$ be a line running along the cylinder.
The two groundstates are generated by the action of the `Wilson loop operator'
$$V(\eta) \equiv \prod_{l~\text{crossed by }~ \eta} \ttau_l^z$$ 
}
\parbox{.3\textwidth}{\includegraphics[width=.3\textwidth]{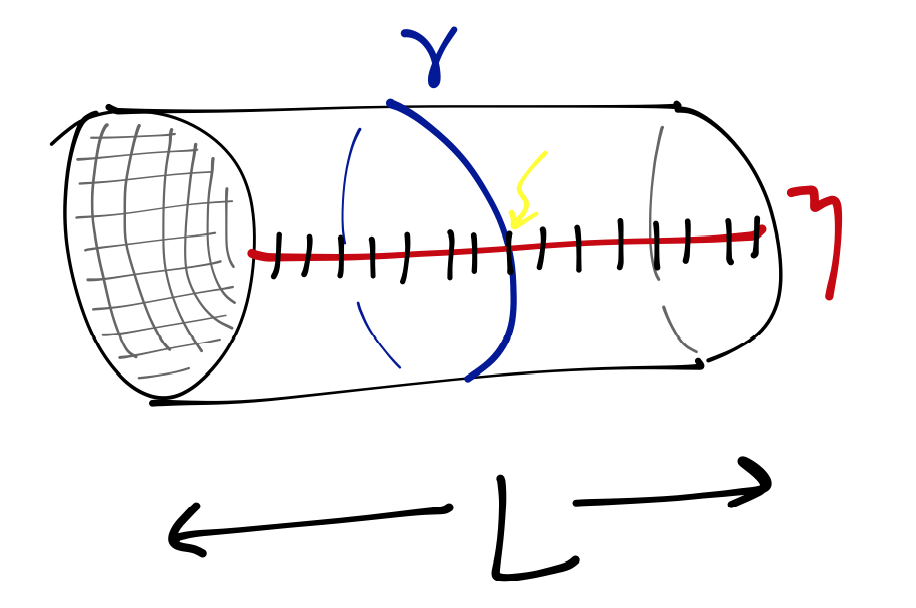} }
in the sense that 
$$ \ket{\gs_2} = V(\eta) \ket{\gs_1} ~.$$
This is also a groundstate (at $g=\infty$) since there is no plaquette which violates $\BB_p=1$
(more simply: $[ \HH_{g=h=0}, \WW_x(\eta)] = 0 $).
They are {\it distinguished} by 
$W(\gamma) \equiv  \prod_{l \in \gamma} \ttau^x_l$ in the sense that 
the two groundstates are eigenstates of this operator with distinct eigenvalues:
$$W(\gamma)  \ket{\gs_{\alpha} } = (-1)^\alpha \ket{\gs_{\alpha}} , ~~\alpha=1,2. $$
This follows since $\{ W(\eta), V(\gamma) \} = 0 $ -- the two curves share a single link (the one 
pointed to by the yellow arrow in the figure).

At finite $g, h$ (and in finite volume), there is tunneling between the topologically degenerate groundstates, 
since in that case
$$ [\HH, \prod_{l \in \gamma} \ttau^z_{l} ] \neq 0 . $$
This means
$$ \bra{\gs_2} \HH \ket{\gs_1} \equiv \Gamma \neq 0 . $$
However, the amplitude $\Gamma$ 
requires the creation of magnetic flux on some plaquette (\ie~a plaquette $P$ with $B_P = - 1$, which costs energy $ 2g$), 
which then must hop (using the $  - g \ttau^z$ term in $\HH$) all the way along the path $\eta$, of length $L$, to cancel the action of $V(\eta)$.
The amplitude for this process goes like
$$ \Gamma \sim 
{\bra{\gs_2} \( -  \ttau^z_1 \) \( -  \ttau^z_2  \) \cdots 
\( -  \ttau^z_{L} \) \ket{\gs_1} 
\over 2g  \cdot 2g  \cdot \dots 2g  } 
\sim \(  { 1 \over 2 g  } \)^{L }=  e^{ - L \log 2g }  $$
which is {\it extremely tiny} in the thermodynamic limit.
The way to think about this is that the Hamiltonian 
is itself a local operator,
and cannot distinguish the groundstates from each other. 
It takes a non-perturbative process, exponentially suppressed
in system size,
to create the splitting.

\sidebare

\hskip-.23in
\parbox{.69\textwidth}{Here is why {\bf deconfined flux lines}
mean
{\bf long-range entanglement}:
The picture at right shows why 
-- in a state described by fluctuating {\it closed} strings -- 
there is a contribution
to the EE of region $A$ which is independent of the size of $A$: 
if a string enters $A$ and it is known to be closed, 
then it must leave again somewhere else; this is 
one missing bit of freedom, so $S \sim L/\eps - \log 2 $. 

So far everything I've said works on any graph (actually: cell complex).  
And so far I've described the solvable limit.  But 
this physics actually pertains to an open subset of the phase diagram -- it is a phase of matter.
}
~\parbox{.3\textwidth}{\includegraphics[width=.3\textwidth]{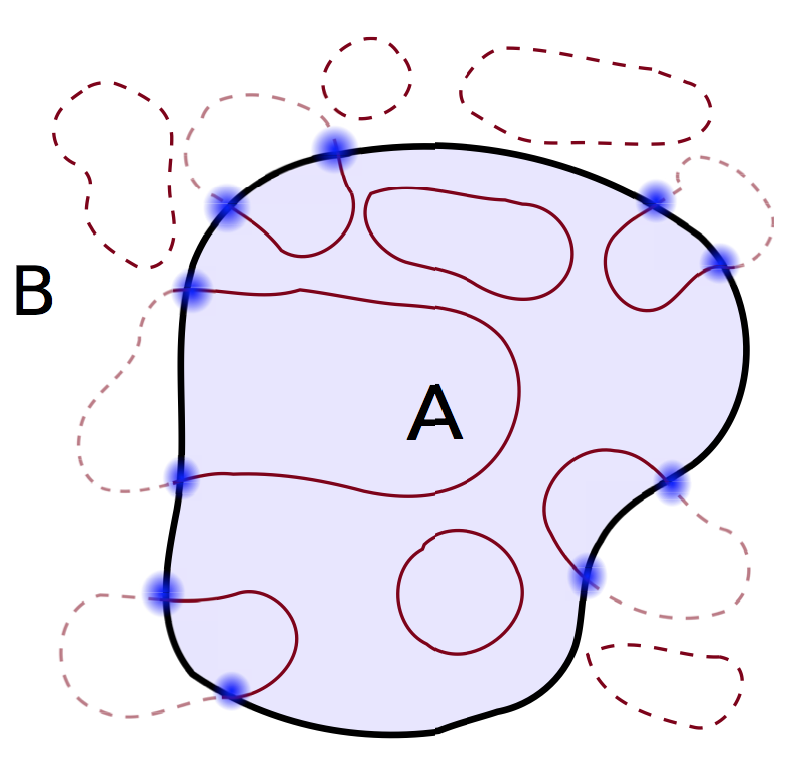}\\
~~~~~\ttref{[fig: Tarun Grover]}
}

\vskip.1in

\hskip-.23in
\parbox{.68\textwidth}{
Perturbations $ \Delta H = \sum_l \( h \sigma^x_l + g \sigma^z_l \) $  produce
a nonzero correlation length.
Let's focus on $D=2+1$ for what follows.
These couplings $g$ and $h$ are respectively a string tension and a fugacity for the electric flux string endpoints: charges.  
Make these too big and the model is confined or higgsed, respectively.  
These are actually adiabatically connected \cite{FS7982}:
Both are connected to the trivial state where e.g. $ H = \sum_l \sigma^x_l$ 
whose groundstate is a product $ \otimes_l \ket{\rightarrow_l}$.
}
~~~\parbox{.29\textwidth}{\includegraphics[width=.2\textwidth]{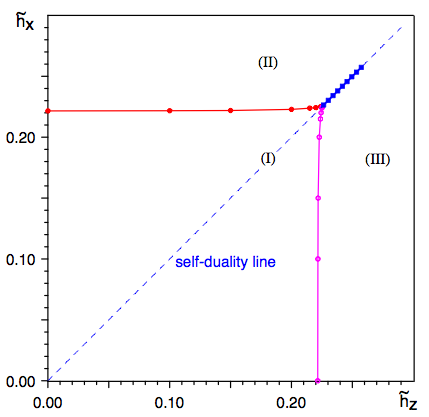} \\
{\ttref{from \cite{2010PhRvB..82h5114T}}}
}

\sidebarb

{\bf Large $g$ is confinement.}  Consider for a moment the limit where $g$ is large and the gauss law term is large, both compared to 
the plaquette term.
In that case, we can make the big terms happy just by setting $ \ttau^z = 1 $: no electric flux.
Inserting a pair of charges is accomplished by violating the star term at two sites --
this forces an odd number of the nearby links to have $ \ttau^z=-1$.
What's the lowest energy state with this property, as a function of the separation between the two charges?

\hskip-.23in\parbox{.5\textwidth}{
To find the potential between static charges $V(x)$, we need to minimize 
\bea \HH(g = \infty ) &=& - g \sum_l \ttau^x_l 
\cr &=&  E_0 +2g \mathsf{L}(\text{string}) ~.\nonumber\eea
}
\parfig{.49}{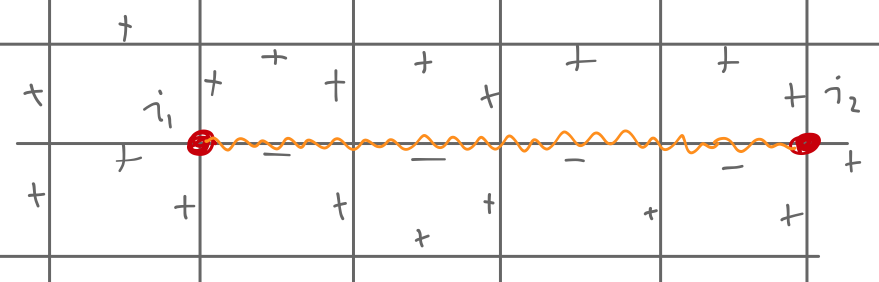} 

Here $E_0 = - g 2 N$ is the energy of the state with no electric flux and no external charges,
where $2N$ is the number of sites.
$\mathsf{L}(\text{string})$ is the length of the electric flux string: 
the string can be said to have a nonzero tension (energy per unit length), $2g$.
Clearly this minimization is accomplished by a straight line, and the potential between the charges is 
$$ V(x) = + 2 g x $$
which is linearly rising with separation, and implies a constant attractive force 
$$ F = - \partial_x V = 2g . $$

\sidebare

Here's a way to understand how this comes about:
{\bf EFT in terms of CS theory.}  
Let's include a field which creates $e$ particles
and one which creates $m$ particles.
They are just bosons, so these are scalars.

But the $e$ and $m$ particles are mutual semions
($W V = - V W $ if the two paths cross an odd number of times).
How do we encode this?  We can couple them to CS gauge fields.
Two hints for how to do this:
(1) We don't want to change the particles' {\it self-}statistics, 
so we want to attach a flux to $e$ to which it doesn't couple minimally.
(2) The original model didn't break parity.
You may know how to get around both these problems from the ABJM model: doubled CS theory.
The result is:
\be\label{eq:TCEFT} L = { k \over 4 \pi } \( a \partial b  + b \partial a \) 
+ | ( \partial + a ) e |^2 + | ( \partial + b) m |^2  - V(e,m)\ee
$$ V(e,m)  = r_e | e|^2  + u_e |e|^4 + r_m | m|^2 + u_m |m|^4 
+ v |e|^2 |m|^2 $$
$r_e$ is the energy cost for adding an $e$ particle at rest, so that should be $2\Gamma_e$, 
the coefficient of the star term.
The $u_e$ terms encode the fact that two $e$ particles don't want to 
(in fact, cannot) sit on top of each other.
The last term represents a contact interaction between the two kinds of defects.
The kinetic terms for $ e$ and $m$ come from the $\ssigma^x$ and $\ssigma^z$ perturbations.\footnote{Actually I've lied a little bit by implying 
relativistic kinetic terms for $e, m$.  And I should add chemical potentials for $e, m$ which also depend on $h,g$.}
This EFT illuminates the structure of the phase diagram!

(An alternate route connecting $\IZ_2$ gauge theory
to doubled-Chern-simons theory can be found 
in \cite{Maldacena:2001ss}.)


The TC can be made to look even more innocuous 
(\ie~almost like something you might find describing spins in an insulating piece of Earth rock)
by some relabelling of variables \cite{WenTC}.
The above model emerges $\IZ_2$ gauge theory.  
The generalization to $\IZ_N$ is simple but requires orienting the links.
A generalization which produces a photon appears \eg~here \cite{PhysRevLett.88.011602}.
The generalization to p-form gauge fields is also interesting
and is described in \S\ref{sec:p-form-appendix}.

The model I've described (in $D=2+1$) has a lovely self-duality which is important for what I want to say next.  
Consider the map which produces a new lattice from $\Gamma$
by defining a site for each plaquette, a link for each link and a plaquette for each site.
(See \S\ref{sec:lattice-duality} for the generalization to other $p$, other $d$.)
Then we can think of the variables as living on the links of the 
dual lattice, and the result is simply to interchange which of $A$ and $B$ we think
of as star and plaquette terms, and 
which are the electric and which are the magnetic defects.

\subsection{Quantum critical points with mutually-nonlocal fields}

The study of conformal field theories (CFTs)
has many motivations.  
They govern the physics of continuous phase transitions.
They can be used to define quantum field theory non-perturbatively.
They may have a place in high-energy physics beyond the Standard Model.
They are central to the AdS/CFT correspondence by which
we now understand some quantum gravity theories.
It is reasonable also to anticipate that CFTs in $D$ dimensions provide
the necessary data for topological order in $D+1$ dimensions
(as they do in $D=2$).

However, known examples of CFTs in dimensions above 1+1 are sparse.
The known examples mostly come from perturbing gaussian theories,
and it seems likely that the paucity of known examples is an artifact of our lack of tools.
In the special case of supersymmetric theories, many examples are known
and some of these are qualitatively different from 
the perturbative examples.
In particular there are critical points of gauge theories where
both electric and magnetic degrees of freedom become light
\cite{Argyres:1995jj}.
In four dimensions, it is not even known how to 
write a Lagrangian description of such a model.
Such a phenomenon is similar to the physics of deconfined quantum critical points \cite{2004Sci...303.1490S,2004PhRvB..70n4407S} which can arise 
at an interface between Higgs phase (condensing electric charge) and confining phase (condensing magnetic charge).
The extra ingredient required is that the defects which 
destroy the Higgs condensate (vortices or vortex strings) 
must carry the magnetic charge whose condesnation produces confinement, 
so the two transitions are forced to occur together.
It is likely that there are many more examples of such critical points
awaiting the development of the correct tools to study them.

Two such critical points can be found in the
toric code phase diagram.
Motivated by an attempt to understand the stability of the topological phase, 
Monte Carlo simulations \cite{2010PhRvB..82h5114T} identified two new (multi-)critical points
in its phase diagram.
These critical points occur on the self-dual line,
a value of the parameters for which there is a symmetry exchanging  electric and magnetic excitations, 
and the critical points occur at the onset of condensation of these degrees of freedom,
where they are simultaneously becoming light.
The questions of 
the nature of these critical points,
and of their field theory description was left open.
Such a field theory description must involve gapless excitations
with both electric and magnetic quantum numbers.

The effective field theory above provides a starting point.
The last ($v$) term is required to get the line of first-order transitions\footnote
{Note also that the couplings in the effective action \eqref{eq:TCEFT}
are some functions of the microscopic couplings $g, h$.
As we vary $r_{e/m}$ in the figure,
$v$ varies as well. 
This understanding was developed with Alex Kuczala.}.
Sachdev and Xu \cite{Xu:2008jx}  studied a closely-related class of field theories,
and identified rich physics of symmetry-broken phases.
Their interest was in a different phase diagram 
which did not contain the second critical endpoint.
There are some results about critical exponents 
from large $N$ and large $k$.  
There are close connections to Chern-Simons matter theories studied recently 
in the high-energy theory literature
(\eg~\cite{Giombi:2011kc, Aharony:2012nh, Aharony:2012ns}).

\subsection{Where did those gauge fields come from? part 2: parton construction}

\label{sec:partons}
A practical point of view on what I'm going to describe 
here is a way to guess variational wavefunctions for fractionalized groundstates.
A more ambitious interpretation is to think of 
the parton construction as a duality
between a model of interacting electrons \redcom{(or spins or bosons or ...)} 
and a gauge theory of (candidate) `partons' or `slave particles'.
Like any duality, it is a guess for useful low-energy degrees of freedom.
The goal is to describe states {\it in the same Hilbert space} as the original model,
in terms of other \redcom{(hopefully better!)} variables.
The appearance of gauge fields (perhaps only discrete ones) is an inevitable side effect 
\redcom{(?)} when there is fractionalization of quantum numbers
\greencom{(spin-charge separation, fractional charge ...)} in $D>1+1$.

I will describe the construction in two steps.
For references on this subject,
see \eg~\cite{Kim1999130, wen04, RevModPhys.78.17, Lee:2010fy}.

{\bf Parton construction: step 1 of 2} (kinematics) 

Relabel states of the many-body ${\cal H}$ with new, auxiliary variables.
For example, suppose $c$ the annihilation operator
for an electron (spinless, \eg~because it is polarized by a large magnetic field).  
Suppose we are interested in the (difficult) model with 
$$ H = \sum_{\vev{ij}} \( t_{ij} c^\dagger_i c_j^\nd  + h.c. \) 
+ \sum_{\vev{ij}}  V n_i n_j $$
Two comments: (1) We suppose that the hopping terms $ t_{ij} $ include some lattice version
of the magnetic field, so $ t_{ij} = t e^{ \ii \CA_{ij}}$.
(2) This kind of `Hubbard-V interaction' is the shortest range interaction we can have 
for spinless fermions (since the density $n_i = c_i^\dagger c_i $  is zero or one and so satisfies $n_i^2 = n_i$).

For example, a parton ansatz appropriate to the $\nu = {1\over 3 } $ Laughlin FQH state is
$$\greencom{\eg} ~~~ c = f_1 f_2 f_3 = {1\over 3!} \epsilon_{\alpha\beta\gamma} f_\alpha
f_\beta f_\gamma $$
$f$s are complex fermion annihilation operators (they must be fermionic in order 
that three of them make up a grassmann operator).

Not all states made by $f$s are in ${\cal H}$.
There is a redundancy: if we change  \\
$f_1 \to e^{ i \varphi(x)} f_1, 
f_2 \to e^{ -i \varphi(x)} f_2, f_3 \to f_3, ... $\\
then the physical variable $c$ is unchanged.
In fact, there is an $SU(3)$ redundancy
$f_\alpha \to U_\alpha^\beta f_\beta, ~~ c\to \text{det}U c $.    We are making the ansatz that $c$ is a baryon.

The Lagrange multiplier 
imposing 
\be\label{eq:gauss-parton}
f_1^\dagger f_1 = f_2^\dagger f_2 = c^\dagger c = \text{number of } e^-; f_2^\dagger f_2 = f_3^\dagger f_3 \ee
is the time component $a_0$ of a gauge field.
(The relation to the number of actual electrons arises since $c$ creates one of each.)

To write an action for the $f$s which is covariant under this redundancy, introduce the 
spatial components of the gauge field,
$a_i$.   
Perhaps you don't like this idea since 
it seems like we added degrees of freedom. 
Alternatively, we can think of it as arising from $e^-$ bilinears, in decoupling 
the $c_x^\dagger c_x c_{x+i}^\dagger c_{x+i} $ interaction
by the Hubbard-Stratonovich trick.
What I mean by this is:
\bea e^{ \ii V \sum_{\vev{ij}}  \int dt n_i(t) n_j(t)  } 
&=& e^{ \ii V c_i^\dagger(t)  c_i^\nd(t)   c_j^\dagger(t)  c_j^\nd(t)  } \cr 
&\buildrel{\eqref{eq:gauss-parton}}\over{=}&
e^{ \ii V\sum_\alpha f_{i\alpha}^\dagger(t)  f_{i\alpha}^\nd(t)   \sum_\alpha f_{j\alpha}^\dagger(t)  f_{j\alpha}^\nd(t)}
\cr &=& \int [ d \eta^{\alpha\beta}_{ij} ] e^{    \ii \int dt ~
\sum_{\alpha\beta, \vev{ij}}
\( { | \eta^{\alpha\beta}_{ij} |^2\over V } 
+  f_{i\alpha}^\dagger(t) f_{j\alpha}^\nd(t) \eta^{\alpha\beta}_{ij}  + h.c. \) } \eea
where $\eta$ is a new complex (auxiliary) bosonic field on each link.
Now let $ \eta_{ij} = |\eta_{ij} | e^{ \ii a_{ij } } $ (for each $\alpha\beta$) and ignore the (massive)
fluctuations of the magnitude  $|\eta_{ij}| = t_{ij}$.  
{\it Voil\`a} the gauge field, and the parton kinetic term\footnote{I learned from Tom DeGrand that
closely-related ideas 
(with different motivations) were studied by the lattice community
\cite{Hasenfratz:1992jv, Hasenfratz:1993az}.}.


How does the practical 
viewpoint of constructing possible wavefunctions arise?
Guess weakly interacting partons:
$ H_\text{partons}  = - \sum_{ij} t_{ij} f_i^\dagger e^{ i a_{ij} } f_j +h.c.$
Then fill bands of $f$ and project onto the gauge invariant subspace.

But what about the fluctuations of $a$ (i.e. we still have to do the $a$ integral)?  Microscopically, $a$ has no kinetic term; in that 
sense the partons are surely confined at {\it short} distances. 
$a$ only gets a kinetic term from
the parton fluctuations, by processes like this:
\parbox{3cm}{\includegraphics[width=3cm]{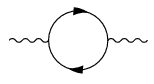}}
.  The hope is that with enough other partons around, they can be shared and juggled amongst the
electrons, so that which parton is in which electron fluctuates.

{\bf Parton construction: step 2 of 2} (Dynamics)

Such a rewrite is always possible, and there are many possibilities.
The default result of such a rewriting
is that the gauge theory {\bf confines} at low energies.
By a confining state, I mean one in which 
the energy cost to separate partons  
is much larger than
other scales in the problem,
namely,
the gap, or the inverse lattice spacing, or energies associated with chemistry (gasp).  
This means there is no fractionalization,
and no topological order and usually leads us back to the microscopic description in terms of 
the microscopic degrees of freedom.  
(It doesn't mean the parton description is useless however; 
see \S\ref{sec:SPT}).

Pure 2+1d gauge theory likes to do this.
Recall that the Maxwell or Yang-Mills kinetic term is an irrelevant operator
according to naive dimensional analysis, if we treat the gauge field as a connection
(\ie~something we can add to a spatial derivative).
This is true even of (compact) $U(1)$ gauge theory \cite{Polyakov:1976fu}:
In terms of the dual photon $\sigma$, defined by 
$\partial_\mu \sigma \equiv \epsilon_{\mu\nu\rho} \partial _\nu a_\rho $,
the proliferation of monopoles produces an effective potential of the form
$$ V_\text{eff} = \Lambda^3 e^{ i \sigma} + h.c. ~~~
\implies \text{mass for $\sigma$} .
$$

\parbox{.56\textwidth}{
Let me emphasize again that it's {\it deconfined} states of parton gauge theories that are most interesting here.
So we are looking for gauge theories which behave oppositely to QCD, 
really like {\it anti-}QCD, where the partons are deconfined {\it below} the confinement scale
$\Lambda_{\text{anti-QCD}}$, as in the figure at right.
Interesting states we can make this way correspond to interesting phases of gauge theory, 
a worthy subject.

Our discussion in this section has followed this diagram starting from 
the highest energies (chemistry!) and guessing the lower-energy degrees of freedom
that result from the interactions of the constituents.  (This dialectic between high-energy physics and condensed matter physics,
of GUT and anti-GUT,
is described vividly by Volovik \cite{volovik2003universe, Volovik:2007vs}.)
}\parbox{.37\textwidth}{\includegraphics[width=.39\textwidth]{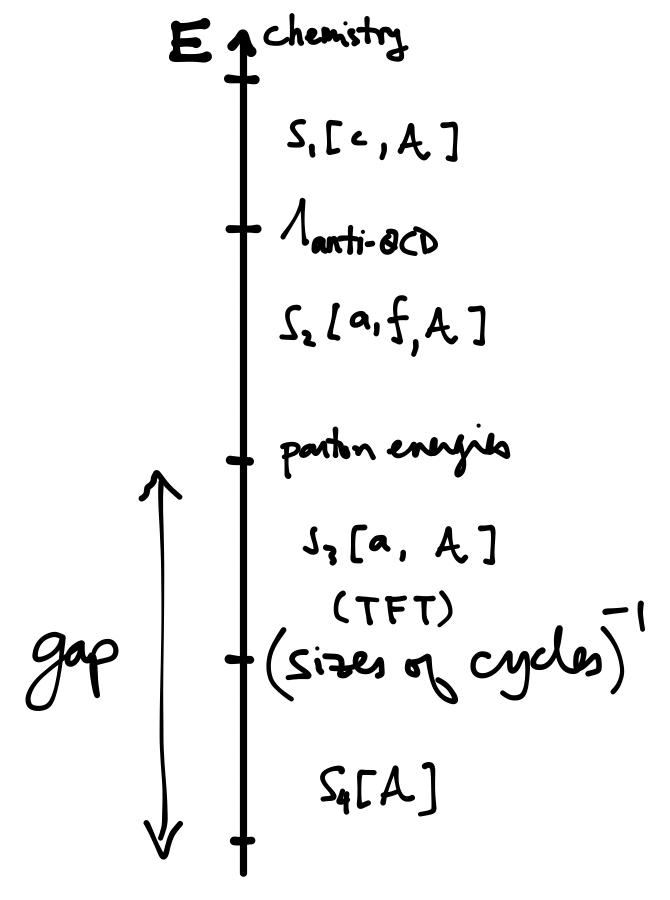}}

Deconfinement requires an explanation.
Known exceptions which allow for this:  
\begin{itemize}
\item enough dimensions that the Maxwell term becomes marginal and we can have a Coulomb phase.
\item partial Higgsing to $\IZ_n$. Condensing electric charge makes monopoles heavy.
\item lots of charged degrees of freedom at low energy.  
One way to describe their effects is that 
they produce zeromodes on the monopole configuration, 
and the monopole operator only contributes to higher-dimension operators 
involving insertions of the light fields.
Interesting constraints on how many modes is enough, 
from strong-subadditivity of the entanglement entropy,
were derived in \cite{Grover:2012sp}.
Partons which are gapless at points in $k$-space 
are called algebraic (something) liquids;
the `something' is whatever visible degrees of freedom they carry, \eg~spin; 
if they happened in this model it would be charge.
If the partons form a Fermi surface, that is certainly enough (Sung-Sik Lee reviews his proof 
of this in \cite{Lee:2010fy}).  
This is the kind of spin liquid which may have been observed in 
various materials in the past decade or so.

\item in $D=2+1$: the Chern-Simons term $a \wedge da$ is marginal, and can gap out gauge dynamics, 
as we saw in \S\ref{sec:abelian-CS}.
\end{itemize}
If I've forgotten some please tell me.

Let's pursue the Laughlin example a bit further,
but let's retreat to the continuum.
So consider a pile of electrons 
in 2+1 dims, in large uniform $B$ such that 
the number of electrons per lowest Landau level state is 
$$ {1\over 3} = \nu_e \equiv {N_e \over N_\Phi(e) }= {N_e \over e B A / (hc)} .$$
The fact that this band is partially filled means 
that if the electrons are free, 
the system is gapless.  But this degeneracy is fragile.

If on the other hand, the electron fractionalizes as 
$ c = f_1 f_2 f_3, $  then $f_\alpha$ carries charge $1/3$.
Consider then each $f_\alpha$ in same $B$, and suppose the partons are free (as a first approximation). 
Their filling fraction is:
$$ \nu_f = 
{N_f \over N_\Phi(e/3) }= {N_e \over N_\Phi(e/3) } = 3 \nu_e = 1~~.$$

The wonderful thing about this guess is that 
the partons can now form a gapped state: 
that is, we can pretend they are free and fill their bands,
so that they are a band insulator (an integer quantum Hall state).
Then integrating out the gapped partons produces a 
(nonsingular but nontrivial) contribution to the effective 
action for the gauge field: 
the IQH nature of the bands means that 
there is a Hall response for any gauge fields to which they are coupled.
This is encapsulated precisely by the CS term!
\footnote{We showed that QHE means a CS term earlier.  
The massive dirac fermion in $2+1$ dimensions also has a Hall response (recall that $m$ breaks parity in $D=2+1$).
The more microscopic calculation can be done in just the same manner as the
path integral calculation of the chiral anomaly, and the $\epsilon_{\mu\nu\rho}$ arises
for the same reason.
Let me show the calculation in a related model of massive Dirac fermions: 
\bea \nonumber 
\log \int D\psi D\psi^\dagger ~ e^{ - \int d^3 x \bar \psi \( \ii \Dslash - m \) \psi  }
&=& \tr \log \( \ii \Dslash - m \) 
\cr 
&\equiv& \tr \log \( 1 - \ii \Dslash/m \) e^{ - \Box/M^2 } + \text{cst}~~~~~
\Box \equiv \( \ii \Dslash\)^2 = - ( \partial + a )^2 - \half \Sigma_{\mu\nu} F^{\mu\nu} 
\cr 
&=& - \tr \sum_{n=1}^\infty {1\over n} \( { \ii \Dslash \over m  } \)^n e^{ - \Box/M^2 } ~.
\eea
where $ \Sigma_{\mu\nu} \equiv \half [ \gamma_\mu, \gamma_\nu] $ is the rotation generator.
Now expand the regulator exponential as well 
and the $a^3$ term goes like 
$$ \tr { \ii a_\rho \gamma^\rho\over m } {1\over M^2 } \( \( \partial + a \) ^2 + \half \Sigma_{\mu\nu} F^{\mu\nu} \)  =  \underbrace{\tr \gamma_\rho \Sigma_{\mu\nu} }_{= \eps_{\rho\mu\nu}}   \half \ii a^\rho F^{\mu\nu} $$ 
}
Hence we arrive at \eqref{eq:CSEFT}, for some particular choice of $K$ 
determined 
by the QH response of the partons, 
\ie~by their charges the Chern numbers of their bands.

$$\bluecom{\text{integrate out gapped partons: }}~~~~ \int [Df]  ~ e^{ i \int f^\dagger \( \partial + a \) f } = e^{ i { k \over 2\pi} \text{CS}(a) + \cdots } $$

The low-energy effective field theory is 
$SU(3)_1$ CS theory \greencom{with gapped fermionic quasiparticles}
$\simeq U(1)_3$ Laughlin state.
\greencom{(Laughlin qp = hole in $f$ with Wilson line)}
$$ \Psi
= \bra{0} \prod_i c(r_i) \ket{\Phi_{mf}}
 = \( 
\underbrace{
\prod_{ij}^N z_{ij}  e^{ - \sum_i^N |z_i|^2 / (4 \ell_B^2(e/3)) }}_\text{$\nu=1$ slater det of charge $1/3$ fermions} \)^{3} $$

$$ \sigma^{xy} = {(e/3)^2 \over h} \times 3 = {1\over 3}  e^2 /h. $$

%
%
%

$D=2+1$ is kind of cheating from the point of view of emergent gauge fields.
This is because the Chern-Simons term
is a self-coupling of gauge fields
which gives the photon a mass without breaking gauge symmetry
(that is: without the addition of degrees of freedom).
We have seen above that this does not necessarily require breaking parity symmetry.

For a long time I thought that gauge fields were
only interesting for condensed matter physics when deconfinement could be 
somehow achieved, \ie, when there is topological order.
We'll see some examples in \S\ref{sec:SPT} 
where even confined emergent gauge fields can do something interesting!

For more about parton gauge theory I heartily recommend 
Sung-Sik Lee's TASI 2010 lectures \cite{Lee:2010fy}.
In his lectures 2 and 3, he applies this method to 
bosons and to spins and provides a great deal of insight.

\begin{shaded}
{\bf Attempted parable.}

The parton construction is 
a method for `solving' non-holonomic constraints,
like inequalities.
Here is an example:
I can solve the condition $ y>0$ by writing $ y = x^2$.
So we can do a 0-dimensional path integral (integral) 
over $ y>0$ in terms of an unconstrained variable $x$ by writing
$$ \int_0^\infty  dy ~ e^{- S(y)} 
= \half  \int_{-\infty}^\infty dx~ e^{ \log x - S(x^2) } . $$
In this model, the operation $x \mapsto - x$ is a gauge redundancy.
In this case, it is a finite dimensional gauge group 
and we account for it by the factor of $\half$ out front.

The extra $\log(x)$ term in the action 
from the Jacobian is
like a contribution from the gauge fluctuations.
If I were clever enough I would illustrate deconfinement here,
but I guess that isn't going to happen in zero dimensions.

\end{shaded}

The parton construction makes possible\\
\bluecom{
\noindent
$\bullet$ new mean field ansatze, \\
$\bullet$ candidate many-body groundstate wavefunctions, \\
$\bullet$ good guesses for low-energy effective theory, \\
$\bullet$ accounting of topological ground-state degeneracy and edge states,\\
$\bullet$ an understanding of transitions to nearby states.  (Here are some illustrations
\cite{PhysRevB.89.235116, Barkeshli:2013bma}.)
}

It has the following difficulties: \\
\redcom{
\noindent
$\bullet$ making contact with microscopic description, \\
$\bullet$ its use sometimes requires deciding the IR fate of strongly coupled gauge theories.
}

\breakS

\section{States which can be labelled by their boundary physics}
\label{sec:SPT}

This section provides a low-tech quantum field theory point of view on
symmetry-protected topological (SPT) states. 
Sources for this discussion include \cite{Senthil:2014ooa, Vishwanath:2012tq, Xu:2013bi}.
A useful and brief review is the second part of Ref.~\incite{Turner:2013kp}.
See also \htmladdnormallink{this}{http://www.condmatjournalclub.org/?p=2001} Journal Club for Condensed Matter Physics 
commentary by Matthew Fisher.

We return to our list of possible ways to distinguish phases
(without having to check every possible adiabatic path between them
in the infinite-dimensional space of Hamiltonians).
Here is a simple yet still-interesting question (which has been very fruitful in the past few years): 
how do we distinguish phases of matter 
which preserve $\gG$ and don't have TO?

\hskip-.23in
\parbox{.75\textwidth}{
One answer: put them on a space with boundary,
\ie~an interface with vacuum or with each other.
Quantized (hence topology) properties of the surface states can be characteristic of distinct phases.

The rough idea is: just like varying the Hamiltonian in time to another phase
requires closing the gap $\HH = \HH_1 + g(t) \HH_2$, so does varying the Hamiltonian in space 
 $\HH = \HH_1 + g(x) \HH_2$.
}~~~~\parfig{.15}{figs/fig-interface.png}

This section can be called `holography without gravity'.  
But rather than describing a {\it duality} between bulk and boundary physics,
the boundary physics is going to provide a {\it label} on the bulk physics.



\vskip.2in
Def: a gapped groundstate of some Hamiltonian $\HH$ preserving $\gG$ (without TO) which is distinct from any trivial product state 
is called a SPT state with respect to $\gG$.

\hskip-.23in
\parbox{.59\textwidth}{These states form a group
under composition.
-A is the mirror image.  
A cartoon proof of this is given at right.

Note that with topological order, even if we can gap out the edge states, there 
is still stuff going on 
(\eg~fractional charges) in the bulk.  Not a group.
}~~~
\parbox{.35\textwidth}{
\includegraphics[width=.1\textwidth]{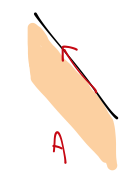} ~~~~~
\includegraphics[width=.1\textwidth]{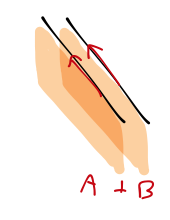}  \\
\parbox{.30\textwidth}{\includegraphics[width=.25\textwidth]{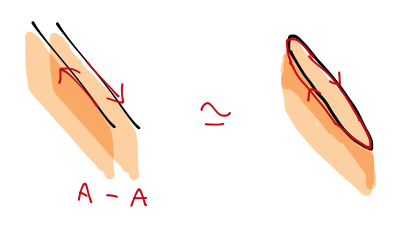}} $\simeq \emptyset$
}

How to characterize these states more precisely?  Many possibilities have been explored so far.
To get started:
\begin{enumerate}
\item If $\gG \supset \gU(1)$, we can study the response to
an external electromagnetic (EM) field.  Just like the integer quantum Hall effect,
this will involve some quantized coefficients.
\end{enumerate}
Without a $\gU(1)$, we need something else.  (And two states with the same quantized EM response
may be distinct for some other reason.)
\begin{enumerate}[resume]
\item What happens if you gauge $\gG$?   In general this produces a new state, with TO (or gapless).  That state is a label. 
 (This works both ways: labelling TO or gapless states is the hard part.)  
 I'm not going to say more about this here.
\item Weird quantized stuff at the surface.  e.g. the surface can have fractionalization and TO.  
\end{enumerate}

{\bf Goals:} SPT states of bosons (especially in $D=3+1$).
Interesting partly because it {\it requires} interactions.
States of bosons also means states of spins by a mapping which I will not review here.

The question being asked here is basically: 
how can symmetries be realized on many-body systems.
We'll maintain a vigorous focus on examples, and not worry about classification.
One outcome of this direction which 
is of interest to quantum field theorists
(but which I will not discuss)
is that the surface of an SPT facilitates 
the emergence of supersymmetry
from a lattice model
\cite{2014Sci...344..280G}.
The basic idea is that the supersymmetric fixed point
has two relevant deformations,
but one of them is forbidden by a funny (SPT-edge) realization 
of time-reversal symmetry.

\subsection{EM response of SPT states protected by $\gG \supset \gU(1)$}

\subsubsection{$D=2+1$, $\gG=\gU(1)$}

This is what we did in \S\ref{sec:3dU1}.
Our conclusion was 
$ \sigma^{xy} = \lim_{\omega \to 0} {\vev{j^x j^y}|_{k=0}\over\ii \omega}  = \nu { e^2 \over h}  = {\nu \over 2\pi},$    
and $\nu$ is quantized if there is no fractionalization (and even in a model of bosons without fractionalization).

\subsubsection{$D=3+1$ and $\gG = \gU(1) \sdtimes \IZ_2^T$}
The effective field theory for any 3+1d insulator, below the energy gap,
has the following form
\be\label{eq:TI-EFT} S_\text{eff}[\vE, \vB] = \int \dd^3 x \dd t \( \eps \vE^2 - {1\over \mu} \vB^2 + 2 \alpha \theta \vE \cdot \vB  + \CO(\vE, \vB)^4 \) \ee
where $\eps, \mu$ are the dielectric constant and permittivity, and $\alpha$ is the fine structure constant.
(In saying that the next corrections go like the fourth power of $E,B$ I am assuming 
that we can approximate the material as isotropic and using rotation invariance.)
Flux quantization implies that 
$${ \alpha\over 32 \pi^2 } \int_{M_4} \vE \cdot \vB = {1 \over 16\pi^2}\int_{M_4} F \wedge F  \in \IZ$$ 
is an integer for any closed 4-manifold $M_4, \partial M_4 = \emptyset$.  
This means that the partition function is periodic
$$ Z(\theta + 2\pi ) = Z(\theta) $$
and hence the spectrum on a closed 3-manifold is periodic in $\theta$.
(As we will discuss, shifting $\theta$ by $2\pi$ is not so innocuous on a space with boundary or for the wavefunction.)

Time reversal acts by 
$$ \CT: (\vE, \vB) \to (\vE, - \vB) $$
which means $ \theta \to - \theta $,
which preserves the spectrum only for $\theta \in \pi \IZ$.
So time-reversal invariant insulators are labelled by a quantized
`magnetoelectric response' $\theta/\pi$ \cite{QH0824}.

Now consider what happens on a space with boundary.
The interface with vacuum is a domain wall in $\theta$, between
a region where $\theta = \pi$ (TI) and a region where $\theta =0$ (vacuum).
The electromagnetic current derived from \eqref{eq:TI-EFT} is\footnote
{A comment about notation: here and in many places below
I refuse to assign names to dummy indices when they are not required.
The $\cdot$s indicate the presence of indices which need to be contracted.
If you must, imagine that I have given them names,
but written them in a font which is too small to read.
}
\be\label{eq:bulkEM} j_{EM}^\mu = { e^2 \over 2 \pi h } \eps^{\mu \cdot \cdot \cdot} 
\partial_\cdot \theta \partial_\cdot A_\cdot + \cdots \ee
where the $ \cdots$ indicate contributions to the current coming from
degrees of freedom at the surface which are not included in \eqref{eq:TI-EFT}.
If we may ignore the $\cdots$ (for example because the edge is gapped),
then we find a surface Hall conductivity
\be\label{eq:surfaceHall}
 \sigma_{xy} = { e^2 \over h} { \Delta \theta \over 2 \pi } = { e^2 \over h } \( \half + n \) \ee
where $\Delta \theta $, the change in $\theta$ between the two sides of the interface,
is a half-integer multiple of $2\pi$.
To be able to gap out the edge states,
and thereby ignore the $\cdots$, 
it is sufficient to break $\CT$ symmetry,
for example by applying a magnetic field.


\hskip-.23in\parbox{.69 \textwidth}{There are two different ways of breaking $\CT$.
The 1+1d domain wall between these on the surface 
supports a {\it chiral} edge mode.

The periodicity in $\theta \simeq \theta + 2\pi$ for 
the fermion TI can be understood 
from the ability to deposit an (intrinsically 2+1 dimensional)
integer quantum Hall system on the surface.  
This changes the integer $n$ in the surface Hall response \eqref{eq:surfaceHall}.
Following \cite{Senthil:2012tm} we can argue
that a non-fractionalized system of bosons in 2+1d must have a Hall response
which is an even integer;
therefore a 3+1d boson TI has a $\theta$ parameter
with period $4\pi$.

}
\parfig{.3}{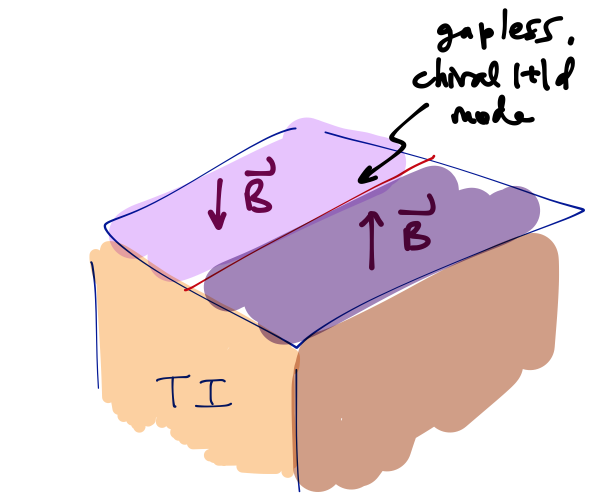}

\sidebarb

\hskip-.23in
\parbox{.78\textwidth}{
Free fermion TIs exist and are a realization of this physics with $\theta= \pi$.
The simplest short-distance completion 
of this model is a single massive Dirac fermion:
$$ S[A, \psi] = \int \dd^3 x \dd t \bar \Psi \( \ii \gamma^\mu D_\mu - m - \tilde m \gamma^5 \) \Psi  . $$
}
~~\parfig{.19}{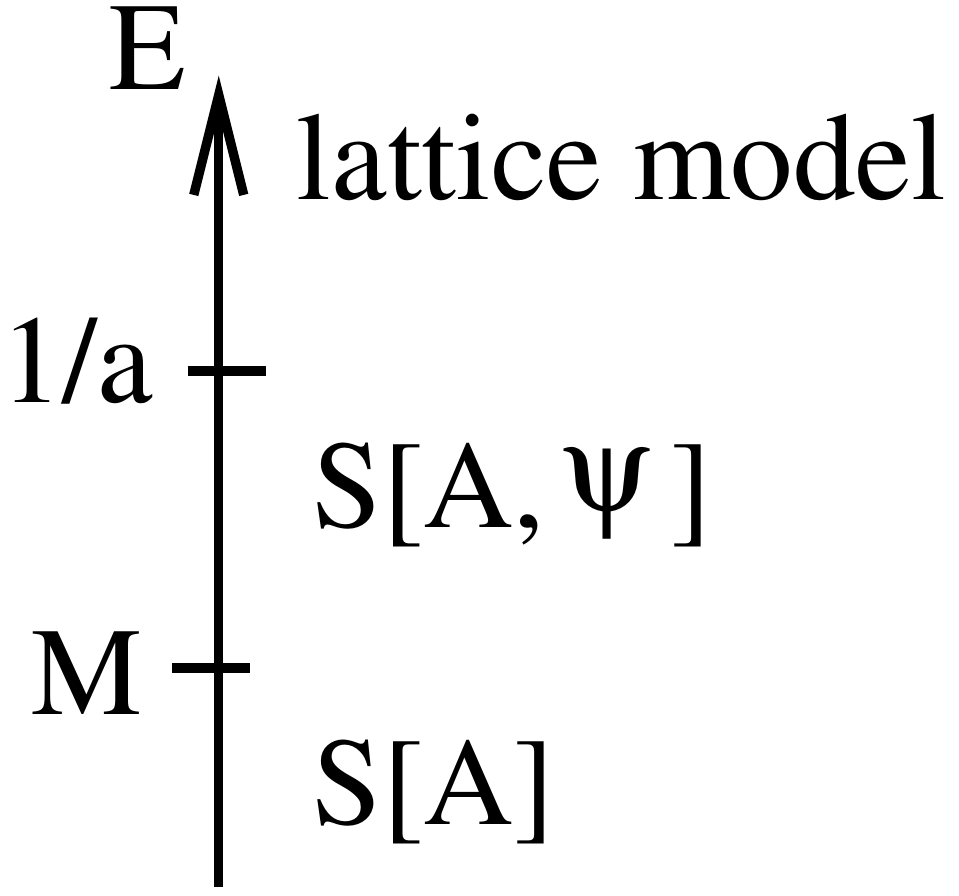}
It is convenient to denote $ M \equiv m + \ii \tilde m$.
$$\CT: M \to M^\star $$
so time reversal demands real $M$.
Integrating out the massive $\Psi$ produces an effective action for the background gauge field
(and $M$) of the form above:
$$ \log \int [D\Psi] e^{ \ii S_{3+1}[A, \psi]} 
= { M \over |M| } \int { \dd^4 x \over 32 \pi^2} \epsilon^{abcd} F_{ab} F_{cd} + \cdots$$
The sign of $M$ determines the theta angle.

\hskip-.23in
\parbox{.69\textwidth}{
An interface between the TI and vacuum is a domain wall in $M$ between
a positive value and a negative value.
Such a domain wall hosts a 2+1d massless Dirac fermion \cite{Jackiw:1975fn}.
(The $\CT$-breaking perturbation is just its mass,
and the chiral edge mode in its mass domain wall
has the same topology as the chiral fermion zeromode in the 
core of a vortex \cite{Jackiw:1981ee}.)
}
\parfig{.3}{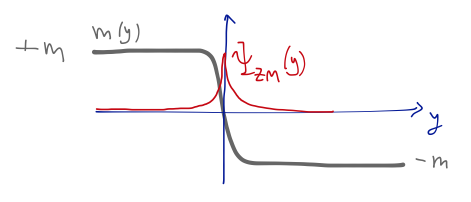}
 
A further short-distance completion of this massive Dirac fermion 
(as in the figure) comes from filling an integer number
of bands with a nontrivial Chern-Simons invariant of the Berry curvature
\cite{Roy-CS, FuKaneMele3d, QH0824}.

\sidebare

With interactions and disorder other edge states are possible
within the same bulk phase, including gapped edge preserving $\CT$ \cite{PhysRevB.88.115137, 2013arXiv1306.3238W,2013arXiv1306.3286M,2013arXiv1306.3230B,2013arXiv1306.3250C}.
New approaches to the edge physics of this system
have appeared recently in \cite{2015arXiv150505141W, 2015arXiv150505142M}.

\subsection{K-matrix construction of SPT states in D=2+1}


(This discussion is from \cite{Lu:2012dt}.) Recall our TO example of abelian FQH, 
with effective action 
$$ S[a_I] = \sum_{IJ} {K_{IJ}\over 4\pi} \int a_I \wedge \dd a_J .$$
The real particle (electron or boson) current is $ j_\mu = \eps_{\mu \nu \rho} \partial_\nu a^I_{\rho} t_I $. 
That is: coupling to the external gauge field is  $ \Delta L = \CA  t^I \eps \partial a^I$. 
We showed that this produces a Hall response
$$ \sigma^{xy} ={1\over 2\pi}  t^{-1} K^{-1} t .$$

The number of groundstates $= \det K^g $.
If we choose $K$ with $\det K=1$, this suggests that there is no topological order.

Actually one more set of data is required to specify the EFT: 
quasiparticles are labelled by $ l^I$ couple to $l_I a^I$.\\
self (exchange) statistics: $\theta = \pi l^T K^{-1} l $ . \\
mutual statistics: $\theta_{12} = 2 \pi l_1^T K^{-1} l_2 $.\\
external U(1) charge of qp $l$: $ Q=t^TK^{-1}l $ .

To make an SPT state, we must ensure that  all these quantum numbers
are multiples (not fractions!) of those of the microscopic constituents.

To describe a boson IQH state, consider $K=  \sigma^x$.  Think of the two states as like two 'layers' or species of bosons, 
so we can take statistics vectors $l_1 = (1,0), l_2 = (0,1)$. These are self bosons and mutual bosons.
If we take the charge vector to be $ t = (1,1)$ (both species carry the charge)
then this state 
has $\nu = 2$.

\sidebarb

\parbox{.79\textwidth}{
{\bf Edge physics.}  Consider abelian CS theory on the lower-half plane.
$$ S = { k \over 4 \pi} \int_{\IR \times \text{LHP}} a \wedge \dd a $$
}
\parfig{.2}{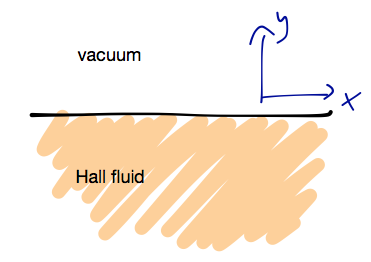}
The equations of motion for $ a_0$ say $ 0 = f =da$.
This is solved by  $a = \ii g^{-1} \dd g =  \dd \phi$, $ \phi \simeq \phi + 2\pi$.
\cite{W89121, Elitzur:1989nr, wen04, Belov:2005ze}.

\vskip.05in

Only gauge transformations which approach $\Ione$ at the boundary preserve $S_{CS}$
This implies that $\phi$ is dynamical on the boundary.
(Or equivalently, we must add a degree of freedom identical to $\phi$ 
 to cancel the gauge variation of the action.)

{A good choice of boundary condition is:} $0 = a - v( \star_2 a) $
i.e. $ a_t = v a_x $.  \greencom{$v$ is UV data.}
We can think of it as arising from a gauge invariant boundary term.

$$ S_{CS}[a= \dd\phi] = { k \over 2 \pi } \int dt dx \( \partial_t \phi \partial_x \phi + v \(\partial_x \phi \)^2 \) . $$
Conclusion: $\phi$ is a chiral boson. 
 $kv >0$ required for stability.
The sign of $k$ determines the chirality.

\hskip-.23in\parbox{.69\textwidth}{
For the case of IQHE ($k=1$),  the microscopic picture in terms of free fermions is at right.
For free fermions in a magnetic field, the velocity of the edge states
is determined by the slope of the potential which is holding the electrons together.
(This can be understood by considering the motion of a classical charged particle in 
a large enough magnetic field that the inertial term can be ignored:
$ q \vec v \times \vec B = - \grad V$, solve for $v$.)
}\parfig{.3}{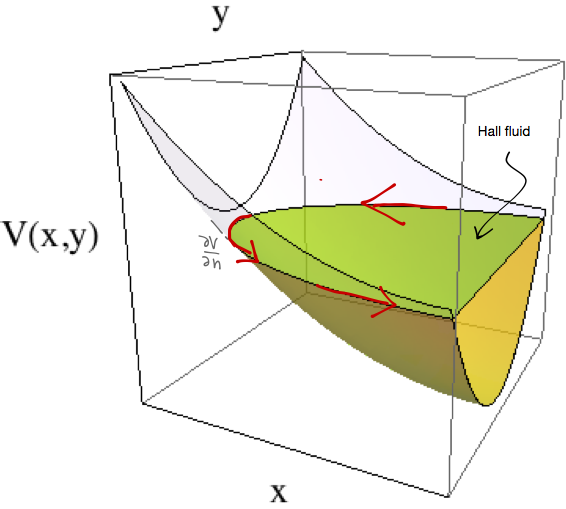}
It is clearly not universal information.

The Hamiltonian $H$ depends on the boundary conditions; the Hilbert space ${\cal H}$ does not.

Put back indices:
$$ S = { K^{IJ} \over 4 \pi} \int_{\IR \times \text{LHP}} a_I \wedge \dd a_J $$
$$ S_{CS}[a^I= \dd\phi^I] = {1\over 4 \pi } \int dt dx  \( K^{IJ} \partial_t \phi^I \partial_x \phi^J + v_{IJ} \partial_x \phi^I \partial_x \phi^J \) . $$
($v$ is a positive matrix, non-universal.)

This is a collection of chiral bosons.  The number of left-/right-movers is the number 
of positive/negative eigenvalues of $K$.

\sidebare

For $\nu=2$ boson IQH \cite{Senthil:2012tm}: $K= \ssigma^x$.
$$ S_{CS}[a^I= \dd\phi^I] = 
{1\over 4 \pi } \int dt dx  \( \partial_t \phi^+ \partial_x \phi^+
- \partial_t \phi^- \partial_x \phi^- +  v \( \partial_x \phi^\pm\)^2 \) $$
$\implies$ $ \phi^\pm \equiv {1\over \sqrt{2}}  \cob \( \phi^1 \pm \phi^2 \)  $ 
are left/rightmoving.

\hskip-.23in\parbox{.69\textwidth}{
Conclusion: it's just a non-chiral free boson (at the \gSU(2) radius).  
Ordinary, at least for string theorists.

But, {\it only} the rightmover is charged!  
The difference arises in the coupling to the external gauge field: since $t = (1,1)$,
}
\parfig{.3}{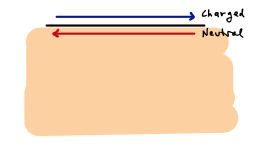}
$$L \ni  \CA^\mu \partial_\mu \( \phi^1 + \phi^2\) \propto \CA^\mu \partial_\mu \phi^+ ~.$$
Specifically, although $c_L = c_R$, 
only the left mover $ \phi_+$ carries the $\gU(1)$ charge.
This means that preserving $\gU(1)$, we can't backscatter, 
that is we can't add to the action (local) terms like $ \Delta S = g_{\pm } \cos\( \phi^+ \pm \phi^- + \alpha\) $
($\alpha$ is a constant)
which would lift the edge states.  
(Such terms made from just $ \phi^+$ could not be local.)
This means the $\gU(1)$ {\it protects} the edge states.

In its realization as the gapless low-energy theory of the spin-$\half$ AF Heisenberg chain,
only the $\gSU(2)_\text{diag}$ is manifest, not the chiral symmetries, which are emergent.

\subsection{BF theory for 3+1d boson SPTs}

Consider the following $D=3+1$ analog 
of CS theory \cite{Vishwanath:2012tq}
$$ S[B, a] = \sum_I  {1\over 2\pi} \eps^{\cdot\cdot\cdot\cdot}
B_{\cdot\cdot}^I \partial_\cdot a_\cdot^I 
+ \vartheta \sum_{IJ} {K_{IJ}\over 4 \pi} \partial_\cdot a^I_\cdot  \partial_\cdot  a^J_\cdot \eps^{\cdot\cdot\cdot\cdot} . $$
Note that the theta angle $\vartheta$ here is not the same as the $\theta$
in the magnetoelectric resonse.

It is topological, like CS theory, in that we didn't need to introduce the metric
to integrate the action covariantly.  In $D=3+1$ we need the form degrees  to add up $2+1+1 =  4$.
We can add analogs of Maxwell terms (for both $B$ and $a$), but 
just like in $D=2+1$ they are irrelevant, \ie\ they merely introduce new UV physics,
they don't change the IR.

(Note that the more general seeming thing with a more general matrix
coupling $F$ and $B$ can be removed by an integer-valued field redefinition
which changes nothing.)


Focus on the case $K = \ssigma^x$.
One virtue of this effective action is that it reproduces the 
EM response we expect of a topological insulator.  If we couple to an external $\gU(1)$ gauge field $\CA$ by
$$ \Delta \CL = \CA_\mu \( j_1^\mu + j_2^\mu\) $$
then 
$$ \log \int [Da DB] e^{ \ii S[a, B, \CA]}  = \int { 2 \vartheta \over 16 \pi^2 } \dd \CA \wedge \dd \CA + \cdots $$
that is, the magneto-electric response is $\theta_{EM} = 2 \vartheta$.
So $ \vartheta = \pi$ will be a nontrivial boson TI.

Briefly, who are these variables?  in 2+1:  the flux of the CS gauge field was some charge density.
Here, each copy is the 3+1d version of charge-vortex duality, where
for each boson current
$$ j_\mu^{I=1,2} = {1\over 2\pi} \eps_{\mu\cdot\cdot\cdot} \partial_\cdot B_{\cdot\cdot}^I $$
which has $ \partial^\cdot j_\cdot^I = 0$ as long as $B$ is single-valued.

The point of $a$ is to say that $B$ is flat.
The magnetic field lines of $a_\mu$ are the vortex lines of the original bosons $b_I$.

\subsection{Ways of slicing the path integral}

Now let's think about the path integral for a QFT with a theta term.
Examples include the BF theory above, and many non-linear sigma models 
which arise by coherent-state quantization of spin systems.
In general what I mean by a theta term is a term in the action
which is a total derivative, and where the object multiplied by 
theta evaluates to an integer on closed manifolds.
The following point of view has been vigorously emphasized by 
Cenke Xu \cite{Xu:2011sj, Xu:2013bi}.

When spacetime is closed $Z(\theta + 2\pi) = Z(\theta)$.  
on a closed spacetime manifold $M_D$
$$ Z_\theta(M_D) \equiv \int [D\text{stuff}] e^{ - S} = \sum_{ n \in \pi_2(S^2)} e^{ \ii \theta n } Z_n $$
and $ Z_\theta(M_D) = Z_{\theta + 2\pi}(M_D)$.
In particular, we can take $M_D = S^1 \times N_{D-1}$ to compute
 the partition function on any spatial manifold  $N_{D-1}$.
 This means the bulk spectrum is periodic in $\theta$ with period $2\pi$.

With boundaries, it not so in general. 
A boundary in space produces edge states.  We'll come back to these.

A boundary in time in the path integral means we are computing wavefunctions.
For quantum mechanics of a single variable $q(t)$, this is manifested in the Feynman-Kac formula
$$ \psi(q) = \int _{q(t_0) = q} \prod_{t \in (-\infty, t_0)} \dd q(t) ~e^{ - S_\text{euclidean}[q] } ~.$$
For a field theory, 
`position-space wavefunction' means a wavefunctional $ \Psi[\phi(x)] $,
in 
$$ \ket{\Psi} = \int [ D \phi(x) ] \Psi[\phi(x)] \ket{\phi(x)} $$
where $x$ labels {\it spatial} positions, and $ \ket{\phi(x)}$ are coherent states for the field operator $\hat \phi(x)$.
Which wavefunction? If the path integral is over a large euclidean time $T$ 
before reaching the boundary, 
this is a {\it groundstate} wavefunction,
since the euclidean time propagator $e^{ - T \HH } $
is a (un-normalized) projector onto lowest-energy states.

\sidebarb

Semi-philosophical digression: An important guiding concept
in the study of interesting gapped states
is that it is the {\it same stuff}
living at a spatial boundary (edge modes) as at
a temporal boundary (the wavefunction)
\cite{W89121, Elitzur:1989nr, MR9162, 2011PhRvL.107j6803S}.
This perspective first arose (I think)
in the context of quantum Hall states
where famously \cite{MR9162}
one can write groundstate and several-quasiparticle wavefunctions
as correlation functions of certain operators 
in a 1+1d CFT, which is the same CFT that 
arises at a spatial edge.
Why should this be true?
It's because the bulk 
can be described by a path integral
for a Chern-Simons gauge theory
which has a certain WZW model living at its
boundaries, wherever they are.
For a spatial boundary, it produces a copy of that CFT at the boundary
(roughly the group-valued CFT field $g$ is related to the CS gauge field by
$ A = g^{-1} \dd g $).
\begin{center}
\includegraphics[height=.2\textwidth]{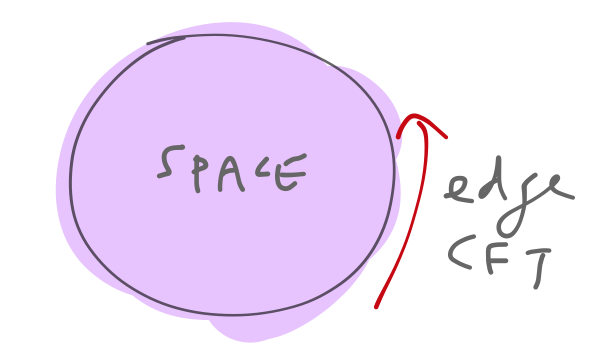}
~~\includegraphics[height=.2\textwidth]{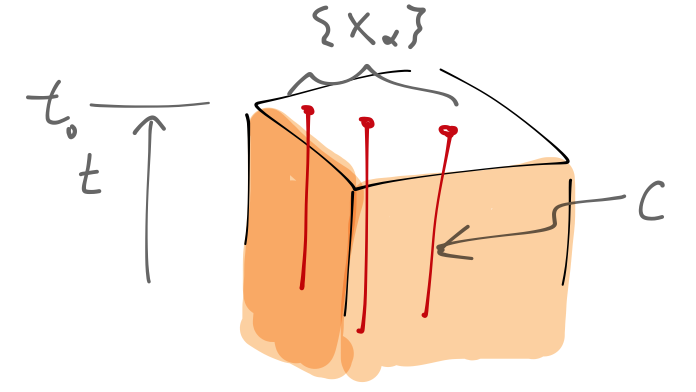}
\end{center}
For a temporal boundary, the path integral expression for the wavefunctional
(with some Wilson line insertions at the positions of the electrons) takes the form
\be\label{eq:MR}  \Psi[g(x)] = \int_{A(t_0, x) = g^{-1} \dd g } e^{ \ii S[A]}  W[C] = \vev{ \prod_\alpha V_\alpha(x_\alpha) }_{WZW} .\ee
A too-brief explanation of this rich formula: 
the wilson line insertion is $W[C] = \tr_R {\cal P} e^{ \ii \oint_C A} $
where $R$ is a representation of the gauge group $G$ and ${\cal P}$ 
is path ordering. 
In \eqref{eq:MR}, $ x_\alpha$ are the locations where the curve $C$ intersects the fixed-$t=t_0$ surface,
and $V_\alpha$ are some operators in the CFT the appropriate representations $R$ of $G$\footnote{
For more on this connection in the context of the quantum Hall effect 
I recommend David Tong's new QHE notes
\cite{Tong:2016kpv}, 
in particular the section headed ``What the Hell Just Happened?"}.

For spin chains, this point of view
is used in \cite{anne-bang-nielsen}
to construct spin chains whose continuum limit is the SU$(2)_k$ WZW model with $k>1$.

For 3+1d boson SPT states, the analogous bulk EFT
is, instead of CS gauge theory, 
some weird BF theory or strongly-coupled sigma model,
both of which we'll discuss below.
At a spatial edge, we have some vortex excitations in $D= 2+1$.
Correspondingly, the bulk wavefunctions
will turn out to have a nice representation in a basis of states
labelled by vortex loop configurations in $D=3+1$.

\sidebare

Side remark: the canonical application of this story is to the 
{\it Haldane chain} -- a chain where each site carries a representation of SO$(3)$.
At low energies, such chains are described by an NLSM with a theta term.
$\theta = 0$ is trivial and gapped.  $\theta =2\pi$ is gapped and trivial in the bulk
but the edge states are spin $\half$s -- a projective representation of SO$(3)$.

\vskip.1in
\hrule 
\vskip.1in

Let's apply this picture to BF theory.
The basic manipulation we are doing here is formally the same as in \cite{Xu:2013bi}.

In contrast to the case of a closed manifold, if we compute the path integral 
on an (infinite) cylinder (\ie~with two boundaries, at $\tau = \pm \infty$),
then $\theta$ does matter, not just mod $2\pi$.

Choose $ A_0 = 0$ gauge.  Since $A$ and $B$ are conjugate
variables, the analog of position space here is $ \ket{\vec A(x)}$.
For the same reason, we can only specify BCs on one or the other:
\be\label{eq:feynmankacsigma} \int_{\scriptsize 
\begin{matrix} \vec A(x, \tau = \infty) = \vec A(x) \\
\vec A(x, \tau = -\infty) = \vec A'(x) 
\end{matrix}
} [D\vec A(x, \tau) DB(x,\tau)] e^{ - S[\vec A(x, \tau), B]}
= \vev{\vec A(x)  \Big| 0}\vev{0\Big| \vec A'(x)} 
\ee
Notice that in expressions for functionals like $S[A(x, \tau)]$
I am writing the arguments of the function $A$ to
emphasize whether it is a function at fixed euclidean time or not.
The fact that the theta term is a total derivative 
$$ F^I \wedge F^J = \dd\( A^I \wedge F^J \) \equiv \dd  w(A) $$
means that the euclidean action here is
$$ S[\vec A(x,\tau)] = 
 \int_{M_D} 
{1\over 4\pi}  B \wedge F 
+ \ii \theta \int_{\partial M_{D}} \dd^{D-1} x \( w(\vec A(x)) - w(\vec A'(x))\)  . $$
The $\theta$ term only depends on the boundary values, and comes out of the integral 
in \eqref{eq:feynmankacsigma}.

The integral over $B^I$ is 
$$ \int [DB] e^{ \ii  {1\over 4\pi}\int B^I \wedge F^I } = \delta [F^I]. $$
The delta functional on the RHS here sets to zero
the flux of the gauge field for points in the interior of the cylinder.

After doing the integral over $B$,
there is nothing left in the integral
and we can factorize the expression \eqref{eq:feynmankacsigma} to determine:
\be\label{eq:bosonGS}
 \Psi[\vec A_I(x)] = \underbrace{\exp{  \ii { \vartheta \over 8 \pi^2 } \int_\text{space} A_\cdot^I \partial_\cdot A^J_\cdot \eps^{\cdot\cdot\cdot}  K^{IJ}  } }_{\buildrel{K=\ssigma^x}\over{=} \ii {\vartheta\over 2}
 \( \text{linking $\#$ of $2\pi$ magnetic flux lines} \) 
 }
\ee

What does this mean? 
Label configurations of $A$ by the flux loops (i.e. the field lines of the vector field).
This wavefunction is $ (-1)^\text{linking number of the $1$-loops and the $2$-loops} $.

If we break the U$(1) \times\text{U}(1)$ symmetry, the flux lines of $1$ and $2$ 
will collimate (by the Meissner effect).

Claim: in the presence of an edge, these flux lines can end.
The ends of these flux lines are fermions.
(Warning: doing this right requires a framing of the flux lines -- \ie~they shouldn't collide.)
Conclusion: on the surface of this SPT state of bosons there are
{\it fermionic} vortices.

\begin{figure}[h!]
\begin{center}
\includegraphics[width=\textwidth]{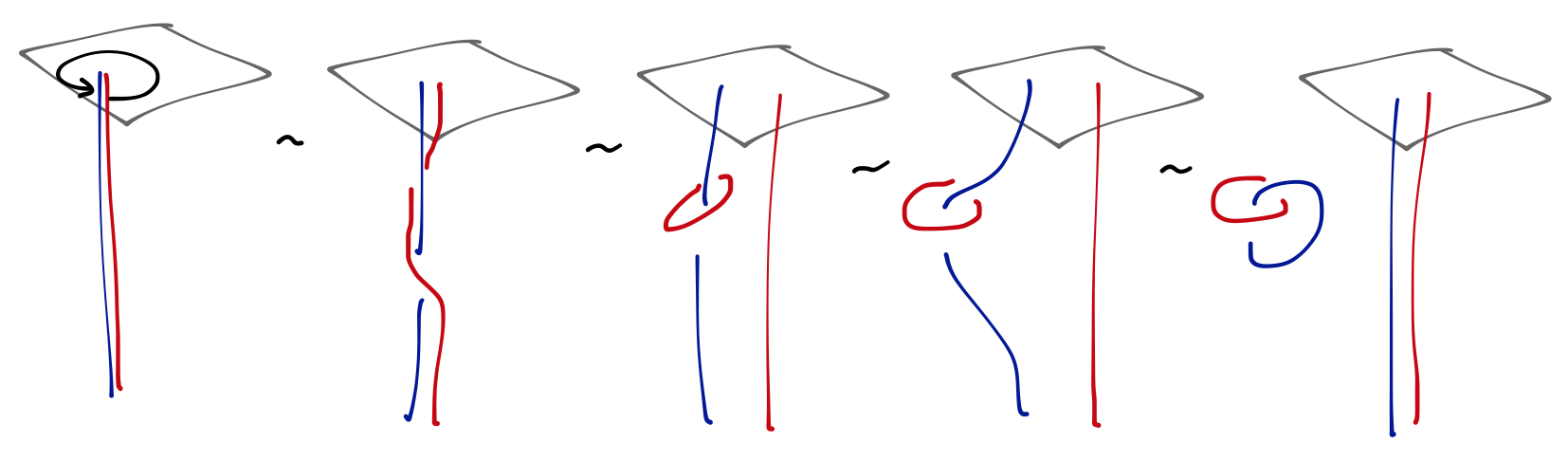}
\caption{The end of the ribbon is a fermion, from \cite{Xu:2013bi}.  
In the first step, we rotate the red string around the blue one.
The squiggles mean that 
the states associated with these configurations
have the same amplitude in the groundstate, according to \eqref{eq:bosonGS}.}
\end{center}
\end{figure}

Note that the BF theory describes a very strongly confined abelian gauge theory in the following sense:
the flux gets set to zero by the $B$ term.
(with a string source for $B$ the flux gets localized to the source.)

{\bf Comment on Kodama state of gravity.}
This wavefunction \eqref{eq:bosonGS} actually solves the Schr\"odinger equation
for quantum Maxwell theory at {\it finite} coupling.
There is even a non-Abelian version of it for which this is also true.
There is even an analog for gravity called the `Kodama state'!
What's the catch?  It's not normalizable as a wavefunction 
for photon fields \cite{Witten:2003mb};
attempting to quantize the model about this groundstate
gives negative energy for one of the two circular polarization states.
But as a wavefunction for the confining phase of the gauge theory it's fine.

Note that \cite{Xu:2013bi} does an analogous thing 
for very strongly coupled sigma models with theta terms;
they just set the kinetic term to zero (!) and find wavefunctions
closely analogous to \eqref{eq:bosonGS}.
They would have the same problem as Witten points out if they thought of their wavefunctions
as wavefunctions for gapless magnons.  But for the disordered phase
of the sigma model (gapped and analogous to confinement) it is just fine.

Speaking of confining gauge theories in $D=3+1$: for a long time
it was believed (more defensible:  I believed) that only deconfined phases of 
parton gauge theories were interesting for condensed matter --
why would you want to introduce a gauge field 
if it is just going to be confined?
An exception 
appears in \cite{RV0821, BM1293, Grover:2012wp, 2012arXiv1210.0909L}
where 
confined states of parton gauge theories are used to describe superfluid phases in $D=2+1$ in terms
of variables that allow access to nearby fractionalized phases
(and the degrees of freedom that become light at the intervening phase transitions).
More recently, confined states of parton gauge theories have been used 
to discuss $D=3+1$ SPT states in \cite{Cho:2012mg, Metlitski:2013uqa, Ye:2013wma, 2013arXiv1306.3286M}.

\subsection{Comments}

\subsubsubsection{Coupled-layer construction.} An important omission 
from the above discussion is the `coupled-layer' construction of SPT states,
which allows a construction of the bulk state 
directly using many copies of the boundary excitations.

\hskip-.23in
\parbox{.4\textwidth}{
An example for the case of the boson IQHE \cite{Vishwanath:2012tq}
is shown in the picture at right.
This is quite holographic in spirit.
The way the edge excitations emerge then is just like in the classic picture
of edge charge from polarization of an insulator;
the role of the polarization angle is played by $ \arctan\left({t_e \over t_0}\right) $,
where $t_{e/o}$ are the couplings between the layers
on the even and odd links respectively.
}
\parbox{.59\textwidth}{
\includegraphics[width=.59\textwidth]{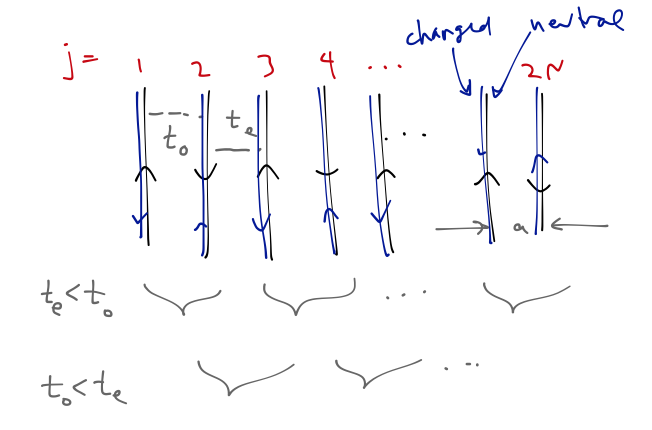}}

\subsubsubsection{Snake monsters.} 
I would like to mention 
a lovely recent construction
in 
\incite{wns2015}.  
They showed that 
a groundstate which is a uniform-magnitude superposition
of fluctuating closed strings does not necessarily mean topological order.
In particular, the signs matter.  
In this paper \incite{Ben-Zion:2015cyz},
we use this idea to make solvable
models of boson SPT states, 
by attaching symmetry-carrying degrees of freedom to the 
lines of electric flux of the models of \incite{wns2015}.

\subsubsubsection{Why you should care about SPTs.}  
For me, the point of studying SPT states is twofold.  
One reason is that by understanding all the ways in which
symmetries can be realized in quantum systems,
we can learn how to gauge them,
and thereby make new states with topological order.  
Although quite a bit has been learned about 
what SPTs there are, and about their classifications
(\eg~\cite{Kitaev-2011, Kitaev-unpublished, ChenGuWen, Kapustin:2015uma, Kapustin:2015uma}), 
this program is not complete.  

A second reason is that each SPT state 
represents an obstruction to
regularizing a QFT with certain properties.  
Such a QFT does not come from a sod-like model.
Some new anomalies have been found this way (\eg~\cite{Kravec:2013pua, Thorngren:2014pza, Kravec:2014aza}).
This perspective has added some new vigor to the quest 
to define the Standard Model (a chiral gauge theory) on the lattice \cite{Kaplan:2009yg, Eichten:1985ft, Wen:2013ppa, 2014arXiv1402.4151Y, You:2014vea}.

\breakS
\section{Entanglement renormalization by an expanding universe}
\label{sec:s-sourcery}

A deep lesson of late 20$^\text{th}$ century physics is the renormalization group philosophy: 
our thinking about extensive systems should be organized scale-by-scale.  
This lesson has been well assimilated into our understanding of
classical statistical physics and in perturbative quantum field theory,
and it is geometrized by AdS/CFT.
In strongly-correlated quantum systems, however, we still have a great deal to learn from this viewpoint,
in particular
about eigenstates and even groundstates of local model Hamiltonians.

\subsection{Basic expectations about entanglement
in groundstates of local hamiltonians}

\label{subsec:area-law}

\parbox{.6\textwidth}{
We began our discussion by saying that $\CH = \otimes_x \CH_x$ is the Hilbert space. 
However, most of many-body Hilbert space is fictional, in the sense that it cannot be reached 
from a product state
by time evolution 
with local Hamiltonians.
The representation of the wavefunction 
$$ \ket{\psi} = \prod_i \sum_{s_i = \pm } c_{s_1...s_{2^{L^d}}} \ket{ s_1...s_{2^{L^d}} } $$
}~~
\parbox{.38\textwidth}{
    \includegraphics[width=0.38\textwidth]{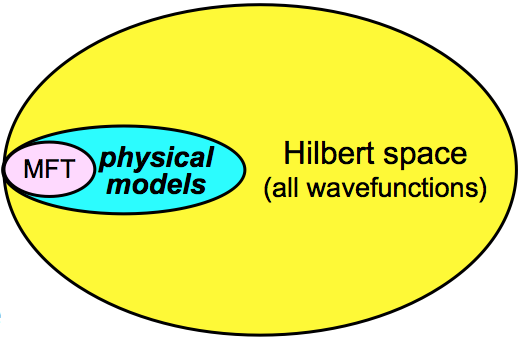} \\
    {\tiny Fig courtesy of Dan Arovas, Master of Keynote}
 }
 as a vector of 
$ e^{L^d \log 2} $ complex numbers $c_{s_1...s_{2^{L^d}}}$ is not useful.
You can't get there from here.
(For more rhetoric along these lines, I recommend \eg~\cite{Poulin:2011zz}.)
This result is closely related to 
the {\it Small Incremental Entangling} Theorem \cite{2013PhRvL.111q0501V},
which we will use below.
It bounds the rate at which 
time-evolution by a local Hamiltonian
can change the EE of a subregion\footnote{
I think this statement does not contradict 
opposite-sounding statements that have been made by other lecturers
at this school about QFT.
For example, David Simmons-Duffin proved that any state of a CFT
can be made by acting with convergent sums of 
primaries and descendants on the vacuum and time evolving.
I believe the important loophole is that the duration of the time evolution is not finite.}.

\hskip-.21in
\parbox{.89\textwidth}{Mean field theory means unentangled states of the sites, 
of the form
$\ket{\psi_\text{MF}} =  \otimes_i \( \sum_{s_i = \pm } c_{s_i} \ket{ s_i} \) $.
This writes the state in terms of only $L^d$ numbers $c_{s_i}$;
this is too far in the other direction.   
}~~\parfig{.1}{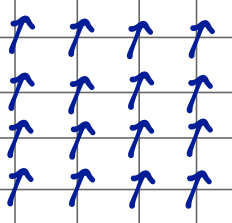}

There is more in the world:   entangled groundstates (even if only short-range-entangled) mean new phenomena,
\eg~SPT states (whose entanglement
looks something like this: \parbox{2cm}{\includegraphics[width=.4\textwidth]{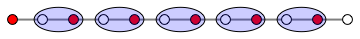}}~~~~~~~~~~~~~~~~~~~~~~~~~~~~~~)
topological order,  quantum critical points, ...

 \begin{framed}
Problem: characterize the physical corner of $\CH$ 
 by entanglement properties and parameterize it efficiently,
 in particular in a way which allows observables to be efficiently calculated 
 given the wavefunction.
 \end{framed}

An energy gap implies correlations of local operators are short-ranged.
\greencom{$ \vev{\CO(x) \CO(0) } \sim e^{ - |x|/\xi} $}.
But the entanglement needn't be!  
(The toric code is an example.)
The time has come to quantify entanglement between $A$ and its complement $\bar A$:
\bea 
\rrho_A &\equiv& \tr_{\bar A} \ket{\psi}\bra{\psi}  \cr
 &&\cr
S(A) &\equiv& - \tr \rrho_A \log \rrho_A  
\nonumber
\eea

Groundstates of local Hamiltonians are  special: 
generically (with few, well-understood exceptions) 
the entanglement entropy of large-enough subregions satisfies an area law. 
(Large-enough means large compared to the correlation length, so that the 
massive particles can't get involved.)
This means $S_A$ scales like the area of the boundary of the region in question.  
The idea is just that strongest entanglement is between nearest neighbors.
I will state it as a \\
{\bf Basic expectation}
({\bf area law}):   
In groundstates of local, motivic $\HH$,  
$$S_A = a R^{d-1} + \text{smaller} $$
where $R$ is the linear size 
of the subregion $A$, say its diameter.
This statement is supported by a great deal of evidence and 
has been rigorously proved for gapped systems in 1d \cite{2007JSMTE..08...24H, 2013arXiv1301.1162A}.
We will give a more general explanation below, following \cite{s-sourcery}.
It has been essential in identifying efficient numerical representations of groundstates 
in terms of tensor networks,
and in the development algorithms for finding them 
(a useful introduction is \cite{Orus:2013kga, Orus:2014poa} and just recently 
\cite{2016arXiv160303039B}).

The area law is a coarse statement which says essentially
that each degree of freedom is most entangled with its closest neighbors.  
In $d=1$ space dimensions, quite a bit is known rigorously: 
An energy gap implies area law for groundstate \cite{2007JSMTE..08...24H}.
In turn, the area law implies a MPS (matrix product state) representation of the groundstate 
(for a review of this subject, I recommend Ref.~\incite{Orus:2013kga}).
This is the form of the groundstate that is output by DMRG algorithms 
(for more explanation, see \eg~\cite{2011AnPhy.326...96S}).
$$ 
\parbox{1.7cm}{\vskip0in \mbox{\includegraphics[height=.5cm]{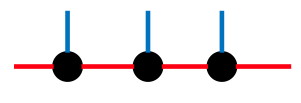}}}
= \sum_{{ \cor a_{1,2...} =1}}^{{\cog \chi}}  M^{{\color{lightblue} \sigma_1}}_{\cor a_1 a_2 } M^{{\color{lightblue}\sigma_2 }}_{\cor a_2 a_3 }  \cdots \ket{
{\color{lightblue}\sigma_1, \sigma_2 \cdots }} 
$$
{\cog $\chi$}, the range of the auxiliary index, is called the {\it bond dimension}.
This encodes the groundstate in $  L^{d=1} \( \chi^{2} \) $ numbers.
In such a state, each site is manifestly entangled only with its neighbors.

In $d>1$ more can happen.  
Let me remind you that the essential beyond-mean-field-theory phenomenon is
{\bf emergence of gauge theory}.  
A direct generalization of an MPS to $d>1$ is called a `PEPS' 
and looks like this: \parbox{2cm}{\includegraphics[width=2cm]{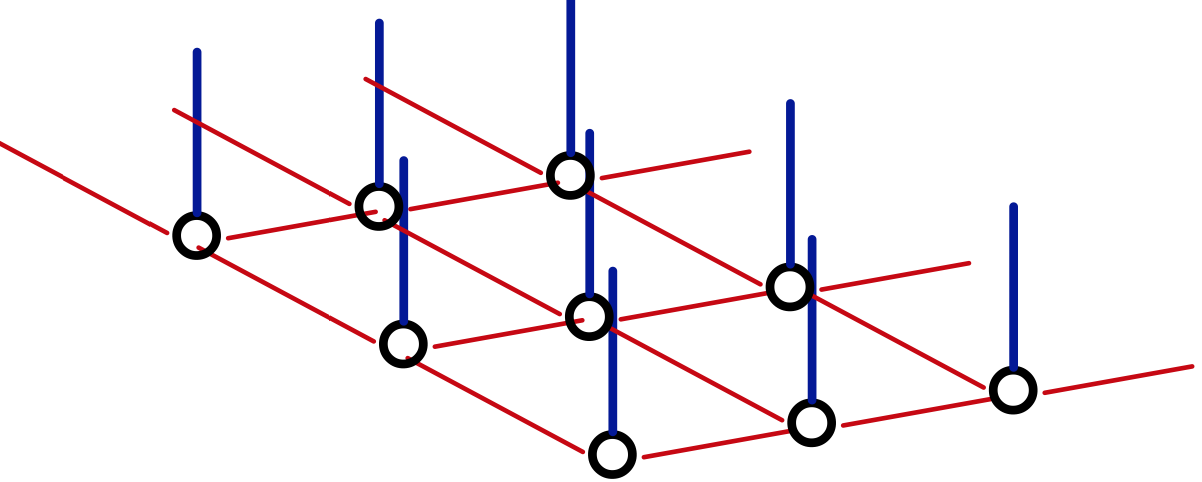}}.
Even if the bond dimension $\chi$ is small,
such a network is slow to contract, which one needs to do to compute matrix elements,
and to determine the values of the tensors, for example in a variational calculation.
Worse, even with a gap, 
rigorous results only 
show that there exists a PEPS with $\chi \sim  e^{ \log^d(L) } $,
growing with linear system size $L$.

%

Numerical methods which incorporate only this datum (the area law) struggle
with gapless states in $d=1$ and even with gapped states in $d >1$.
A more refined statement takes into account how much entanglement
there is at each length scale.
A payoff of incorporating this extra data allows one to make efficiently-contractible networks.
The process of organizing our understanding of the entanglement in the state scale-by-scale 
is called {\it entanglement renormalization}.
The best-developed implementation of this idea is MERA (the multicale entanglement renormalization ansatz) \cite{mera}, which is the state-of-the-art method for the study of 1d quantum critical points 
\cite{Pfeifer:2008jt, 2011arXiv1109.5334E}.

{\bf Brief comments about groundstate EE for gapless case and Fermi surfaces.} 
Groundstates of CFTs in one dimension famously violate the aree law (\eg~\cite{Calabrese:2009qy}): 
$$S(A) = { c\over 3} \log R/\eps + ... $$
for a region $A$ which is an interval of length $R$, 
$\eps$ is a short-distance cutoff, 
and $c$ is the Virasoro central charge.  
Critical points above one dimension, in contrast,
are expected to satisfy the area law.  More about this in \S\ref{sec:gaplessEE}.

The entanglement entropy of 
groundstates with a Fermi surface 
violates the area law,
by a logarithmic factor: $S(A) \sim ( k_F R)^{d-1} \log \( k_F R\) $.
This was shown for free fermions in 
\cite{PhysRevLett.96.100503, 
PhysRevLett.96.010404},
which conjectured a nice expression
for general shape of FS and of the region called 
the Widom formula.
An appealing picture of the violation 
in terms of 
1d systems at each point on the Fermi surface 
was developed in
\cite{Swingle:2009bf,
Swingle:2010yi}.
A strong numerical test
of this picture was made in 
\cite{2011PhRvL.107f7202Z}.
This allows for an extension 
to non-Fermi liquids (where the CFT at each $\vec k_F$ is
not just free fermions), 
a result which was 
confirmed numerically in
\cite{2011PhRvL.107f7202Z}.

\noindent {\bf Geometry is made of entanglement.}
 Above I have emphasized 
the practical consequences 
of entanglement renormalization for numerical algorithms for quantum lattice systems of 
interest to condensed matter physics.
A further motivation for its study is the fact that this point of view 
provides a deep connection to gauge/gravity duality \cite{Swingle:2009bg} and hopes of understanding its origins.
In particular, there is a great deal of evidence that 
the robustness of the geometry of the dual gravity system
is directly related to the structure of entanglement in the quantum state it describes \cite{Swingle:2009bg, 2009arXiv0907.2939V}.

As other talks at this school will have made clear,
this discussion is therefore a step in a larger program to understand the emergence of space
in gauge/gravity duality: 
entanglement determines \redcom{(much of)}\footnote
{An interesting exception seems to be behind horizons, where time is emergent: it seems extra data 
about the {\it complexity} of the state is required, 
discussion of which can be found in Refs.~\incite{Brown:2015lvg, Brown:2015bva}.} 
the bulk geometry.
The entanglement of a subregion is bounded by the minimum number of bonds which must be cut to remove it from the graph.
(Hyperbolic tensor networks which saturate this bound were recently constructed 
in Ref.~\incite{Swingle:2012wq, Pastawski:2015qua}.)
A slogan for extracting the geometry is the equation relating the 
line element to differences of entropies:
$$ \parbox{.65\textwidth}{\includegraphics[width=.65\textwidth]{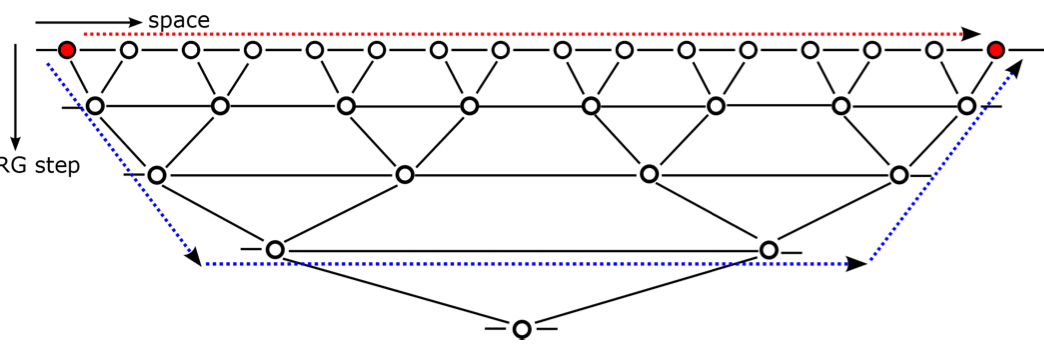}} 
ds^2 \sim dS^2 $$

An outline of the program looks like this:
$$ 
\CH, \HH
\buildrel{ {\text{RG circuits}}}\over{\to}
\parbox{.2\textwidth}{\includegraphics[width=.2\textwidth]{figs/fig-emergent-geometry.png} }
\buildrel{\text{entropy $\sim$ area}}
\over{\to}
G_{\mu\nu} 
= T_{\mu\nu}  
$$
The second part of the program is the part which has a received a lot of attention, using
the Ryu-Takayanagi formula related area to entanglement entropy.
Here we will focus on the first step.

\subsection{$\cog s$-sourcery}

%
%

{Here's a brief summary of what we'll do in this subsection.}
Using such a renormalization group point of view, 
 Brian Swingle and I gave a proof of the area law 
for a large class of gapped systems in arbitrary dimensions \cite{s-sourcery}.
The key idea is a construction called an {\it RG circuit}.
Briefly, it is an approximately-local unitary map from the groundstate at linear size $L$ (times some decoupled
product states called `ancillas') to 
the groundstate at size $2L$.
The work in \cite{s-sourcery}
demonstrated the virtues of a (quasi)-local unitary map realizing an RG step which doubles the system size.
It can be used to prove rigourously the area law for the entanglement entropy of subregions.
It can be used to build an efficiently-contractable tensor network representation of the state (a MERA).

For some systems no such map exists 
from {\it one} copy of the groundstate.
Rather, one must act on $s>1$ copies of the state at size $L$ 
to obtain unitarily one copy at size $2L$.
We call such a state an $s$-source RG fixed point.
The first nontrivial gapped example of $s>1$ was found by Haah \cite{2014PhRvB..89g5119H}. 
$s \leq 1 $ is a necessary condition for the low-energy physics to be well-described by a 
continuum quantum field theory.

For gapped states, we can demonstrate the existence of an RG circuit by various methods 
which rely on the gap.  In particular, from a continuous Hamiltonian path, 
powerful filtering techniques \cite{2005PhRvB..72d5141H, 2010arXiv1008.5137H} exist which produce an exact quasilocal unitary map.


{\bf Context.}
As always, we assume a local many body system, with $ H = \sum_x H_x$  a hamiltonian `motif',
and that support of $H_x$ is localized.
We will consider families of systems labelled by (linear) system size $L$:
$ H_L $ with groundstate(s) $ \{ \ket{\psi_L} \}$.

We seek an efficiently-findable ${\cor \UU}$ which constructs the groundstate 
{\it from smaller subsystems which are not entangled with each other} : 
$$ \ket{\psi_L } \buildrel{?}\over{=} {\cor \UU} \ket{0} ^{\otimes L} $$

\hskip\parsnip
\parfig{.39}{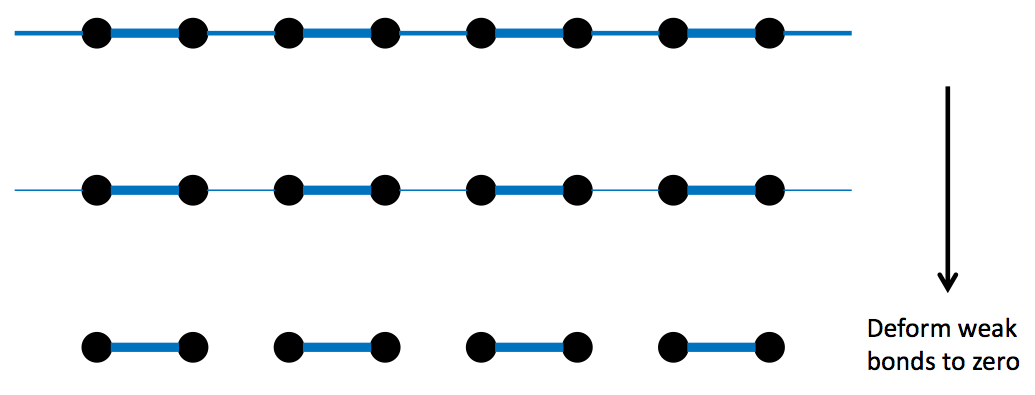}
~\parbox{.58\textwidth}{
Depicted at left is a  {\bf warmup example} (which we will learn to call $d=1, \cog s=0\cob$): adiabatically deform a 1d band insulator to  a product state.
}
By this I mean: consider a chain of fermions
$\{c_i, c_j^\dagger\} = \delta_{ij} $.
\footnote{Actually fermions are not sod-like, in that their microscopic Hilbert space is {\it not} a
tensor product: fermions at distant sites anticommute.  Let's not worry about this right now.}
Take a quadratic hamiltonian, so we can solve it:
$$H = \sum_{i=1}^L t_i c_i^\dagger c_{i+1} + h.c. .$$
Gradually turn off $t_i$ for odd $i$.
When they get to zero, the system is a collection of two-site molecules,
and the groundstate is the product of their groundstates.

This suggests the following strategy
to find our $\cor \UU$:  Find a hamiltonian $H(0)$ whose groundstate
is the product state $\ket{\psi(0)} \equiv \ket{0} ^{\otimes L} $.

\hskip-.23in
\parbox{.4\textwidth}{

Then {\it adiabatically} interpolate between $H(0)$ and the desired hamiltonian $H= H(1)$:
$$ H(\eta) \equiv (1-\eta) H(0) + \eta H .$$
}
\parbox{.59\textwidth}{\includegraphics[width=.6\textwidth]{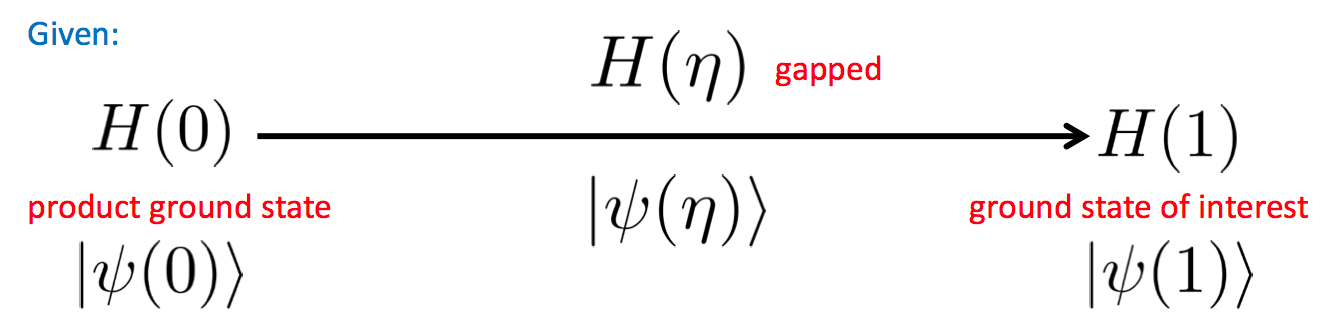}
}
Then, we can just use the (fake) time evolution operator as our unitary:
$${\cor \UU} \buildrel{?}\over{=} \text{P} e^{ \ii \int^1_0 d \eta \HH(\eta) }   .$$

{There are just two problems with this plan, in general:}

\begin{enumerate}

\item The first problem is a technical, solvable one. Adiabatic evolution has a nonzero failure probability
(per unit time, per unit volume).

\begin{shaded}
{The solution \cite{2005PhRvB..72d5141H} is:  }\\
$\text{Find quasilocal } \KK~\text{such that} $
$$\ii  \partial_\eta\ket{\psi(\eta)} = \KK(\eta) \ket{\psi(\eta)}$$
Then use this quasilocal generator to product a 
quasi-local $ \cor \UU\cob = e^{\ii \int_0^1d\eta \KK(\eta) } 
$
\cob.
Quasilocal means each term in $K$ is almost exponentially localized:
$$  ~~ K = \sum_x K_x, ~~ K_x = \sum_r K_{x, r} ,$$
where $K_{x,r}$ is supported on disk of radius $r$, $ \norm{ K_{x,r} } \leq e^{ - r^{1-d} } $.

The idea is to filter out the low-energy stuff: 
$$ K =- \ii \int\limits_{-\infty}^\infty dt F(t) e^{ \ii H(\eta) t }
\partial_\eta H(\eta) e^{ - \ii H(\eta) t} $$
$ F(t) $ odd, rapidly decaying, $\tilde F(0) =0$, $\tilde F(\omega) = - { 1\over \omega}, |\omega| \geq \Delta $.
Where $\Delta$ is the energy gap.

\end{shaded}

\color{black}

\item 
The second problem with the plan above is a crucial physical one
whose resolution is the key result:
Nontrivial states of matter are {\it defined} by the inability to find such a gapped path to a product state!

\end{enumerate}
\color{black}

{\bf Expanding Universe Strategy.}
Instead, we are going to {\it grow} the system $ \ket{\psi_L} \to \ket{\psi_{2L}} $ with local unitaries.
The resulting ${\cor \UU}$ which constructs 
$\psi_L$ directly from a product state will in general not have finite depth.
But ${\cor \UU}$ will have an RG structure.
A crucial question to keep in mind here is: which properties of ${\cor \UU}$ are universal, \ie~properties 
of the phase, and not its particular representative?

{\bf Assumptions:} 
The raw material for this construction includes
a bath of `ancillas' $ \color{lightblue} \otimes \ket{0}^M$, which we assume to be freely available.
All the rigorous results require an energy gap $\Delta$ for all excitations.
There may be groundstate degeneracy $G(H_L)$,
but we assume the groundstates are {\it locally indistinguishable}.
This is a necessary condition for the state to be stable to 
generic small perturbations of $H$, since otherwise 
adding the operator which distinguishes the states would split the degeneracy.




\hskip-.23in
\parbox{.48\textwidth}{
\begin{shaded}
{\bf Def \cite{s-sourcery}:} \cobl An \color{darkgreen}$s$-\cobl source RG fixed point
(in $d$ dimensions)
is a system whose groundstate on $(2L)^d$ sites
can be made from the groundstate on $L^d$ sites  
(plus \color{lightblue} unentangled ancillas \cob) 
using a quasilocal unitary.
\end{shaded}
\cob
Notice that the process is not like crystal growth, in that the new degrees of freedom
are not accreted at a surface, but rather {\it intercalated} hierarchically.
} 
~~~~
\parbox{.45\textwidth}{
\cob
A $d=1, \cog s \cob = 1$ example:\\
\includegraphics[width=.45\textwidth]{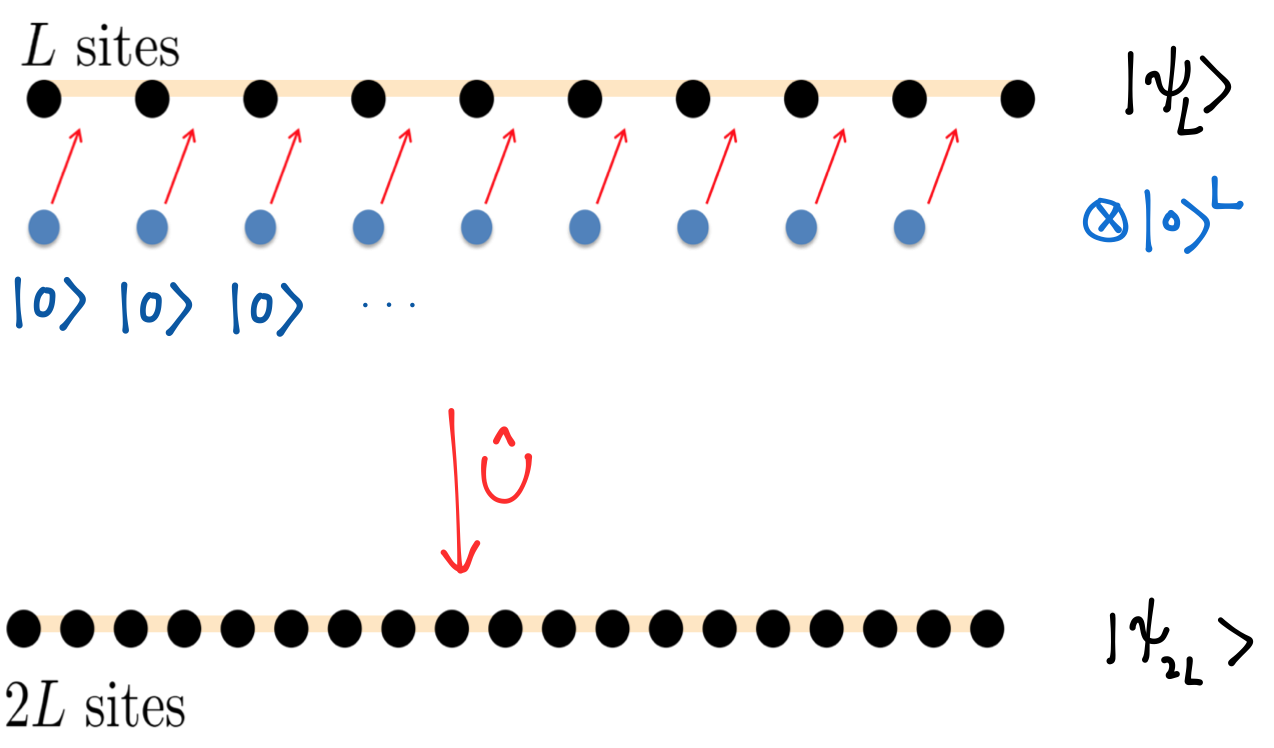}

$$ \ket{\psi_{2L} } = \cor \UU \cob \( \ket{\psi_L }\otimes 
\color{lightblue}
\ket{0}^L \cob\) . $$
}

\cob

More generally, the desired RG circuit $ \cor \UU \cob $ acts on $ \grs$ copies,
and the right number of ancilla qbits to double the system size:
$$ \ket{\psi_{2L} } = \cor \UU \cob \(
\underbrace{ \ket{\psi_L } \cdots \ket{\psi_L} }_{\grs} \otimes 
\color{lightblue}
\ket{0}^{M} \cob\)  
~~~~~~~\greencom{M  = L^d ( 2^d - s ) } $$

\hskip-.23in
\parbox{.48\textwidth}{
In more than one dimension, we need to start with more ancillas per original degree of freedom, 
as in the figure at right, for $d=2, s=1$:
}~
\parbox{.48\textwidth}{
\includegraphics[width=.48\textwidth]{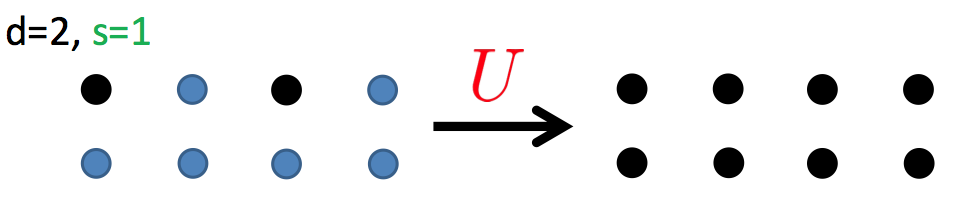}
}

\cob
Notice that we are using `fixed point' metonymically to describe the whole phase. 
For a gapped phase, there is some point in the phase diagram 
where the correlation length goes to zero, which can literally be a fixed point of the procedure.
\footnote{An interesting question I was asked during the lectures was: how do I know there 
is only {\it one} such fixed point in the phase?  This is crucial for this nomenclature to be self-consistent!
I know this because by `phase' I mean in particular 
that the physics is stable in an open set in the phase diagram.
Less tautologically, we are talking about gapped fixed points, 
which have no lowest-energy modes, and hence nowhere to go.
(It might be good to make this more precise.)
This means that the nearby RG flow lines must all go {\it towards} the $ \xi = 0$ fixed point in the IR.
If there were two such fixed points, there would have to be a separatrix 
between them at which the gap must close (It helps to draw some pictures here).  Hence they can't be in the same phase.
}

All this begs the question:
how to construct {\cor $\UU$}?
For a gapped state, we can construct $\cor \UU$ by quasiadiabatic evolution, 
following \cite{2005PhRvB..72d5141H}.

\hskip-.23in
\parbox{.4\textwidth}{
It is just like our dream of finding $\cor \UU$ by adiabatic evolution from product states, 
but 
for $\grs = 1$, we must start with $\grs = 1$ copy at size $L$, 
so we are not building the macroscopic state directly from product states.

}
\parbox{.59\textwidth}{\includegraphics[width=.6\textwidth]{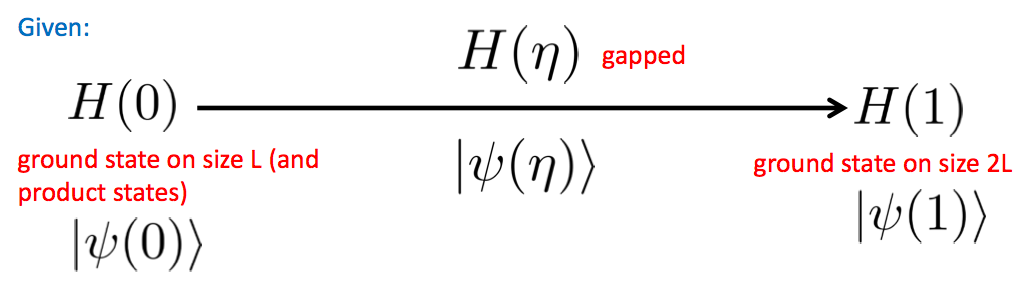}}

Such a procedure will produce a quasi-local $ \cor \UU$ 
with the desired properties.


{\bf Why is this useful?}
Now I will give a long series of reasons this is a useful perspective to take on 
extensive quantum systems.

\begin{enumerate}

\item Such a circuit controls the growth of entanglement with system size.
%
The existence of ${\cor \UU}$ with the given locality properties
implies recursive entropy bounds: 
\bea
S(2R) & \leq &  \grs S(R) + k R^{d-1}   \cr 
S(2R) &\geq & \grs S(R) - k' R^{d-1}  
\nonumber
\eea
for some constants $k, k'$.
This uses the Small Incremental Entangling result of \cite{2013PhRvL.111q0501V}
(which builds on earlier work of Kitaev and Bravyi).

From this we can deduce an
Area law {\it  theorem}: 
any $s\leq 1$ fixed point in $d>1$ enjoys an area law for EE of subregions.
$$ S(A) \equiv - \tr \rho_A \log \rho_A \leq k |\partial A| = k R^{d-1} .
~~~~\parbox{.2\textwidth}{\vskip-.2cm\includegraphics[width=.2\textwidth]{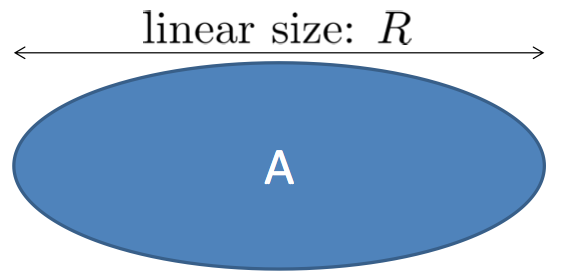}}
 $$

\item  
The circuit produces a counting of groundstates as a function of system size.
The groundstate degeneracy satisfies:  $ G(2L)  = G(L)^{{\cog s}} $.
This means that states with $\grs >1$
must necessarily have strange growth of $G$ with system size.

The examples I know of gapped states with $s>1$ 
are layers of FQHE (kind of cheating)
and 
Haah's cubic code ($d=3, s=2$) \cite{2014PhRvB..89g5119H},
for which Haah explicitly constructed the RG circuit (!).
This implies that these models have no continuum description
in terms of a quantum field theory, 
since as we will see below, any gapped field theory has $ \grs \leq 1$.

\item The smallest possible \cog $s$ \cob  is a property of the phase, 
and can be used as a classification axis, quantifying the amount of entanglement in the groundstate.

\item The circuit implies a MERA.


\end{enumerate}

Before saying more about the last point,
it is important to emphasize that all known
(at least known to me)
extensive quantum systems 
fall into this $\grs$-source classification.  
Here are some examples.

\begin{itemize}
\item 
\parbox{.8\textwidth}{Mean field symmetry-breaking states  have $\grs=0$ ,
since they actually do have product-state representatives within the phase.
}~~~\parbox{.2\textwidth}{\includegraphics[width=.2\textwidth]{figs/fig-1d-band-deform.png}}

\item Chern insulators and integer quantum Hall states have $\grs=1$.

\hskip-.23in
\parbox{.8\textwidth}{In 
\cite{s-sourcery}
we gave an explicit construction of a gapped path in the space of Hamiltonians 
which doubles the system size of some Chern bands.
It involves some simple maneuvers of varying the nearest-neighbor 
(solid lines) and next-nearest neighbor
(dashed line) hopping terms, 
so as to couple in the new sites.
}~~~\parbox{.1\textwidth}{\includegraphics[width=.1\textwidth]{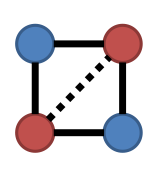} }

\hskip-.23in
\parbox{.56\textwidth}{
In momentum space, we are folding the Brillouin zone in half.
You can watch the movie 
of the band-folding in the Supplemental Material
\htmladdnormallink{here}{http://journals.aps.org/prb/abstract/10.1103/PhysRevB.93.045127}
(I haven't figured out how to embed a gif in pdf yet).}
~\parbox{.35\textwidth}{
\includegraphics[width=.38\textwidth]{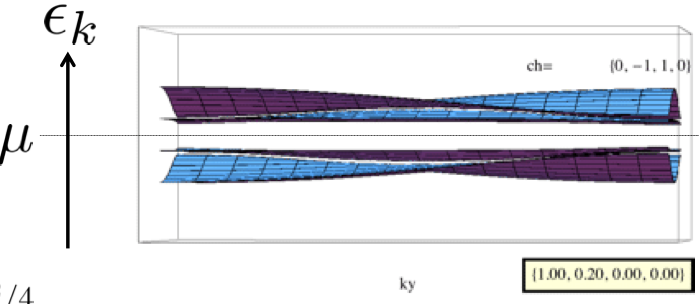}  }

\item Topological states (discrete gauge theory, fractional QH states) have $\grs=1$.

\end{itemize}

\begin{shaded}
\parbox{.48\textwidth}{
\includegraphics[width=.49\textwidth]{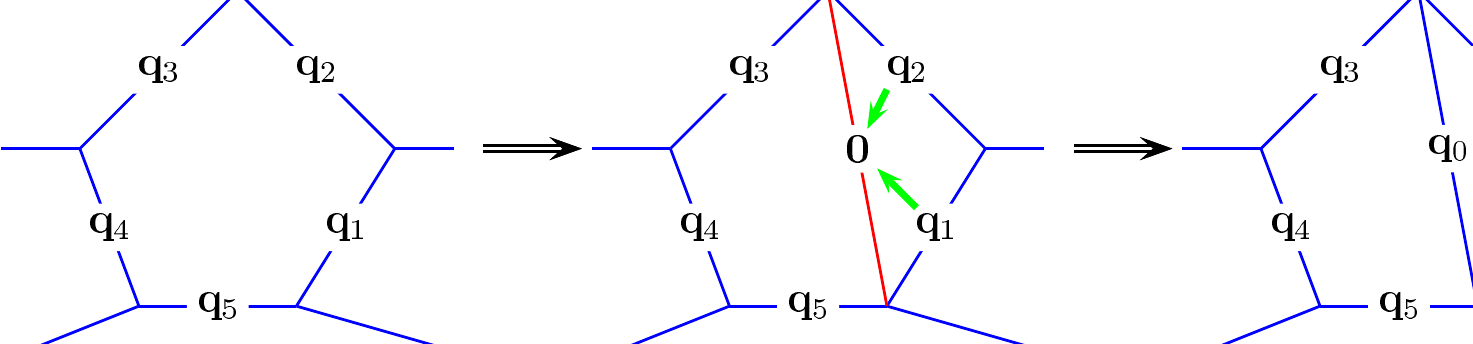}\\
\includegraphics[width=.49\textwidth]{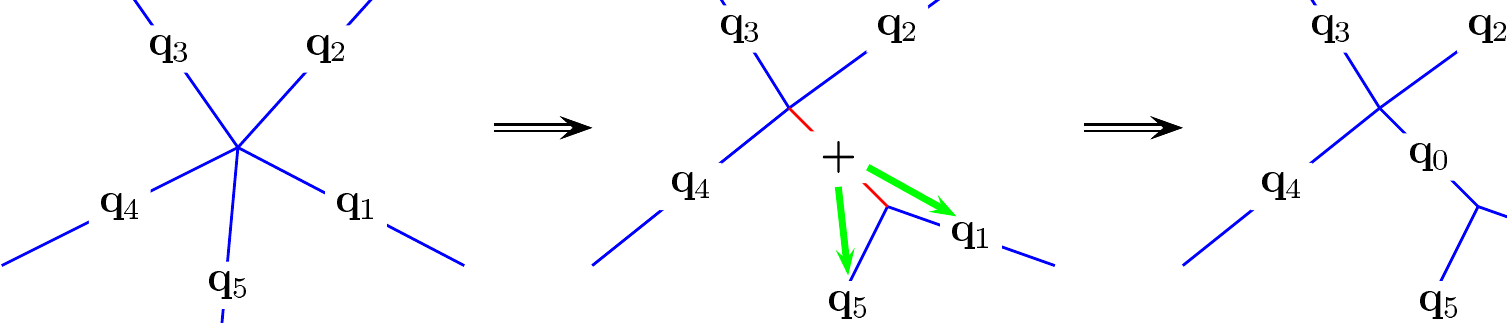}\\
\ttref{[Fig.~from \cite{2009PhRvB..79h5118G}]}}
~~
\parbox{.43\textwidth}{The toric code really is a fixed point of this kind of operation \cite{2008PhRvL.100g0404A, 2009PhRvB..79h5118G}, not just metonymically.

The figure is describing a process of adding an ancilla and 
incorporating it into the lattice model. }
We act on the hamiltonian with a (brief!) series of 2-qbit gates:
the green arrows are CNOT gates, which act on 2-qbits as 
$ P_C(0) \otimes \Ione_T + P_T(1) \otimes X_T $ 
where C is for `control' and T is for `target'.
Conjugating operators, 
The CNOT gate acts as 
\bea
1_C Z_T  &\leftrightarrow& Z_C Z_T \cr 
1_C X_T &\leftrightarrow& 1_C X_T \cr
Z_C 1_T &\leftrightarrow& Z_C 1_T \cr
X_C 1_T &\leftrightarrow& X_C X_T 
\nonumber
\eea
It is a fun exercise to convince yourself that 
this maps the TC Hamiltonian on the initial graph
to a Hamiltonian with the `stabilizer algebra' of the final graph.
(That little outpouring of jargon
was necessary because the terms in the resulting $H$ are not exactly the same;
rather we get terms like $B_{p_1} B_{p_2} + B_{p_1} $ where 
$p_1$ and $p_2$ are the new plaquettes.  But the set of groundstates is the same.)

\end{shaded}

\begin{itemize}[resume]
\item Let us define the notion of a {\it topological quantum liquid}
to be any state which is insensitive to smooth deformations of space.
This includes any gapped QFT, and in particular the gapped topological 
states just mentioned.

Any such state has $s=1$.  Here's why: 
place it in an expanding universe, with metric
$ ds^2 = - d \eta^2 + a(\eta)^2 d \vec x^2 $.
This is a smooth deformation of space.
If we choose $a(\eta)$ to double the system size between
initial and final times, the adiabatic evolution
in $\eta$ produces the desired unitary.

\item

Finally, taking seriously the previous point,
an experimentally-verified example
of an $\grs=1$ fixed point 
is quantum chromodynamics, to the extent that it describes the strong interactions in the real world
(it does).
Here's why this is true:
Our universe is expanding, with a doubling time $t_\text{doubling} \sim 10^{10} \text{years}$.
Since the temperature has dropped below $ \Lambda_{\text{QCD}}$,
the system is pretty well in its groundstate.
During this process, the QCD gap has stayed open, since $m_\pi, m_{\text{proton}} > 0 $.
So this is a gapped path from $\ket{\psi_{L}}$ to $\ket{\psi_{2L}}$ .
This implies the existence of a quasilocal unitary which constructs
the QCD groundstate from a small cluster plus ancillas.
(\ie~QCD  has $\grs=1$).
 This suggests a new approach to simulating its groundstate 
which is in principle very efficient.

\end{itemize}

%

{\bf MERA representations of $s=1$ fixed points.}
Perhaps the most exciting reason to try to produce
explicit RG circuits, is 
that given the form of ${\cor \UU}$, 
one can use it to produce a MERA.

\hskip-.23in
\parbox{.7\textwidth}{
A MERA 
\cite{mera}
is a representation of the groundstate which: 
\begin{itemize}
\item allows efficient computation of observables (few contractions)
\item organizes the information by scale, (like Wilson and AdS/CFT taught us to do)
\item geometrizes the entanglement structure.
\end{itemize}
}\parbox{.3\textwidth}{\includegraphics[width=.5\textwidth]{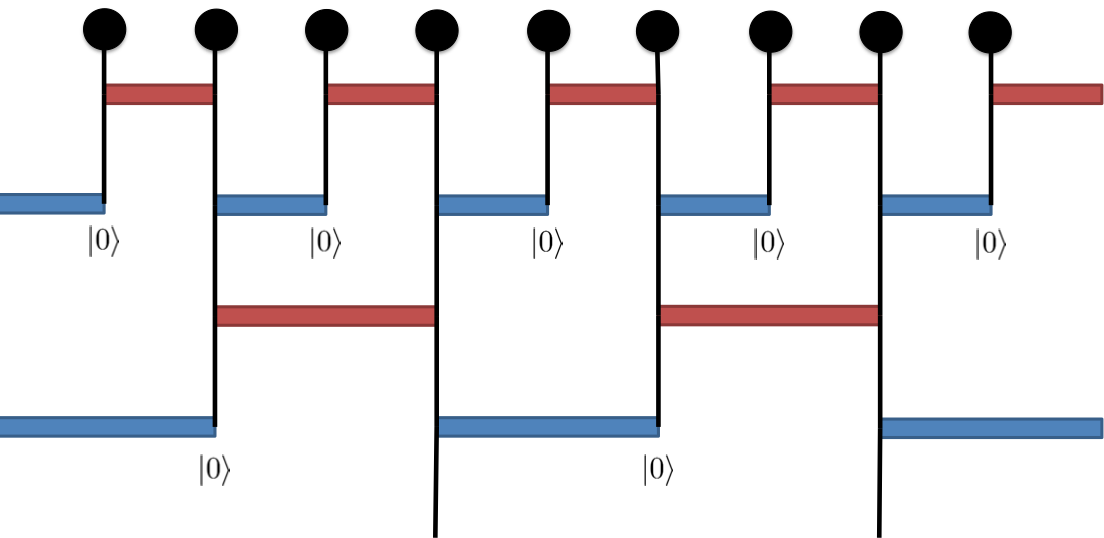}}
It is the state-of-the-art representation of 1+1d critical states \cite{2011arXiv1109.5334E}, 
but is very hard to find the tensors for systems in $d>1$ 
(a claim for which I will cite \cite{2009PhRvL.102r0406E} and some painful personal experiences).

\hskip-.23in
\parbox{.6\textwidth}{A quasilocal $\cor \UU$ can be `chunked' into a low-depth circuit, 
$$ \ket{\psi_L} \simeq \cor \UU_\text{circuit} \cob \ket{\psi_{L/2}} 
\color{lightblue} \ket{0}^{L/2}\cob $$
where the `chunks' (the individual gates) have size which goes like 
$$ \hat \ell \sim \log^{1+\delta}(L) ~~$$
and bond dimension
$$
\chi \sim e^{ \hat \ell^d } \sim e^{ c \log^{d(1+\delta) } (L) } ~.$$
}~\parbox{.4\textwidth}{\includegraphics[width=.4\textwidth]{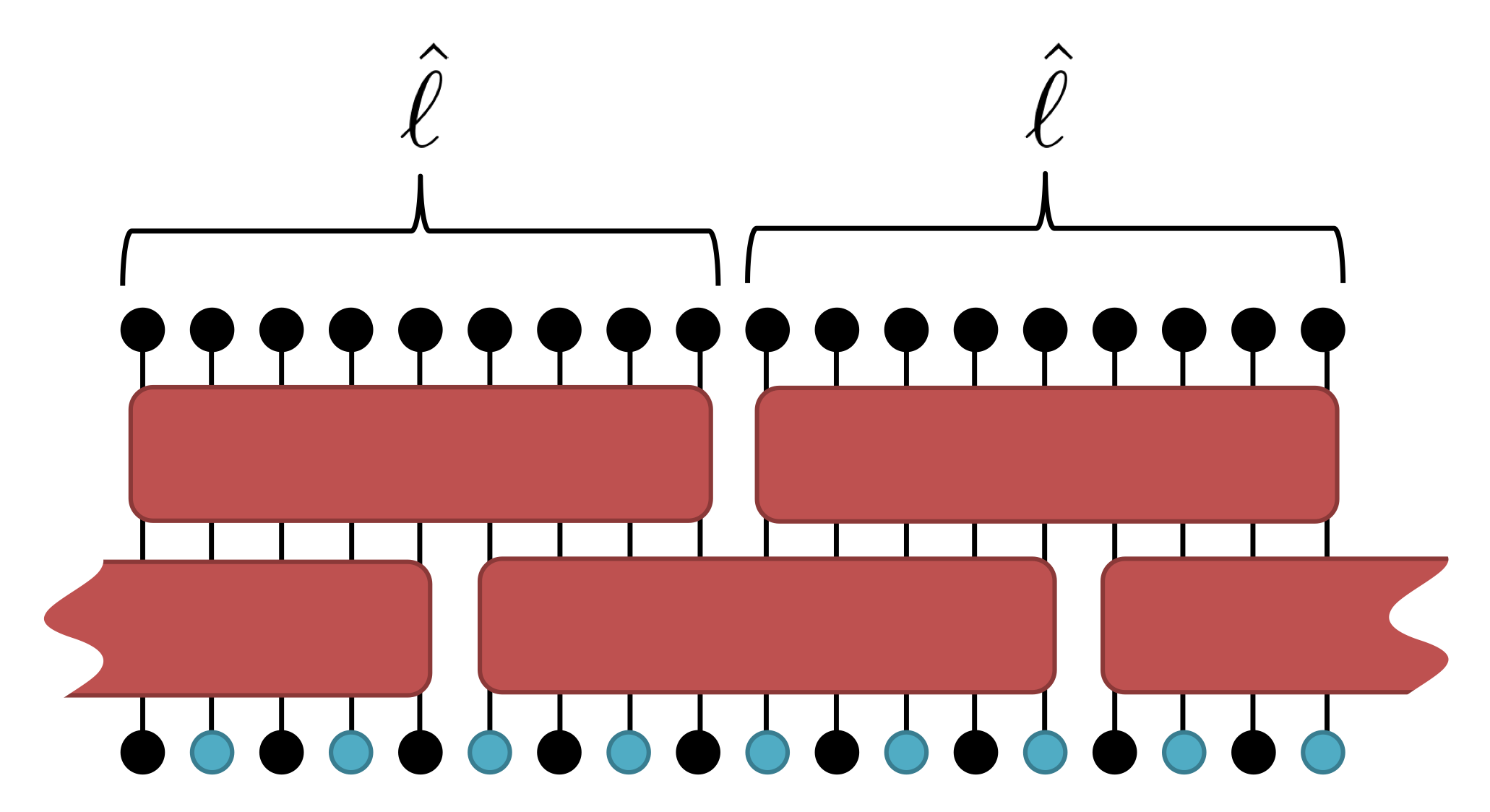}}

This idea awaits numerical implementation.
The crucial point which makes this an exciting possibility is that 
this construction of $ \cor \UU \cob $ 
requires no variational sweeps 
on large system.  The only other input required is the groundstate for
a tiny molecule of the stuff, findable by exact diagonalization.




Finally, the study of $\grs$-sourcery 
has something to say about 
how to think about SPTs
 (the subject of \S\ref{sec:SPT}).
In particular, a robust notion of the condition
that a state is `short-range-entangled'  
is that it is {\it invertible}.
$\ket{\psi}$ is invertible 
if $\exists \ket{\psi^{-1}}, \cor  \UU\cob  $ such that 
$$  \ket{\psi } \otimes \ket{\psi^{-1} } =   \cor \UU \cob \ket{0}^{\otimes 2L^d} 
\text{ has } \grs=0 .$$
That is, when combined with its inverse, it has $\grs =0$.
Closely related ideas have been discussed in 
\cite{Kitaev-2013, 2014arXiv1406.7278F}.

A result we call the `weak area law' says that 
a system with
a unique groundstate on any closed manifold 
(this means no topological order, but such states can still be interesting SPTs)
implies 
the existence of an inverse state
and the area law.

Here is a comic-strip proof of the weak area law:

\parbox{.49\textwidth}
{
\includegraphics[width=.49\textwidth]{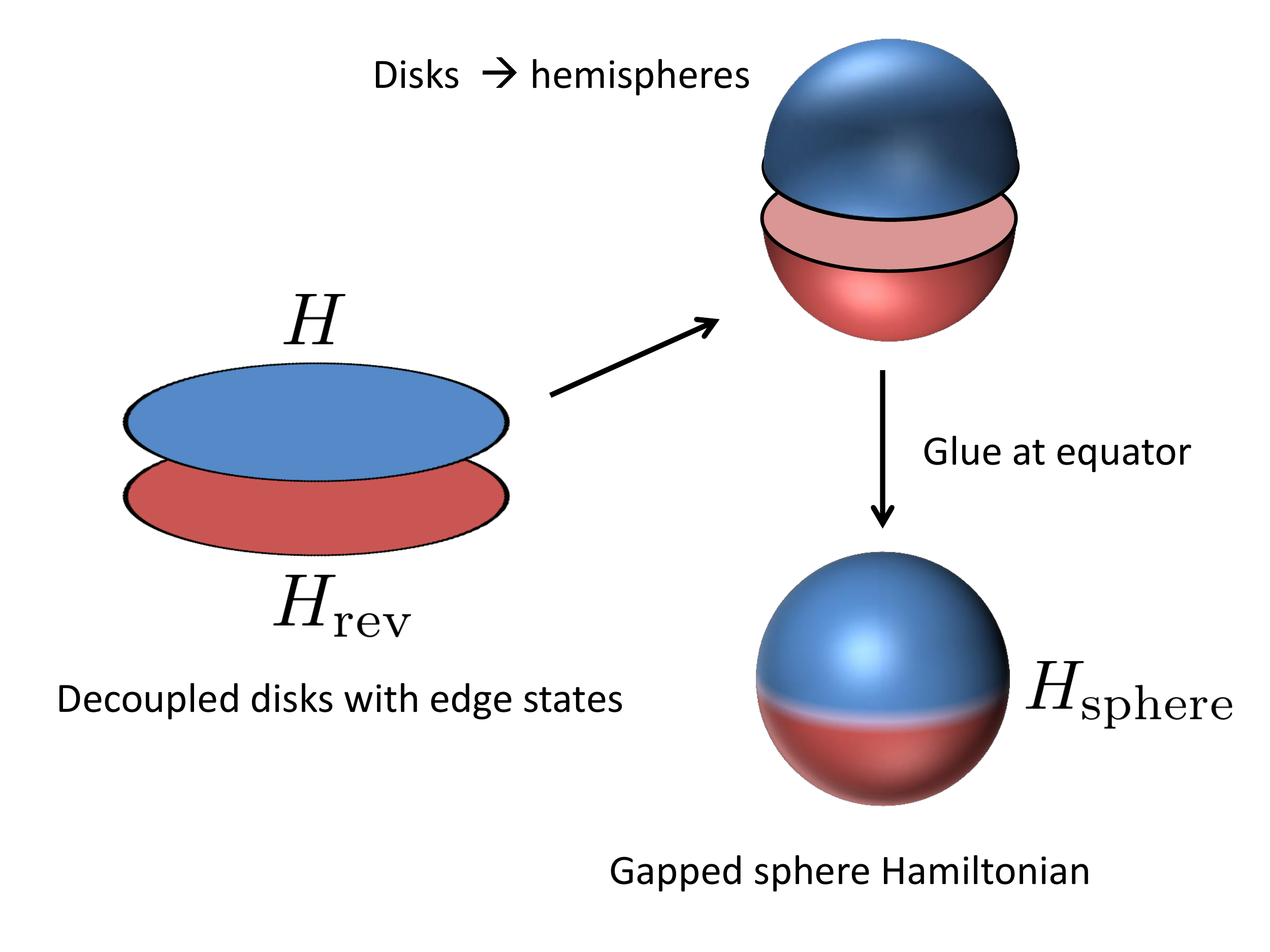} 

Step 1: Take a copy of the system and its parity image $H_\text{rev}$ each on a disk.
Glue the two disks along their boundary to make a $d$-sphere.  
Since there must be a unique groundstate, any
edge states on the disk are killed by the gluing.
}~~
\parbox{.49\textwidth}
{\includegraphics[width=.49\textwidth]{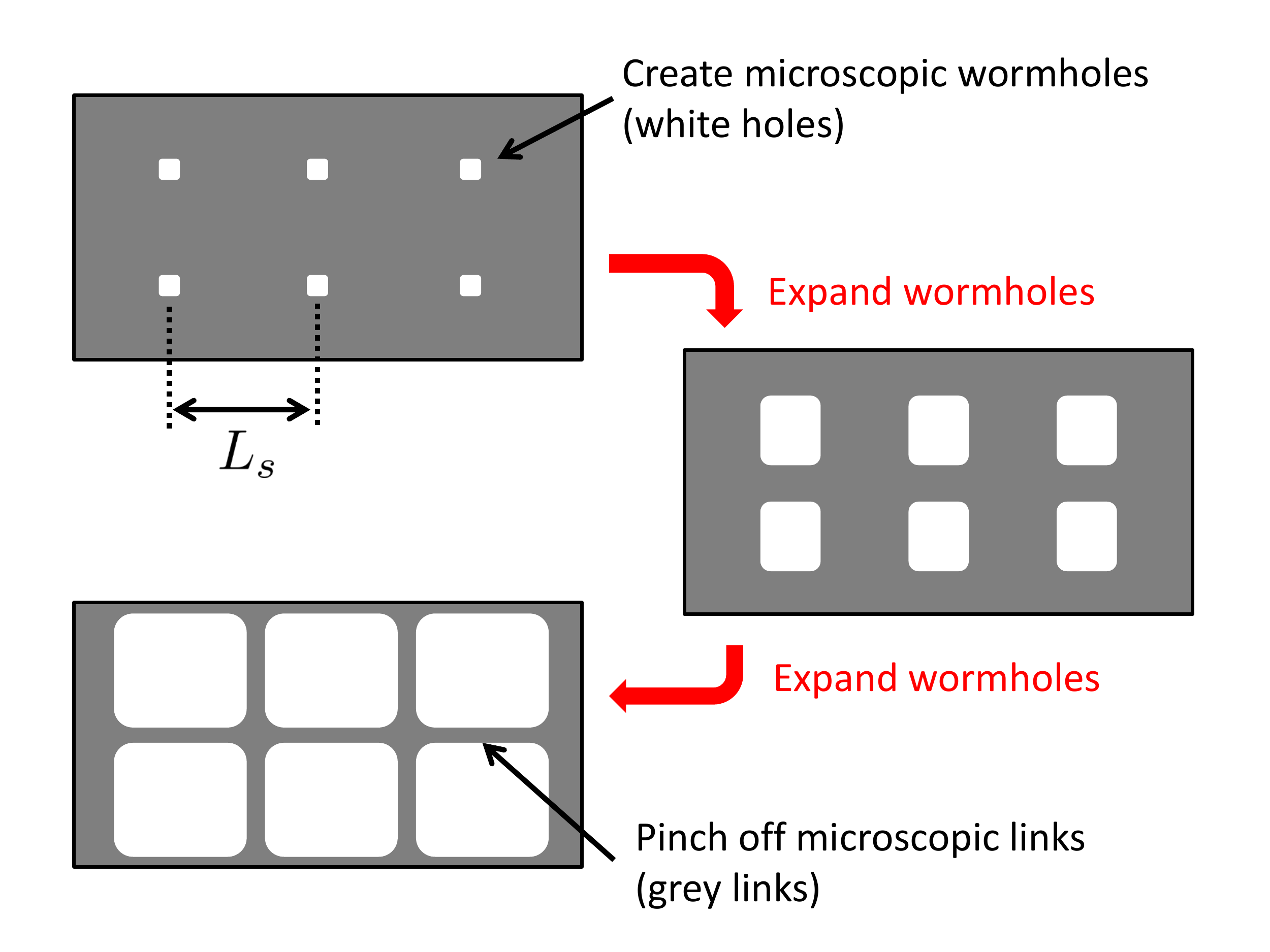}
~~~~~~~~~Step 2: 
By the depicted sequence of deformations of the space, 
make an adiabatic path to $\ket{0}^\otimes$ on $T^d$
}
\parbox{.5\textwidth}{
Here is a side view of 
the pictures at right, the two copies
and the array of wormholes connecting them:
}
\parbox{.48\textwidth}{
 \includegraphics[width=.5\textwidth]{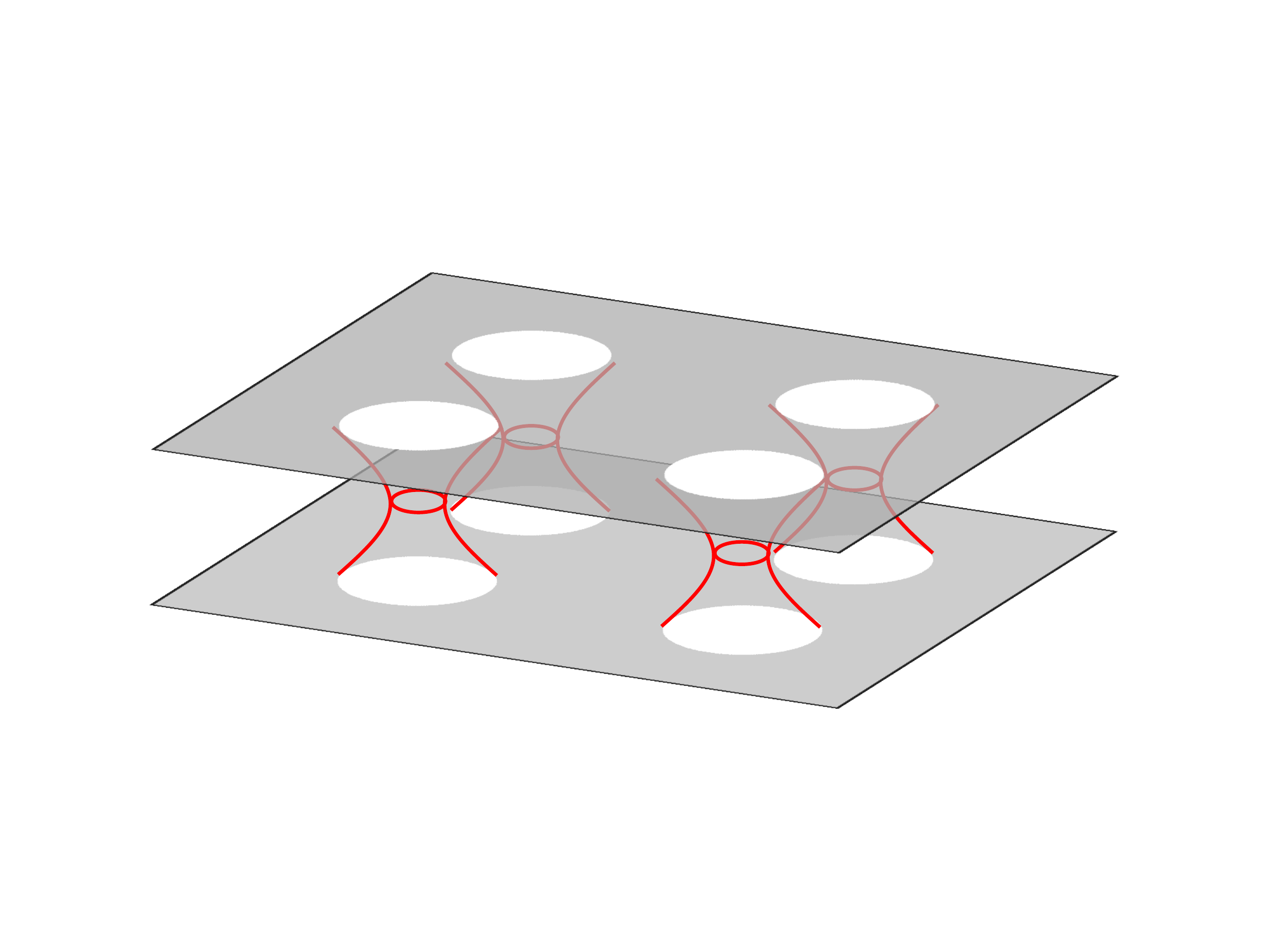}}


%


\cob

\subsection{Gapless entanglement bounds}
\label{sec:gaplessEE}


For many gapless states in more than one dimension -- in particular for critical points and for semi-metals -- 
we expect these same properties to hold.  In particular, the area law should still be true, 
and we expect a good MERA approximation with favorable bond dimension. 
However, the well-developed techniques used above do not apply.  

Moreover, the area law is actually violated in groundstates of metals: \bluecom{$ S \sim R^{d-1} \log k_F R $.}
This violation is a symptom of many low-energy {\it extended} modes.
These modes can be seen in thermodynamics.  
This leads to the idea that constraints on the 
thermodynamic behavior can constrain the groundstate entanglement\footnote{An attempt
to implement a similar idea is \cite{2014arXiv1409.5946B}.}.  

Here is the result \cite{Swingle:2015ipa}:
Parametrize the thermal entropy density of a scale-invariant state as $s(T) \sim T^{{ d - \theta \over z }} $  
(where $z$ is the dynamical exponent,
and $\theta$ is the hyperscaling violation exponent
(the anomalous dimension of $T_{tt}$)).
Then: the area law obeyed when $ \theta < d-1 $ 
(and $0 < z < \infty$).
Recall that metals have $ \theta = d - 1 $ (so that $s \sim c_V \sim T$ in any dimension),
and lie just outside the hypothesis of this result.

The idea by which we can show this can be called {\it entanglement thermodynamics}.
 Recast the entanglement entropy 
 calculation as a local thermodynamics problem (local means $T  = T_x$):
Find $\ssigma_A \simeq Z^{-1} e^{ - \sum_x {1\over T_x} \HH_x  }$
such that $S(\ssigma_A) \geq S(\rrho_A) $.
(The actual Hamiltonian is $\HH \equiv \sum_x \HH_x$, so $\ssigma_A$ is a local Gibbs state.)

Who is $ \ssigma_A$? $\ssigma_A$ is the 
state of maximum entropy which is consistent 
with $ \vev{\HH_x}$, for all patches $x$ completely inside $A$.
 (From this definition, $S(\ssigma_A) \geq S(\rrho_A)$ is automatic.)
An argument just like the Bayesian argument
for the Boltzmann distribution
shows \cite{2010NatCo...1E.149C, 2014arXiv1407.2658S} that
$$
\ssigma_A = 
{e^{ - \sum_x \beta_x  \HH_x  }  \over Z}$$
where the position-dependent coolnesses $\beta_x $ begin their lives as Lagrange multipliers.

\hskip-.23in
\parbox{.6\textwidth}{
Let $ \HH_A \equiv \sum_{x \in A {\tiny{(completely)}}} \HH_x$.
This means that $\ssigma_A  $ is the Boltzmann distribution for $\HH_A +...$ with $T_x = 1/\beta_x$.
\bea 
~\tr \HH_A \ssigma_A 
&=& \tr \HH_A \rrho_A  
\cr 
&=& \underbrace{E_{g,A}}_{\equiv \text{gs energy of $H_A$}} + \CO\( |\partial A|r_\text{patch}| \HH_x | \)
\nonumber
\eea
This means that $ \ssigma_A$ is a state with excitations localized at $\partial A$, $T_x \to 0 $ in interior of $A$.
}~~
\parbox{.4\textwidth}{
    \includegraphics[width=0.4\textwidth]{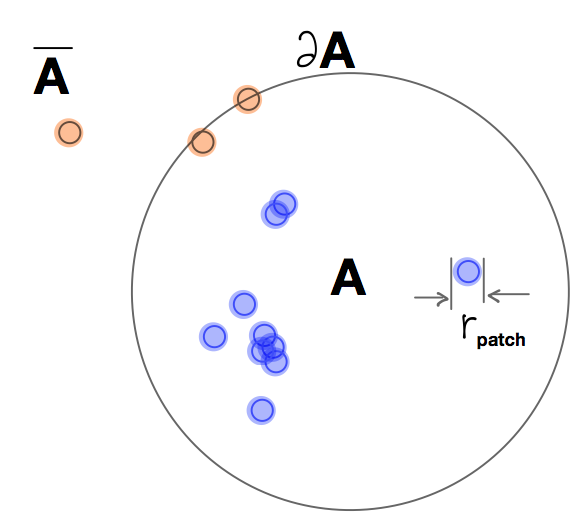}

    {\tiny{(In the figure, $r_\text{patch}$ is the range of $\HH_x$.)}}
}

The crucial fact is that for size-scaling purposes, 
we can approximate
\bea \tr \HH_A \ssigma_A &\simeq& E_{g, A} + \int_A d^dx ~e(T_x) 
\cr 
- \tr \ssigma_A \log \ssigma_A &\simeq&  \int_A d^dx~ s(T_x) 
\nonumber
\eea
where $ e(T_x)  = T s(T_x) $ are the bulk thermodynamic densities at temperature $T_x$.
Here is why.
These {\it local thermodynamics} estimates are true if $1 \gg {\nabla T_x\over T_x } \cdot \xi_x $ 
\greencom{(for all $x$) }
where $\xi_x$ is the local correlation length.  
But now let~~~ $ \ssigma_A({\cor\tau}) \equiv Z({\cor\tau})^{-1} e^{ -  {1\over {\cor\tau}} \cob \sum_x \tilde \HH_x/T_x } 
~~~\buildrel{{\cor\tau} \to 1 } \over{\to } \ssigma_A $.
This state has temperature $T_x({\cor\tau}) = {\cor\tau} T_x $, 
but the local correlation length scales like $ \xi_x({\cor\tau}) \sim T_x({\cor\tau})^{- 1/z} \propto {\cor\tau}^{1/z} $
So \redcom{(unless $z=\infty$!)} the figure of merit for local thermo in state $\ssigma_A({\cor\tau})$ is
$$1 \gg \underbrace{ {\nabla T_x({\cor\tau}) \over T_x({\cor\tau}) }}_{\sim {\cor \tau}^0} \cdot 
\underbrace{\xi_x({\cor \tau})}_{\sim {\cor\tau}^{-1/z}} \buildrel{{\cor \tau} \to \infty}\over{\to } 0 .  $$
$ S(\ssigma_A({\cor \tau})) = {\cor \tau}^{{d-\theta\over z}} S(\ssigma_A) $ 
scales the same way with region size.


\hskip-.23in
\parbox{.5\textwidth}{
To use local thermodynamics, we need $T_x$.
Our question is local, so we can choose a convenient geometry.
We choose a geometry which is translation invariant in $d-1$ dims (PBC).  
$R \gg w$.
Scale invariance implies
$$ T_x \sim \begin{cases} 
\color{blue} x^{-z}  & \cr  \cob
\infty & \text{(no)}\cr 
0 & \text{(sometimes: frustration free $\HH$)} 
\end{cases}
$$
}~~~\parfig{.49}{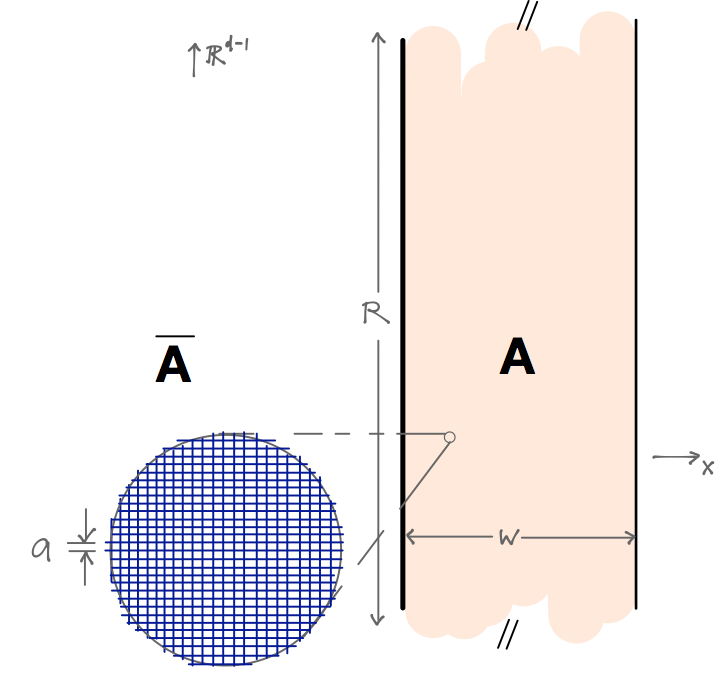}
Basically, the local correlation length is equal to $x$, the distance from the boundary.
Scale invariance determines the dependence of the energy density and entropy density on temperature and hence on $x$ to be:
\be\label{eq:thermo-scaling-theta} ~ e(T_x) \sim x^{-z+\theta -d} , ~~ s(T_x) \sim x^{\theta - d } \ee
So local thermodynamics implies the estimate
\bea S_A \leq - \tr \ssigma_A \ln\ssigma_A 
& \sim  R^{d-1}  \int_a^w dx~x^{-d+\theta}& \cr 
& &\cr
\sim  R^{d-1}& \( a^{-d+\theta + 1 } - w^{-d + \theta + 1 } \) &
 \buildrel{w \to \infty}\over{\to}  \infty \text{ only if } d < 1 + \theta ~.
\nonumber
\eea

\begin{mdframed}

Hence: scale invariant states with $ \theta < d-1 $ obey the area law.

\end{mdframed}

\subsubsubsection{Connection to ${\grs}$-sourcery.}
{\it If} our scaling theory is an $\grs$-source RG fixed point
$$ S(2R) \leq  \grs S(R) + k R^{d-1} ~.$$
Assuming the inequality is saturated (if not, we can use smaller $\grs$) implies
\bea S_A &=& k \( { R \over a }\)^{d-1} \sum_{n=0}^{\log_2\(w/a\)} 
\( \grs \over 2^{d-1} \)^n 
\cr 
&\buildrel{R\gg w \gg a } \over {\simeq} &
k \( {R \over a } \)^{d-1}
\( 1 - \( a \over w\)^{d - 1 - \log_2 \grs } + \cdots \) 
\nonumber\eea
Now Brian had the brilliant idea to compare subleading terms in the EE, with the result:
$$ \boxed{ \grs = 2^\theta } $$
Satisfyingly, a Fermi surface has $\theta = d-1$, hence $\grs= 2^{d-1}$, and indeed marginally violates area law. $\checkmark$

\subsubsubsection{Gapless RG circuits.}
Gapless $\grs$-sourcery is a current frontier.
With Brian Swingle and Shenglong Xu, 
we are developing ideas to construct such RG circuits for some gapless states \cite{sqrt-draft}.
A valuable testbed for these ideas comes from 
certain semi-solvable models, 
which we call ``square-root states", 
whose groundstate wavefunction is the square root of some classical statistical mechanics 
Boltzmann weight.

\breakS

\section{How to kill a Fermi liquid until it stops moving}

Now we muster our courage and try to say something about systems with Fermi surfaces.

\subsection{Non-Fermi Liquid problem}


The Fermi liquid is a (nearly) stable phase of matter.  
This can be understood as follows \cite{1992hep.th...10046P, PhysRevB.42.9967, RevModPhys.66.129}.
Assume that the only degrees of freedom at the lowest energies are a collection of fermion operators
governed by the gaussian fixed point with action
$$ S_0[\psi] = \int dt d^dk~\psi^\dagger \( \ii \partial_t - \eps(k) + \mu \) \psi $$
where $\mu$ is a chemical potential, and $ \eps(k) $ is some single-particle dispersion relation
(think of $ \eps(k) = \vec k^2/2m$).
There is a Fermi surface at the locus
$$ \{  k | \eps(k) = \mu \} $$
which manifests itself as a surface of poles in the single-fermion retarded Green function
\be\label{eq:G-R} G_R(\omega,k) \sim { 1\over \omega  - (\eps(k)  - \mu ) } . \ee
In the free theory, this is a pole on the real frequency axis, indicative of an exactly stable particle state.

If we perturb this fixed point with all possible operators preserving fermion number 
(scaling $ \psi$ so that $S_0$ is dimensionless),
we find that the only relevant operator simply shifts the Fermi surface around a bit.
There is a marginal forward scattering 4-fermion coupling (which leads to a family
of Fermi liquid fixed points labelled by Landau parameters), 
and a marginally relevant BCS interaction which at hierarchically low energies 
leads to superconductivity.

This is great because lots of metals are actually described by this (family of) fixed point(s).
However, some are not.  Such a thing is (unimaginatively) called a non-Fermi liquid (NFL).
This is a state of matter (or a critical point) where a Fermi surface
is discernible (for example, by angle-resolved photoemission,
which measures the imaginary part of $G_R$ in \eqref{eq:G-R}, or more often from transport evidence), 
but where the spectrum is not well-described by long-lived quasiparticles.

The crucial loophole in the inevitability argument above
is the assumption that there are no other gapless degrees of freedom
besides the fermions at the Fermi surface.

In a solid, there are inevitably also phonons, the Goldstone modes 
for the spatial symmetries broken by the lattice of ions.
Because they are Goldstone modes, 
the phonons are derivatively coupled to the Fermi surface
and this limits their destructive capacity.
(In fact, phonons {\it do} mess with the 
well-definedness of the quasiparticles, 
and this was worried over and understood in \cite{PhysRev.134.A566}.)
The lattice vibrations can also determine the temperature dependence 
of the conductivity, if they are the most important momentum
sink for the charge carriers.  More on this below.


Many other possibilities for other modes have been considered.
One possibility is a scalar order parameter
which is gapless because someone is tuning to its ordering critical point.
An exciting possibility is an emergent gauge field,
because a gauge field doesn't need tuning to keep it gapless.  
This actually happens:
the theory of the half-filled Landau level is a (relatively controllable and) real example \cite{HL9312}.\footnote{Our understanding of the half-filled Landau 
has been modified dramatically in the time since
these lectures were delivered \cite{Son:2015xqa, Barkeshli:2015afa, Murthy:2016jnc, Wang:2016fql, Wang:2015qmt, Metlitski:2015eka, 2015arXiv151008455M, 2015arXiv150804140G}, but
this statement is still true.
}

\hskip-.23in
\parbox{.79\textwidth}{
The nonzero probability for the fermion to 
emit a gapless boson 
gives the quasiparticle state with definite $\omega, k$ a nonzero lifetime, 
and more specifically a self-energy which is a power-law in frequency
$  \Sigma(\omega, k) \sim \omega^{2\nu < 1 }$.
By the usual geometric sum of propagator corrections, 
$$ G(\omega,k) \sim {1\over \omega - \eps_k + \mu - \Sigma(\omega,k)}.$$
The lifetime of the quasiparticle is $ \Im \Sigma(\omega, k)$.
The decay happens by the emission of a gauge bosons
as we can see by cutting the self-energy diagram, 
by the optical theorem.
}
\parbox{.2\textwidth}{
\begin{center}
$\Sigma(\omega, k) = $\\
\includegraphics[width=.2\textwidth]{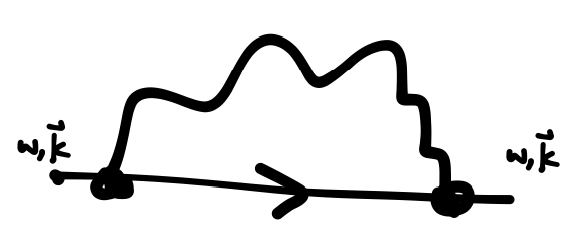}

$$\Im \Sigma \sim $$
$$\sum_{\varepsilon, q} 
\parbox{.1\textwidth}{\includegraphics[width=.1\textwidth]{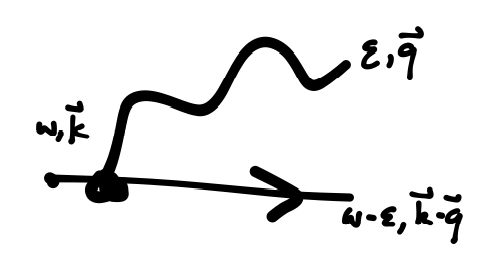} }
$$
\end{center}
}

There has been a lot of work on this type of NFL, in perturbation theory in the coupling
between the Fermi surface and the gapless boson (a small sampling: \cite{AI9448, NW9459, L0902, MS1007, MM1004, Fitzpatrick:2013rfa, Dalidovich:2013qta, Metlitski:2014zsa}).  
But, even when these theories are under control, this can't be the whole story and here's why.\footnote{By the way, a recent paper \cite{Kachru:2015rma}
proposes a deformation of a supersymmetric duality
where one side is closely related to this state,
and the other side does not have a Fermi surface.  
This is some new hope in the theory of NFL.
}


{\bf Just Enough  Discussion of Transport to Make Trouble.}  
Many of the most experimentally accessible properties 
of a material 
are associated with transport of charge and energy.
Generally, these are also the among the most difficult to understand precisely.
Linear response theory relates the electrical conductivity
to the low-momentum limit of the current two-point function.
The story is roughly: 
$$ \vev{j j} \sim 
\parbox{2cm}{\includegraphics[width=2cm]{figs/fig-vacuum-polarization.png}} $$
where the solid lines are now Fermi surface propagators, renormalized
by their interactions with whatever's killing the quasiparticle.
Based on this diagram, 
and your knowledge that $\Sigma(\omega) \sim \omega^{2\nu}$, you might reasonably 
infer that $\rho(T) \sim T^{2\nu}$.

But Ohm's law $ j = \sigma E $ requires dissipation of momentum of the charge carriers;
if their momentum is conserved, an electric field will just accelerate them.
The thing that breaks translation invariance has a big effect on the outcome.

\hskip-.23in\parbox{.79\textwidth}{
One possibility is scattering off of dirt.  
It's a good approximation to say that the dirt is infinitely heavy,
and so this is elastic scattering.  
In this case, the scattering
rate is independent of the quasiparticle frequency, and 
the self-energy is a constant, $\Sigma \sim  { \ii \over \tau_0 } $, 
}
\parfig{.2}{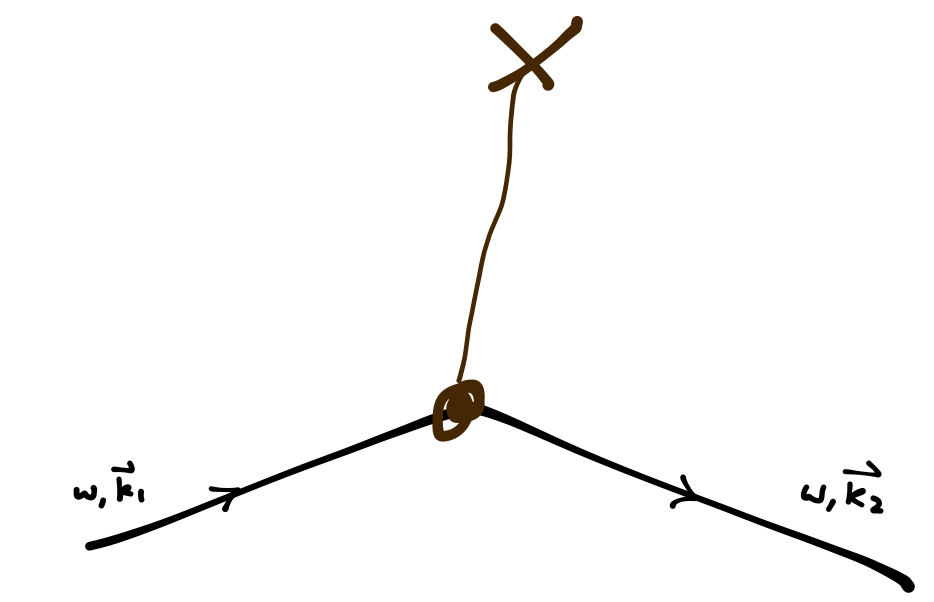}
so
$$ 
G(\omega,k) \sim {1\over \omega - \eps_k + \mu - { \ii \over \tau_0 }}
$$
The single-fermion state with definite momentum decays (into some other momentum state),
with a lifetime independent of frequency (and hence of temperature).
The change in the momentum $|\vec k_1 - \vec k_2 |$ is big.
Therefore the decay of the quasiparticle and the decay of its momentum  
have the same rate.

Consider, on the other hand, the case where the process which kills the quasiparticle
is the emission of a gapless boson with dispersion $ \omega(q) \sim q^z $.
The density of final states is peaked around $ q=0$.
Whatever $q$ is, the final state is different from the initial single-quasiparticle state,
so the quasiparticle decays no matter what.
But at the peak of the distribution in $q$, the momentum of the charge carriers isn't changing at all,
so the current isn't changing at all.
The result is that $\rho(T)$ goes like a power of $T$ larger than $2\nu$.

\parbox{.38\textwidth}{
\greencom{$\omega \sim q^z, z=1$} \parbox{.2\textwidth}{ \includegraphics[width = .2\textwidth]{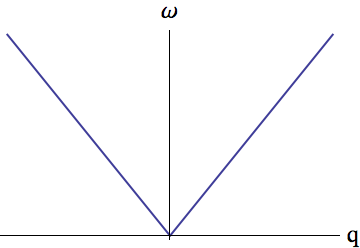} }\\
 $\downarrow $\\
\greencom{$\omega \sim q^z, z\gg 1$}
\parbox{.2\textwidth}{ \includegraphics[width =.2\textwidth]{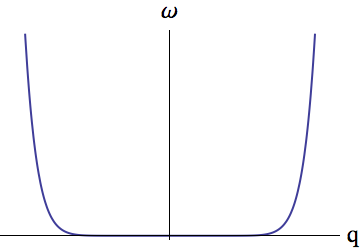}}
}
\parbox{.56\textwidth}{
We can do better if $z$, the dynamical exponent for the boson, is larger.
Then there is a larger range of $q$ with lots of cheap boson states.
Actually finite $z$ isn't good enough.
}

So we are going to discuss the ways known to me of making a $z=\infty$ momentum sink for 
the quasiparticles.

\subsection{A real $ z=\infty$ fixed point}

The first  $z=\infty$ fixed point we must talk about 
is the high temperature limit of the Einstein solid.  
It has some common features 
with the others I'll mention below
(gravity in $AdS_2\times \IR^2$ and some nonlocal 4-fermion Hamiltonians), 
but has the big distinction that it is known to exist in the world, even in Earth rocks.

The Debye frequency is the largest frequency of a normal mode of the lattice of ions in a solid;
when the temperature is above this frequency (in units of $k_B/\hbar$), 
the phonons are all classical: the ions just vibrate independently about their equilibrium positions.
They are localized, but can be excited to an arbitrary frequency (within reason):
this is $z=\infty$.
Phonons really do produce linear-$T$ resistivity, for $T > T_D$.


Furthermore, 
this fixed point (like the others) has an extensive zero-temperature entropy,
at least it would if we could somehow take the Debye temperature to zero:
$$ S(T \to 0, T > T_D)  = d  N k_B \log T , $$
where $N$ is the number of ions and $d$ is the number of directions
in which they can vibrate. 
 (This is the equipartition theorem for $dN$ classical springs, 
 where $ F = - T \log \(Z_1^N \) = - T N d \log(T/T_0) $, $ Z_1 = \int d^dq d^d p~e^{ - \beta \(p^2 + T_0^2 q^2 \) }$.)

The mystery about $\rho(T)$ in the cuprates is rather that the linear-$T$ 
behavior 
(of the resistivity in the normal state around optimal doping)
seems not to notice the Debye temperature at all.  
So we will seek a more exotic explanation.

\subsection{Holographic CFT at finite density and low temperature}

[This part of the story is summarized more amply here  \cite{Faulkner:2011tm}.]
Consider a CFT with gravity dual.
Condensed matter physics is the business 
of putting a bunch of stuff together 
(preferably in the thermodynamic limit) and figuring out what it can do.
So want our CFT to have a notion of stuff.
The vacuum of a CFT is empty; just turning on a finite temperature
produces a weird blackbody plasma, where the density is determined just by the temperature.

Perhaps we can do better if the CFT has a conserved $\gU(1)$ current $j_\mu$.
Then there is a gauge field in the bulk.  
If the current is conserved, this gauge field is massless.
If we can apply our EFT logic to the bulk action, we should 
write a derivative expansion:
\be\label{eq:Einstein-Maxwell} S_\text{bulk}[g, A, ...] = 
\int d^{d}x~dt~dr~\sqrt{g} 
\( {1\over 16 \pi G_N } \( \Lambda + {\cal R}\) 
+ F_{MN}F^{MN} + \cdots \) . \ee
Hence, we are led to study Einstein-Maxwell theory in the bulk.  
Appealing to the calculation outlined around equation \eqref{eq:large-N-disaster}, 
I will call $G_N \sim \color{darkred} N^{-2}$.

The story is more interesting if there are charged fields.  
In particular, we might like to study fermion two-point functions, like $G_R$ above,
to look for a Fermi surface \cite{Lee:2008xf}.
Then we should include in the ellipsis in \eqref{eq:Einstein-Maxwell} a term of the form
$$ S[\psi] = \int \sqrt{g} \bar \Psi \( \ii \Dslash - m \) \Psi .$$

To introduce a finite density of CFT stuff (whatever it is), we should add a chemical potential.  
In the bulk description, this is a boundary condition on $A_t \buildrel{r \to \infty}\over{\to} \mu$.

\hskip-.23in
\parbox{.65\textwidth}{
Then by the magic of holography, we can compute the path integral 
by solving the equations of motion.
One solution with the correct boundary conditions is the AdS-Reissner-Nordstr\"om black hole.
This geometry is a picture of the RG flow induced on the CFT by the chemical potential.
It interpolates between $AdS_{d+2}$ at the UV boundary, 
and a remarkable emergent scaling solution at the IR end, 
which is of the form $ AdS_2 \times \IR^{d}$.  
This geometry is dual 
(and can be used to do calculations in)
a $z=\infty$ IR CFT.  
}
\parfig{.34}{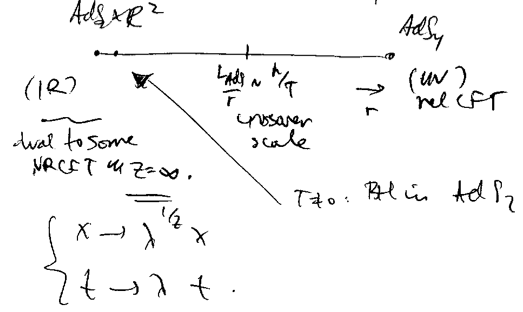}

Then we can use the geometry to calculate the fermion Green's function \cite{Liu:2009dm}.  
The form of the Green's function at low frequency can be understood 
(using the technique of matched asymptotic expansions) to be \cite{Faulkner:2009wj}
$$ G_R(\omega \ll \mu, k) = { a_+(\omega,k)  + \CG(\omega) b_+(\omega,k)
\over  a_-(\omega,k)  + \CG(\omega) b_-(\omega,k) } $$
where $ a_\pm, b_\pm$ are analytic in $\omega$ and $k$, 
and the self-energy is 
$$ \CG(\omega)   = c \omega^{2\nu}  = \vev{\cc^\dagger(\omega) \cc^\nd(\omega)}$$
is the fermion Green's function of a fermionic operator in the IR CFT.
At a value of $k$ with $a_-(0, k_F)  = 0 $, we find a Fermi surface
(in the sense of a surface of zeros of $G_R^{-1}$).
The form of $G_R$ can be reproduced 
by coupling a free Fermi surface (with fermion field $\psi)$
to an IR CFT (with fermion field ${\bf c}$) by an interaction $ \int \psi^\dagger \cc + hc.$
\cite{Faulkner:2009wj, Faulkner:2010tq}.

There are many great features of this construction, perhaps the most significant of which 
is that the destruction of the quasiparticles 
and the dissipation of the current they carry 
are determined by the same power law.
In my opinion, this system is problematic as a toy model of a non-Fermi liquid metal for the following reasons.

First of all, the Fermi surface here is a small part of a much larger system.
I say this because there is no evidence of $k_F$ in the thermodynamics
or transport results at leading order in the $N^2$ expansion, \ie~in classical gravity 
using \eqref{eq:Einstein-Maxwell}.  
I believe it can also be argued that the density of fermions made by the operator 
whose Green function is $G_R$ can be calculated to be $\CO(N^0)$, 
much less than the full charge density, which goes like $ N^2$.  
(Some interesting tension with this last statement is described by 
DeWolfe et al \cite{DeWolfe:2011aa}.
I would love to be wrong about this.)

And speaking of that thermodynamics, it shows the second problem.
The Bekenstein-Hawking entropy of this black hole doesn't go to zero at zero temperature:
$$S(T \to 0, \mu) = S_0 \sim N^2 L^d.$$
There is an extensive zero-temperature entropy, a violation of the third-law of thermodynamics.
The third law of thermodynamics is exactly a manifestation of 
my earlier claim that degeneracy needs a reason, or else it results from fine tuning 
and is unstable.  
(In the case of the Einstein solid above, that low-temperature entropy 
goes away when we include quantum mechanics, and all these 
modes freeze out at $ k_B T  = k_B T_D \sim \hbar \omega_0$.)

This is a black hole information problem!
Semiclassical gravity hides the true quantum groundstate.
It coarse-grains over $ e^{S_0}$ states 
which cannot be distinguished by the weak and ineffectual classical gravity observables.
Like all such confusions about black holes, it is a failure of the limits $N \to \infty$ and $T \to 0$ to commute.

\hskip-.23in
\parbox{.58\textwidth}{
A phase diagram as a function of $1/N$ and temperature then looks like the picture
at right.  
At temperatures above $\mu$, we revert to relativistic CFT or whatever
other UV dragons the system is made of.
The circled regime is where the above analysis of Fermion Green's functions should apply.
$T_c$ indicates a scale at which we might expect to see 
other instabilities, such as a holographic superconductor instability 
\cite{Gubser:2008px, Hartnoll:2008vx}
(if there are charged scalar operators 
of sufficiently low dimension 
\cite{Denef:2009tp}) 
or some spontaneous breaking of translation symmetry (\eg~\cite{Donos:2013gda, Donos:2014uba})
or other phenomena visible in classical gravity (\eg~\cite{Goldstein:2009cv, Kachru:2013voa}). }
\parfig{.39}{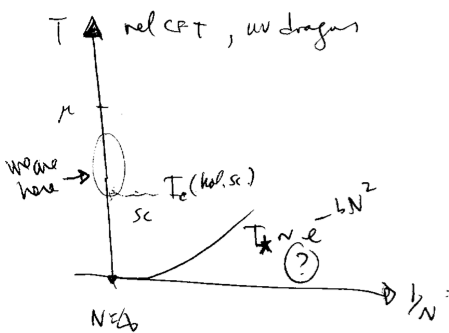} 

If such leading-order-in-$N$ instabilities are somehow absent, 
it seems that the true groundstate can only be seen at temperatures of order $ e^{-b N^2}$,
non-perturbatively small in the gravitational coupling.
Strenuous attempts to find other, more clearly stable
groundstates, for example by including 
the effects of the bulk fermions \cite{Allais:2012ye, Allais:2013lha}, 
have not borne delicious fruit so far.\footnote{More recent work
on the self-destruction of the $AdS_2$ in the IR 
includes \cite{Almheiri:2014cka, Almheiri:2016fws, Engelsoy:2016xyb, Jensen:2016pah,
Maldacena:2016upp}.
}

Maybe I am wrong about this interpretation; that would be great. 
I will mention a tentative hint that I might be in the next subsection.

\subsection{Infinite-range models of $ z=\infty$ fixed points}

The role of $AdS_2 \times \IR^2$ gravity in the previous subsection
was to produce a $z=\infty$ fixed point with a fermion operator $c(t,x)$
of scaling dimension $ \nu + \half$, with $ \nu < \half$ (the free fermion dimension in $D=1$ spacetime dimension).
Another way to accomplish this goal was found by Sachdev, Ye and Kitaev
\cite{2010JSMTE..11..022S, Sachdev:2010um, Kitaev-2015, Sachdev:2015efa, 
Polchinski:2016xgd, Fu:2016yrv, Magan:2016ojb, Jevicki:2016bwu, Maldacena:2016hyu}.
I will not review their calculations
(I recommend the discussion in \cite{Sachdev:2015efa} as a starting point),
but one result is a model with $z=\infty$ scale invariance
(in fact, approximate conformal invariance)
and calculable fermion Green's functions.
(Related work on different models with some 
similar phenomenology includes \cite{2015PhRvB..91t5129Z, Anninos:2016szt}.)

One extra wrinkle is needed to get what we want:
the fermion with $ \nu < \half $ is what Kitaev calls `the bath field', 
$ \tilde c_i = \sum_{jkl} J_{ijkl} c^\dagger_j c^\dagger_k c^\nd_l $.
This is like `alternative quantization' in holography.
At the very least, the model automatically has a relevant
operator which comes from $ \tilde c^\dagger \tilde c^\nd$.


This $z=\infty$ fixed point also has an extensive $T\to 0$ entropy.\footnote{
Does violation of the third law of thermodynamics follow inevitably just from $ z=\infty$?
Looking at the scaling of the energy density with a hyperscaling exponent $\theta$
\eqref{eq:thermo-scaling-theta}, 
$$ C_V = {\partial e \over \partial T} = c_0 T^{ { d- \theta \over z }   } $$
(and using $ S = \int {c_V\over T } dT $)
we see that avoiding a nonzero (or infinite) $S_0$ seems to require
the hyperscaling exponent to diverge.  
Moreover it seems to require 
$ \alpha \equiv \lim_{z \to\infty} {\theta \over z} $
to be finite and negative.
(Thanks to Sean Hartnoll for a conversation about this in Schiphol airport a while ago.)
This is the canonical-ensemble version of the puzzle about 
the scale-invariant density of states described in
\cite{Jensen:2011su, Jensen:2016pah}: 
$ \rho(E)  = e^{ S_0} \delta(E) + e^{ S_1} /E $ 
is the form required by scale invariance.
$S_0 \neq 0$ means degeneracy, but
the second term is not integrable.
}
Is there some sense in which $S_0$ is protected here?
You might think that tuning away the quadratic terms is unnatural.
\cite{You:2016ldz}
shows that exactly this model can arise at the surface of an SPT phase;
the anomalous realization of time reversal symmetry ($ \ii \to - \ii  $ with
{\it no} action on the majorana modes)
completely forbids any quadratic term.
Is this really enough to guarantee an extensive entropy?

The $z=\infty$ fixed point was {\it a} solution of a set of 
dynamical mean field theory equations for a bunch of replicas of the system.  
Often, weird behavior of the entropy in a solution of such equations
is a signal that there are other less-replica-symmetric solutions.  
In similar models, there are quantum glassy solutions \cite{Ye:1992cv, 2000PhRvL..85..840G}.

Kitaev's stated motivation for studying this model was 
as a model that he could show exhibits quantum chaos 
with the same Lyapunov exponent as one gets from a black hole calculation.
So maybe this model has more in common with $AdS_2 \times \IR^2$ than
just $z=\infty$ conformal symmetry. 
And perhaps the entropy is not just a degeneracy waiting to be split:
in a single groundstate (found by exact diagonalization), 
the entanglement entropy of a subsystem of $k$ spins
seems to grow linearly with $k$ \cite{Fu:2016yrv, Tarun:2016}.

\subsection{Final words}

\begin{itemize}

\item{} {\bf Phenomena.}  Much moreso 
than trying to make predictions for experiments using holographic duality, 
a realistic goal is simply to make contact with {\it phenomena},
and to use these toy models to understand mechanisms by which physics happens.  
Thermalization and the onset of chaos are good examples of this
(as confinement and chiral symmetry breaking were for the use of holography
to study the vacuum of gauge theory).

I'm feeling a bit bad about how negative I've been about physical applications of holographic duality above.
To counterbalance this, here are two recent, concrete, positive outcomes of work
in this direction.

The first is the work of Hartnoll and Karch 
to develop a scaling theory of transport in strange metals \cite{Hartnoll:2015sea}.
Before holography, many folks argued against an anomalous 
dimension for the current operator,
but holography makes this possibility clear (in retrospect), 
and it is a crucial ingredient in their
(otherwise completely non-holographic) story.

The second is the remarkable recent suggestion made by 
Son \cite{Son:2015xqa}, which has revised our understanding
of the best-understood non-Fermi liquid in nature
(what I called the `HLR state' above).
Rumor has it that Son's idea was inspired by trying to do holography for Landau levels.
The idea (in the end, holography-independent) is that the composite fermion in the half-filled Landau level 
is a Dirac particle (in the sense that there is a Berry phase of $\pi$ 
around the Fermi surface).
This idea, which seems to be correct \cite{2015arXiv150804140G}, 
makes possible a particle-hole symmetric
description of the state,
and has led to some dramatic insights, including
new field theory dualities 
and new ideas about spin liquid states 
and phase diagrams of thin films \cite{Wang:2016fql, Wang:2015qmt, Metlitski:2015eka, 2015arXiv151008455M, Mulligan:2015zua, Mulligan:2016rno, Karch-Tong-2016, Murugan-Nastase-2016, 
TheEndOfTheWorld, Metlitski:2015yqa}.

\item{} {\bf Long-range interactions.} From the beginning in this discussion, 
I have emphasized the study of well-defined lattice models 
with {\it local} hamiltonians.  This is a huge class of systems about which much 
can be said and which describe many real physical systems and about which there is much more to be learned.
In non-relativistic physics, however, the restriction to local interactions
is not as well-motivated by physics as it is by cowardice:
the coulomb interaction is actually long-ranged.
It is often screened, but not so well in an insulator.  
It is scary because, for example, in the presence of long-range interactions,
a gap does not imply small correlation length.  
To give a real example, the range of the interactions has a big effect on
the physics of the half-filled Landau level (it changes the exponent in $C_V \sim T^{\alpha}$).
There is more physics to be found here.

\item{} {\bf Beyond groundstates.}  Another point of cowardice is the focus on groundstates in these lectures.  
The frontier of excited states, finite energy density states,
and driven systems awaits the application of these methods. 
Even steady states of driven quantum systems are not understood.
To put it simply: 
like many other things we learned in freshman physics (like dimensional analysis), 
Ohm's law is a deep statement.  

\item{} {\bf Gravitational order.} An alternative, more precise, definition of topological order which is often used in the case of gapped systems is:
the system has a set of degenerate groundstates  $ \ket{\alpha} $ 
(the code subspace)
between which no local operator can distinguish (in the thermodynamic limit, $L\to\infty$): 
$$ \bra{\alpha} \CO \ket{\beta} = c \delta_{\alpha\beta} e^{ - L^a } $$
where $c, a$ are some constants.  
(This is the starting point of some rigorous proofs of the stability of topological order \cite{2011CMaPh.307..609B, 2010JMP....51i3512B}.)

Thinking about this definition,
it's tempting to point out that a gravitational theory
(at least a generally covariant one)
satisfies this definition, in a somewhat silly way: 
there {\it are} no local operators $\CO$.  

Some progress on this kind of idea 
(of the phase of GR with a gapless metric variable as 
an ordered phase)
was made recently in 
\cite{Hartnoll:2013sia}.  Specifically, they pursued
the analogy between Polyakov's confinement of 
3d abelian gauge theory by monopole instantons \cite{Polyakov:1976fu},
and the effects of gravitational instantons.

This brings us to the fascinating and mysterious notion of the alternative, a {\it disordered} phase of gravity, 
where $ ds^2=0$ or something like that
(I learned about this idea from the last section of \cite{Witten:1988ze}).
Previous encounters with this idea include Bubbles of Nothing \cite{Witten:1981gj}
(with a capital `N'), and 3d Chern-Simons gravity \cite{Witten:1988hc}.
Some progress on Theories of Nothing was made by 
studying the interface between the Nothing phase and the phase
with gravitational order \cite{Adams:2005rb, McGreevy:2005ci, Hellerman:2006ff}.

I should probably also mention here the idea of induced gravity, 
as reviewed \eg~here \cite{Adler:1982ri}.  
This is where we introduce a dynamical metric 
on a $d$-dimensional space, 
with no kinetic terms 
(confined at short distances) and couple it to some extensive matter; 
the matter fluctuations will inevitably induce an Einstein-Hilbert term,
which can lead to deconfinement of the metric variable at long distances.  
This sounds a lot like the description I gave of the parton construction in \S\ref{sec:partons}.
There are (at least) two important differences I know about: 
I don't know a convincing origin of the metric degree of freedom.
Also, the emergence of general covariance seems more difficult than the emergence of 
a simple local gauge invariance.
But most importantly: this idea is not holographic enough.  
We are beginning with a Hilbert space with too many degrees of freedom -- 
extensive in $d$ dimensions.  Maybe there is some loophole that the required
couplings required to emerge the metric dynamics
necessarily thin out the low-energy degrees of freedom to only $(d-1)$-dimensions'-worth.

The connection with topological field theory (TFT) is interesting also because 
a gravitational theory (which is background independent) 
on some topological space 
is a topological theory (maybe not so much a field theory non-perturbatively)
-- the physics does not depend on a choice of metric.
Such a theory is independent of the metric the hard way --
we have to integrate over metrics.  
There is another class of topological theories 
where the metric never appears in the first place \cite{Witten:1988ze}.
One might have hoped that the kind of TFTs that 
arise by twisting supersymmetric gauge theories
might be useful as possible low-energy theories
of gapped phases of matter, but 
they tend not to be unitary.  
Specifically, in examples I've looked at \cite{Rozansky:1996bq} (with Brian Swingle), 
the partition function on some space $X$
times a circle, which should count groundstates
on $X$, is not an integer.  
So, at least in these examples, there seems to be no suitable Hilbert space interpretation.

\end{itemize}

{\bf Acknowledgements}

These lectures were given at \htmladdnormallink{TASI 2015}{https://sites.google.com/a/colorado.edu/tasi-2015-wiki/}.
Thanks to Joe Polchinski and Pedro Vieira 
for this opportunity to convey my views about
applied holography, 
to Tom DeGrand and Oliver DeWolfe for their hospitality and helpful comments,
and to the TASI 2015 students for their attention and enthusiasm.
I've used figures made by Brian Swingle,
Dan Arovas and Daniel Ben-Zion.
Some of the logic of the discussion of 
topological order in \S\ref{sec:gauge-fields} is inspired by 
lectures by Chetan Nayak and by Mike Hermele.
Some of the logic in \S\ref{sec:SPT} 
follows lectures by Andreas Karch.
Some of the material in \S\ref{sec:SPT} 
was presented at the 2013
Arnold Sommerfeld Summer School
on Gauge/Gravity Duality.
Some of the material is cannibalized from my lecture notes for
courses on \htmladdnormallink{field theory}{http://physics.ucsd.edu/~mcgreevy/s15/},
its \htmladdnormallink{emergence from lattice models}{http://physics.ucsd.edu/~mcgreevy/s14/}
and its \htmladdnormallink{connections with quantum information theory}{http://physics.ucsd.edu/~mcgreevy/s16/}.
Thanks especially to Brian Swingle for his many contributions to my understanding of physics.
My work was supported in part by
funds provided by the U.S. Department of Energy
(D.O.E.) under cooperative research agreement 
DE-SC0009919.

\breakS

\appendix
\renewcommand{\theequation}{\Alph{section}.\arabic{equation}}
\section{Appendix to \S\ref{sec:gt}: $p$-form $\IZ_N$ toric codes}
\label{sec:p-form-appendix}

After learning about the toric code, it is hard not to ask about simple generalizations.  
Here are some.  I am not sure of the correct reference for these constructions.
The $\IZ_N$ generalization of the 1-form  toric code 
can be found in this paper
\cite{Schulz:2011tk}.
The $p$-form toric code in $d$ dimensions is described in this paper
\cite{2014PhRvL.112g0501H}.

\subsection{Simplicial complexes and simplicial homology}

By a $d$-dimensional {\it simplicial complex} $\DDD$ we will mean 
$$ \DDD  = \{ \cup_{p=0}^d \Delta_p , ~ \partial \} $$
Here $\Delta_p$ is a collection of $p$-dimensional polyhedra
which I will call $p$-simplices (though they are not necessarily made from triangles).
$\partial$ is an oriented boundary map:
$$\partial:  \Delta_p \to \Delta_{p-1}, \sigma \to \partial \sigma$$
where the signs make Stokes theorem work.
Notice that $\partial^2 = 0$ -- the boundary of a $p$-simplex
has no boundary.

If we think of this complex as a triangulation
(or really a shape-u-lation for more general shapes)
of a smooth manifold $X$,
then a point in life of this machinery is that 
it computes the {\it homology} of $X$ -- a collection 
of (abelian) groups $H_p(X, \IZ)$ which are topological invariants of $X$.
To define these groups, we should introduce one more 
gadget, which is a collection of vector spaces 
$$ \Omega_p(\DDD, \IZ_N) , ~p= 0...d \equiv \text{dim}(X) $$ 
basis vectors for which are $p$-simplices: 
$$ \Omega_p(\DDD,\IZ_N) = \text{span}_{\IZ_N} \{   \sigma \in \Delta_p \} $$
-- that is, we associate a(n orthonormal) basis vector 
to each $p$-simplex (which I've just called $\sigma$), and these vector spaces
are made by taking linear combinations of these spaces,
with coefficients in $\IZ_N$.
Such a linear combination of $p$-simplices is called a $p$-{\it chain}.
It's important that we can {\it add} (and subtract) $p$-chains, $ C + C' \in \Omega_p$.
A $p$-chain with a negative coefficient can be regarded as having the opposite orientation.
We'll see below how better to interpret the coefficients.

The boundary operation on $\Delta_p$ induces one on $\Omega_p$.
A chain $C$ satisfying $ \partial C = 0 $ is called a {\it cycle}, and is said to be {\it closed}.

So the $p$th is homology group is equivalence classes of $p$-cycles,
modulo boundaries of $p+1$ cycles:
$$ H_p(X, \IZ) \equiv
{ \text{ker}\( \partial:  \Omega_p \to \Delta_{p-1} \) \subset \Omega_p  \over \text{Im}\( \partial: 
\Omega_{p+1} \to \Omega_p \)  } $$
This makes sense because $\partial^2 = 0 $ --
the image of $\partial: 
\Omega_{p+1} \to \Omega_p $ 
is a subset of  $\text{ker}\( \partial:  \Omega_p \to \Omega_{p-1} \)$.
It's a theorem that the dimensions of these groups
are the same for different (faithful-enough) discretizations of $X$.
For proofs, see the great book by Bott and Tu, {\it Differential forms in algebraic topology}.


It will be useful to define
a `vicinity' map $v$ which goes in the opposite direction
from $\partial$ (but is not the inverse):
\bea 
v: \Delta_p &\to& \Delta_{p+1},
\cr  \sigma &\mapsto& v(\sigma) \equiv \{ \mu \in \Delta_{p+1} | \partial \mu  = + \sigma +  \text{anything}  \} 
\eea
-- it picks out the $p+1$-simplices in whose boundary the $p$-simplex appears with a $+1$\footnote
{For this to work out, it will be useful to assume that the coefficients in the boundary map are only $ \pm 1, 0$.}.

\subsection{$p$-form toric code}

Consider putting a spin variable on the $p$-simplices of $\DDD$.
More generally, let's put an $N$-dimensional hilbert space 
$\CH_N \equiv \text{span} \{ \ket{n}, n=1..N \}$
on each $p$-simplex,
on which act the operators
$$ \XX \equiv \sum_{n=1}^N \ket{n}\bra{n} \omega^n 
= \begin{pmatrix} 1 & 0 & 0 & \dots \cr 0 & \omega & 0 & \dots \cr 0 & 0 & \omega^2 & \dots \cr
0 & 0 & 0 & \ddots  \end{pmatrix}, ~~~~
\ZZ \equiv \sum_{n=1}^N \ket{n}\bra{n+1}
=
\begin{pmatrix} 0 & 1 & 0 & 0 \cr 
 0 & 0 & 1  & 0 \cr 
  \vdots & \vdots & \vdots & \ddots \cr 
   1 & 0 & 0 & \hdots \cr \end{pmatrix} 
$$
where $\omega^N=1$ is an $n$th root of unity.
These satisfy the clock-shift algebra:  $\XX\ZZ = \omega \ZZ\XX$.
For $N=2$ these are Pauli matrices and $\omega=-1$.

Consider the Hamiltonian
$$ \HH = - J_{p-1} \sum_{ s\in \Delta_{p-1} } A_{s}
- J_{p+1} \sum_{\mu \in \Delta_{p+1} } B_{\mu}
- \Gamma_p \sum_{ \sigma \in \Delta_p} \ZZ_\sigma ~~$$
with 
$$ A_s \equiv \prod_{\sigma \in v(s) \subset \Delta_p} \ZZ_\sigma $$
$$ B_\mu \equiv \prod_{ \sigma \in \partial \mu} \XX_\sigma ~.$$
 
 I claim that 
 $$ 0 = [A_s, A_{s'}] = [B_\mu, B_{\mu'} ] = [A_s, B_\mu], ~~\forall s, s', \mu, \mu' $$
so that for $ \Gamma_p = 0 $ this is solvable.

Here's the solution:
Suppose $ J_{p-1} \gg J_{p+1}, \Gamma_p$ so that 
we should satisfy $A_s = 1$ first.
This equation is like a gauss law, but instead of 
flux {\it lines} in the $p=1$ case,
we have flux {\it sheets} for $p=2$ or ... whatever they are called for larger $p$.
The condition $A_s=1$ means that these sheets satisfy a conservation law
that the total flux going {\it into} the $p-1$ simplex vanishes.
So a basis for the subspace of states satisfying this condition
is labelled by configuration of closed sheets.
For $N=2$ there is no orientation, and each $p$-simplex 
is either covered ($\ZZ_\sigma = -1$) or not ($ \ZZ_\sigma = 1$)
and the previous statement is literally true.
For $N>2$ we have instead sheet-nets (generalizing string nets),
with $N-1$ non-trivial kinds of sheets labelled by $ k = 1... N-1$
which can split and join as long as they satisfy 
\be\label{eq:star} \sum_{\sigma \in v(s)} k_\sigma  = 0 \text{~mod~} N ,
\forall s .\ee
This is the Gauss law of $p$-form $\IZ_N$ gauge theory.

The analog of the plaquette operator $B_\mu$ acts like a 
kinetic term for these sheets.
In particular, consider its action on a basis state for the $A_s= 1$ subspace
$ \ket{ C} $, where $C$ is some collection of ($N$-colored) closed $p$-sheets --
by an $N$-colored $p$-sheet, I just mean 
that to each $p$-simplex we associate an integer $k_\sigma$ (mod $N$),
and this collection of integers satisfies the equation \eqref{eq:star}.

The action of the plaquette operator in this basis is 
$$ B_\mu \ket{C} = \ket{C + \partial \mu } $$
Here $C + \partial\mu $ 
is another collection of $p$-sheets differing from $C$ by the addition (mod $N$) 
of a sheet on each $p$-simplex appearing in the boundary of $\mu$.
The eigenvalue condition $B_\mu = 1$ 
then demands that the groundstate wavefunctions $\Psi(C) \equiv \vev{C | \text{groundstate}} $ 
have equal values for chains $C$ and $C' = C + \partial \mu$.
But this is just the equivalence relation defining the $p$th homology of $\DDD$.
Distinct, linearly-independent groundstates are the labelled by
$p$-homology classes of $\DDD$.
More precisely, they are labelled by homology with coefficients in $\IZ_N$, $H_p(\DDD, \IZ_N)$.

We can 
reinterpret the toric code above as a 
$p$-form $\IZ_N$ gauge theory
with  `electric' matter
by associating an $\CH_N$ to each $\ell \in \Delta_{p-1}$;
to bring out the similarity with 
\cite{PhysRevB.71.125102},
I'll call its $\ZZ$ operator $\Phi_\ell$.
Notice that $ \Phi_{-\ell} = \Phi_{\ell}^\dagger$.
We can introduce $\IZ_N$ gauge transformations 
$$\Phi_{\ell} \mapsto \omega_\ell \Phi_\ell, ~~~ 
\ZZ_\sigma \mapsto \prod_{\ell \in \partial \sigma} \omega^\ell\ZZ_\sigma $$
(notice that the latter generalizes the transformation of a link variable,
in which case the boundary of the link is the difference of the two sites at its ends),
in which case the coupling
$$ 
\HH_\text{e} = \sum_{\sigma \in \Delta_p}  \prod_{\ell \in \partial \sigma} 
\Phi_\ell \ZZ_\sigma 
$$
is gauge invariant.
Now notice that we may choose {\it unitary gauge} 
where we completely fix the gauge redundancy by setting $ \Phi_\ell = 1$.
This produces the $p$-form toric code above.
(For the case $p=1, N=2$, this is explained in Fradkin's book, 2d edition.)

$p$-form discrete gauge theory 
is described, for example, in the appendix of \cite{Maldacena:2001ss}.

It is interesting to consider other possibilities
the collection of simplices on which the matter resides.
For example, put a spin on every simplex.
With the appropriate hamiltonian,
this should compute the whole homology complex $ H_\bullet(\DDD, \IZ_N) = \oplus_{p=0}^d H_p(\DDD, \IZ_N) $.

\subsection{Lattice duality}
\label{sec:lattice-duality}

Consider the {\it quantum clock model} on one of the simplicial complexes described above.
This model has $\CH = \otimes_{s \in \Delta_0} \CH_N $, $N$ states on each site.
The Hamiltonian is
\be\label{eq:clock} \HH_\text{clock} =  - \beta \sum_{\ell \in \Delta_1, \partial\ell = i-j}  \ZZ_i \ZZ_j^\dagger
- \Gamma \sum_{i \in \Delta_0} \XX_i  + h.c. ~.\ee
For $N=2$, this is the quantum transverse-field Ising model,
discussed in \eg\ \cite{SubirBook}.
\footnote{I learned recently that I should call this a clock model, not a Potts model.
The quantum Potts model is defined on the same Hilbert space, but with the Hamiltonian
$$ 
\HH_\text{Potts} = 
 - \beta \sum_{\ell \in \Delta_1, \partial\ell = i-j}  
 \sum_{p=1}^N \ZZ^p_i \(\ZZ_j^{p}\)^\dagger
- \Gamma \sum_{i \in \Delta_0} \sum_{p=1}^N \XX_i^p + h.c. ~.
$$
I haven't thought about this one yet.
They both reduce to the quantum transverse-field Ising model for $N=2$.
}
Notice that if we diagonalize $\ZZ_i = e^{ \ii 2 \pi k_i /N}, k_i = 1.. N$, then
$$\ZZ_i^\nd \ZZ_j^\dagger + h.c. = 2 \cos{2 \pi \over N}\( k_i - k_j \) ~.$$

In any $d$, the following (hodge or EM) duality 
(due to Wegner \cite{Wegner1971})
maps this model 
to a gauge theory on the {\it dual graph} $\DDD^\vee$, defined by
associating a $(d-p)$-simplex $\sigma^\vee_{d-p}$ to each $p$-simplex $\sigma_p$ of $\DDD$,
roughly in such a way that they combine to form a volume element.
More precisely, the boundary map on $\DDD^\vee$ can be defined 
by
$$ \boxed{\partial  \(\sigma^\vee_p\) \equiv \( v(\sigma_p) \)^\vee  \in \Delta^\vee_{d-p-1} } \cob~.$$

I will describe first the case where the variables live on the sites of $\DDD$
and then we can figure out the general construction.
The idea is to put a variable
on the links of $\DDD^\vee$ which 
keeps track of the {\it change} of $\ZZ$ across the link;
for $N=2$, this is a sign which is $-1$ only if $\ZZ_i$ and $\ZZ_j$ disagree: 
$$\IZ_2: ~~~ \ttau^z_{ij} \equiv \ZZ_i \ZZ_j .$$
For $N>2$, $\ZZ_i$ and $\ZZ_j$ can differ by a phase, so 
the domain wall takes values in $N$-th roots of unity:
$$ \ttau^z_{ij} \equiv \ZZ_i^\nd \ZZ_j^\dagger . $$

Not all configurations of the $\ttau^z_\ell$s are attainable by this map.
If we label the edges of the dual lattice by a color according to $k_\ell$ in $ \ttau_\ell^{z} = \omega^{k_\ell} $, then
these edges form closed loops in the same sense as above (\ie\
literally closed, unoriented loops for $N=2$, and more generally $\IZ_N$-string-nets).
This is just the familiar fact that level-sets of a function are collections of closed curves.
The allowed configurations of $\ttau_\ell$s satisfy 
$$1 =  \prod_{ \ell \in \partial p^\vee} \ttau_\ell^z , \forall p^\vee \in \Delta_d^\vee, 
~~~
\text{or}~~~  1 = \prod_{\ell \in v(s) } \ttau_\ell^z, \forall s \in \Delta_0, 
~~~\text{or}~~~~\sum_{\ell \in v(s) } k_\ell = 0 \text{~mod~}N. $$
This is just the star condition, \ie\ the gauss law,
for a gauge theory.
In 1+1 dimensions, this condition is empty
and the model is self-dual, as observed by Kramers and Wannier for $N=2$.

Further, the $\ttau^z$s do not completely specify the configurations of the $\ZZ_i$: 
if we act by $ \prod_{i \in \Delta_0} \XX_i$,
to rotate {\it every} site by $\omega$, nothing happens to the $\ttau^z_{ij}$.
This is just one global integration constant in $\IZ_N$.

More generally, we will want a duality equation like:
$$ \ttau_{\sigma}^z \equiv \prod_{\ell \in \partial \sigma^\vee} \ZZ_{\ell} . $$

\subsection{$N \to \infty$}

The case where we replace $\IZ_N$ with $U(1)$ merits
separate discussion.
In that case the state space runs over all integers $ \CH_\infty = \text{span}\{\ket{n}, n \in \IZ \} $, 
\ie\ the Hilbert space of a U$(1)$ rotor.
Another useful basis is the theta-vacua aka Bloch waves: 
$$ \ket{\theta} = \sum_n e^{ \ii n \theta} \ket{n} ,~~~~~ \theta \equiv \theta + 2\pi~.$$ 
We can think of $\theta$ as the direction in which the rotor is pointing.


In this case, we need no longer write the `mod $N$'  in
the star condition
$$0 =  \sum_{\ell \in v(s)}  n_\ell  ~.$$
This is more obviously a lattice version of the gauss law condition $ 0 = \grad \cdot \vec E$ 
for E$\&$M.
A term which imposes this condition energetically just as well as $A_s = 1$ 
is the first term in 
$$ \HH = 
- \sum_{s\in \Delta_{p-1}} \( \sum_{ \sigma \in v(s)} n_\sigma \)^2 
- \sum_{\mu \in \Delta_{p+1} } \prod_{\sigma \in \partial \mu} e^{ \ii \theta_\sigma} ~~+ h.c..
$$
where 
$$ [n_\sigma,  e^{ \pm \ii \theta_{\sigma'}}] =  \pm e^{ \ii \theta_{\sigma}} \delta_{\sigma, \sigma'} $$
-- \ie\ $e^{ \pm \ii \theta}$ are raising and lowering operators.
This second term more obviously approaches 
$ \cos \nabla \times \vec A $ 
in the continuum limit.

Notice that for any finite $N$ there
are two conjugate $\IZ_N$ operations we might consider, 
one generated by $ \XX$, which acts by $ \CO \to \XX \CO \XX^\dagger$, 
so in particular
$$ \IZ_N^X:  \ZZ \to \omega \ZZ, \XX \to \XX, $$
and one generated by $\ZZ$ which acts by 
$$ \IZ_N^Z:  \ZZ \to \ZZ, \XX \to \omega \XX.$$
In the limit $N\to\infty$, one of these acts by
the U$(1)$ transformation $ \theta \to \theta + \eps$.

\breakS
\bibliographystyle{ucsd}
\bibliography{collection}
\blankpage
\end{document}